\documentclass{jfm}

\usepackage{graphicx}
\usepackage{newtxtext}
\usepackage{newtxmath}
\usepackage{natbib}
\usepackage{hyperref}
\usepackage{siunitx}
\hypersetup{
    colorlinks = true,
    urlcolor   = blue,
    citecolor  = black,
}

\newcommand{\RomanNumeralCaps}[1]
\linenumbers

\usepackage{orcidlink}
\usepackage{bm}
\usepackage{stmaryrd}
\usepackage{nicefrac}

\newcommand{\bcdoubledot}{\bm{:}}

\DeclareMathAlphabet{\mathsfbi}{OT1}{\sfdefault}{bx}{sl}
\DeclareMathVersion{sfletters}
\SetSymbolFont{letters}{sfletters}{OML}{ntxsfmi}{b}{it}

\makeatletter
\newcommand{\bmsbilow}[1]{%
  \text{\mathversion{sfletters}$\m@th#1$}%
}

\DeclareRobustCommand{\tensor}[1]{%
  \begingroup
  \ifcat\noexpand #1\relax
    \edef\greek@test{\detokenize{#1}}%
    \edef\greek@test{\expandafter\@cdr\greek@test\@nil}%
    \edef\greek@test{\expandafter\@car\greek@test\@nil}%
    \edef\x{\the\lccode\expandafter`\greek@test}%
    \edef\y{\number\expandafter`\greek@test}%
    \ifnum\x=\y\relax
      \bmsbilow{#1}%
    \else
      \mathsfbi{#1}%
    \fi
  \else
    \mathsfbi{#1}%
  \fi
  \endgroup
}

\makeatletter
\def\absfooterflag{}    % removes the page-1 footer flag
\def\pagelimitfooter{}  % removes the page-4 and page-10 footer flags
\makeatother

\title{Nonlinear electrohydrodynamics of a surfactant-laden leaky dielectric drop}

\author{Michael A. McDougall\,\orcidlink{0009-0005-2967-9437}\aff{1}
  \corresp{\email{michael.mcdougall@strath.ac.uk}},
  Stephen K. Wilson\,\orcidlink{0000-0001-7841-9643}\aff{2,1}\corresp{sw3197@bath.ac.uk}
 \and Debasish Das\,\orcidlink{0000-0003-2365-4720}\aff{1}\corresp{debasish.das@strath.ac.uk}
}
\affiliation{\aff{1}Department of Mathematics and Statistics, University of Strathclyde, Livingstone Tower, 26 Richmond Street, Glasgow G1 1XH, United Kingdom
\aff{2}Department of Mathematical Sciences,
University of Bath,
Claverton Down,
Bath BA2 7AY,
United Kingdom}

\begin{document}
\maketitle

\begin{abstract}
A nonlinear three-dimensional small-deformation theory is presented for a leaky dielectric drop coated with a dilute monolayer of insoluble apolar surfactant and subjected to a uniform DC electric field. The theory is developed within the framework of the Taylor--Melcher leaky dielectric model, and builds on previous work by retaining surface charge convection in the charge conservation equation. Solving the problem in three dimensions and retaining charge convection allows us to capture the transition to Quincke rotation, a symmetry breaking instability wherein a drop begins rotating at a steady angular velocity when the applied electric field strength exceeds a critical value.  We derive a system of coupled nonlinear ordinary differential equations for the drop shape, dipole moment, and surfactant distribution, which we solve numerically. We discuss the combined effects of charge convection and surfactant in the Taylor regime -- in which the field strength is too weak to induce Quincke rotation and the drop adopts an axisymmetric spheroidal shape. In the Quincke regime, we find that the presence of a weakly-diffusing surfactant results in a lower critical electric field than that for a drop with uniform surfactant coverage. Varying the elasticity number, which quantifies the variation of the surface tension as a function of the surfactant concentration, can either increase or decrease the critical field strength depending on the diffusivity of the surfactant. Additionally, we find that the experimentally observed hysteresis in the angular velocity of the drop can disappear when surfactant diffusion is sufficiently weak.
\end{abstract}

\begin{keywords}
Authors should not enter keywords on the manuscript, as these must be chosen by the author during the online submission process and will then be added during the typesetting process (see \href{https://www.cambridge.org/core/journals/journal-of-fluid-mechanics/information/list-of-keywords}{Keyword PDF} for the full list).  Other classifications will be added at the same time.
\end{keywords}

{\bf MSC Codes }  {\it(Optional)} Please enter your MSC Codes here

\section{Introduction}
\label{Introduction}
Surfactants (surface active agents) are a class of chemical compounds that reduce the surface tension between a liquid and a second medium, and are widely employed across several industries for their useful interfacial properties. Among their uses are emulsification, often exploited for food processing~\citep{kralova2009surfactants}; wetting, used to develop improved agricultural chemicals like pesticides and fertilisers~\citep{kovalchuk2021surfactant}; foaming, used to develop a range of personal care products~\citep{bureiko2015current} as well as aqueous film-forming foams for firefighting~\citep{moody2000perfluorinated,kovalchuk2021surfactant}; and dispersion, preventing particle agglomeration in suspensions~\citep{vaisman2006role,lotya2010high}. Moreover, the influence of surfactant on the hydrodynamics of drops is a fundamental scientific problem in its own right. Many experimental, theoretical, and computational studies have sought to elucidate the impact of surfactants on drop deformation, breakup, and motion, amongst other phenomena. In a number of cases, a quantitative understanding of the effects of surfactant is particularly desirable because these effects often contribute to discrepancies between theoretical predictions and observations in experiments, where drop systems are rarely truly clean. A classic example is that of a small drop sedimenting in an unbounded fluid, which has been repeatedly shown to obey Stokes' law for translating spheres rather than the Hadamard--Rybczynski equation~\citep{savic1953circulation,griffith1962effect}. This is explained by the presence of small amounts of surfactant that are convected to the trailing tip of the sedimenting drop, where they give rise to a stagnant cap, a region of the interface which is immobilised by the dense packing of surfactant molecules. The gradient in surfactant coverage leads to a gradient in surface tension, in turn leading to Marangoni stresses that suppress the surface flow and thereby decrease the terminal velocity of the drop~\citep{levich,davis1966influence,sadhal1983stokes}.

Surfactants also play an important role in drop deformation and affect fluid flows both inside and outside the drop. A framework for calculating the shape and flow fields of an almost spherical drop in an arbitrary flow field accounting for surfactant was presented by~\citet{haber1972hydrodynamics}. \citet{greenspan1977deformation} calculated the small deformations of a drop in an initially quiescent fluid with a prescribed deposition of surfactant to the interface. \citet{flumerfelt1980effects} developed a small-deformation theory for drops in shear or extensional flows that accounts for varying surface tension as well as surface viscosities, which resolved longstanding discrepancies between theoretical predictions~\citep{taylor1932viscosity,taylor1934formation,cox1969deformation,chaffey1967second,barthes1973deformation} and experimental observations~\citep{nawab1958viscosity,bartok1959particle,rumscheidt1961particle} regarding the fluid circulation inside drops in shear flows. \citet{sadhal1986deformation} calculated the deformation of a sedimenting drop with an arbitrary prescribed surfactant distribution, and in the case of a stagnant cap distribution found that the drop deforms into a prolate spheroid. \citet{stone1990effects} studied the deformation and breakup of a surfactant-laden drop in an extensional flow using a boundary integral method accompanied by a small-deformation theory. This work was later generalised by~\citet{milliken1993effect} to capture transient breakup dynamics and to treat drop-fluid systems with nonidentical viscosities as well as interactions between surfactant molecules. The results of these studies highlighted two competing effects of variable surface tension, the balance of which determines the ultimate effect of the surfactant on the deformation. The first of these, known as tip stretching, occurs when surfactant is convected by the fluid flow away from the equator and towards the tips of the drop. Thus, the surface tension is decreased at the tips, where the interface must assume a greater degree of curvature to balance the normal stresses, leading to increased deformation when compared to a drop with uniform surface tension. This effect is particularly relevant when surfactant convection is strong relative to diffusion (large surface P\'eclet number). The second effect, known as surface dilution, occurs due to the increasing surface area of the drop that accompanies deformation. The surfactant concentration is globally diluted, leading to a global increase in the surface tension and hence reduced deformation when compared to a drop with uniform surface tension, and often a higher critical strain rate for breakup. This effect is particularly relevant when diffusion is strong relative to convection (small surface P\'eclet number) or when the surface tension is highly sensitive to gradients in the surfactant concentration (in which case the resulting strong Marangoni stresses oppose the convective flux).
The work of~\citet{stone1990effects} assumes that the surfactant is insoluble in the bulk fluids and is present in dilute quantities only on the interface. 
In addition, surfactant solubility \citep{milliken1994influence} and nondilute interfacial concentrations \citep{pawar1996marangoni,eggleton1998adsorption,eggleton1999insoluble} have also been investigated.
\citet{vlahovska2005deformation} extended the theory of~\citet{stone1990effects} to third order in the deformation. 
Other related studies include those on compound droplets \citep{mandal2016effect} and the influence of Marangoni stresses on the rheology of emulsions of surfactant-laden drops in linear flows \citep{mandal2017effect}.

Drop dynamics can also be controlled externally by applying an electric field. Depending on the electrical and mechanical properties of the fluids, such fields can induce a wide range of behaviours. To name a few, drop deformation~\citep{taylor1966studies,ajayi1978note}, breakup~\citep{taylor1964disintegration,torza1971electrohydrodynamic,sherwood1988breakup} or coalescence~\citep{allan1961effects,eow2003drop,ristenpart2009non} can be either promoted or inhibited, translational motion can be induced~\citep{baygents1991electrophoresis,eow2003motion,bandopadhyay2016uniform}, and emulsions of drops can exhibit various non-Newtonian behaviours, including significant changes to the apparent viscosity of the emulsion~\citep{pan1997characteristics,vlahovska2011rheology}. A particularly extensive body of literature exists for the case of an
uncharged, weakly conducting (leaky dielectric) drop immersed in another fluid.
Early theories by \citet{okonski_thacher} and \citet{allan1962particle} assumed both fluids to be perfectly insulating and predicted that an initially spherical drop would elongate into a prolate spheroid along the field direction, owing to a balance between the electric stress normal to the interface and surface tension. \citet{taylor1966studies} proposed a leaky dielectric model (LDM) assuming that both fluids are electrically neutral and that free charge resides only at the drop surface. The surface charge density then satisfies a transport equation balancing a transient charging term, the Ohmic current between the bulk fluids, and charge convection by the interfacial velocity.
The resulting charge distribution leads to an electric stress that has a component tangential to the drop surface, which cannot be balanced by uniform surface tension and therefore drives circulatory flows inside and outside the drop.
At leading order in the drop deformation, the drop assumes an axisymmetric spheroidal shape that may be either prolate or oblate, depending on the ratios of the conductivities, permittivities, and viscosities of the two media. The induced fluid flow is also axisymmetric, and at the drop surface may be directed either from the poles of the drop to the equator or from the equator to the poles, depending on the ratios of the electrical properties of the fluids. This distinction leads to the common classification of leaky dielectric drops as one of prolate A (flow from the equator to the poles), prolate B (flow from the poles to the equator), and oblate (in which case the flow is always from the poles to the equator)~\citep{lac2007axisymmetric}.

The theoretical predictions of \citet{taylor1966studies} were accurate to first order in the electric capillary number (which measures the relative strengths of electric and capillary stresses) and thus valid only for small deviations of the drop from spherical shape. They underestimated the deformations observed in the experiments of \citet{torza1971electrohydrodynamic}. \citet{ajayi1978note} extended the small-deformation theory to second order in electric capillary number, but this refinement did not explain the discrepancy.
Shortly thereafter, \citet{vizika1992electrohydrodynamic} conducted new experiments that showed substantially better agreement with the theory of \citet{taylor1966studies} than did the earlier measurements of \citet{torza1971electrohydrodynamic}. Some residual discrepancies persisted for certain material systems, which the authors attributed to possible electric-field inhomogeneities or to surface charge conduction or convection neglected in the theoretical models. Out of these effects, charge convection has been shown to improve predictions of drop deformation. However, it is difficult to include in asymptotic models because it is inherently nonlinear. Hence, many existing works which retain charge convection use computational techniques including finite element~\citep{feng1999electrohydrodynamic,supeene2008deformation}, Volume-of-Fluid~\citep{lopez2011charge,dong2023unsteady}, boundary element~\citep{lanauze2015nonlinear,das2017electrohydrodynamics}, and Lattice Boltzmann methods~\citep{basu2024role}. 
These studies found that charge convection generally increases deformation for prolate drops and decreases deformation for oblate drops. The first of these is explained by the fact that prolate deformation is primarily due to the normal electric stresses on the surface, while viscous stresses associated with the electrohydrodynamic flow are less significant. Charge convection from the drop equator to the poles tends to strengthen the dipole moment of the interfacial charge distribution, leading to an increase in the deformation. Oblate deformation, on the other hand, may be dominated by either viscous or electric stresses. Charge convection in this case weakens both the dipole moment and the pole-to-equator electrohydrodynamic flow, leading to less deformation~\citep{feng1999electrohydrodynamic}. In strong electric fields, several authors have noted the formation of steep charge gradients (shocks) at the drop equator due to charge convection~\citep{lanauze2015nonlinear,das2017electrohydrodynamics,firouznia2023spectral,peng2024equatorial,peng2025bistability}. 

Charge convection must also be included in three-dimensional models in order to capture a symmetry breaking instability known as Quincke rotation. This instability was first observed in solid spherical particles by~\citet{weiler} and~\citet{quincke1896ueber} and leads to the spontaneous rotation of the sphere around an axis perpendicular to the applied field direction~\citep{turcu1987electric}. It was first recognised by~\citet{lampa} that in order for rotation to occur, the charge relaxation time of the sphere $\tau_{\rm{c}}^{-}$ ($\tau_{\rm{c}}^{-}=\epsilon^{-}/\sigma^{-},$ where $\epsilon$ is the absolute permittivity and $\sigma$ is the conductivity of the sphere) must be longer than that of the surrounding fluid $\tau_{\rm{c}}^{+}$ ($\tau_{\rm{c}}^{+}=\epsilon^{+}/\sigma^{+},$ where $\epsilon$ is the absolute permittivity and $\sigma$ is the conductivity of the fluid). In this case, the surface charge distribution of the sphere is antiparallel to the applied field and the sphere is unstable. %(Figure~\ref{fig1}). 
If an infinitesimal perturbation perpendicular to the electric field is applied to the sphere, a destabilising electric torque gives rise to steady rotation of the sphere when the electric field strength exceeds a critical value~\citep{jones1984quincke}.
% $E_{\rm{c,s}},$ first derived by~\citet{jones1984quincke} and given by
% \begin{equation}\label{critfield}
% E_{\rm{c,s}}=\sqrt{\frac{2\mu^{+}(2+Q)(1+2S)}{3\varepsilon^{+}\tau_{\rm{MW}}(SQ-1)}},
% \end{equation}
% where $S=\sigma^{+}/\sigma^{-}$ is the ratio of the conductivities $\sigma^{\pm}$, $Q=\varepsilon^{-}/\varepsilon^{+}$ is the ratio of the permittivities $\varepsilon^{\pm},$ and
% \begin{equation}\label{maxwellwagnertime}\tau_{\rm{MW}}=\frac{\varepsilon^{-}+2\varepsilon^{+}}{\sigma^{-}+2\sigma^{+}}\end{equation}
% is the Maxwell--Wagner relaxation time \citep{jones1995electromechanics}, representing the timescale for polarization of the surface of the sphere upon application of the electric field. The plus superscript denotes properties of the surrounding fluid, while the minus superscript denotes properties of the sphere. It is clear from~\eqref{critfield} that if $SQ<1,$ no critical field exists and Quincke rotation is impossible. 
If the charge relaxation time of the sphere is shorter than that of the surrounding fluid, the dipole moment is parallel to the applied field, and any infinitesimal perturbation leads to an electric torque that returns the sphere to its stable equilibrium. Quincke rotation of solid particles has attracted attention particularly in studies of synthetic active matter \citep{bricard2013,bricard2015,das2019active,han2021low,zhang2021quincke,mauleon2023dynamics,reyes2023magnetic,fitzgerald2025rolling}. Recently, Quincke rotation of liquids has been exploited to generate a diverse range of non-equilibrium patterns, from rolling filaments and fluidic lattices to active emulsions under electrohydrodynamic confinement \citep{raju2021diversity}.

% \begin{figure}
%     \centering
%     \includegraphics[width=\linewidth]{figs/fig1.pdf}
%     \caption{The surface charge distribution on a solid sphere immersed in a viscous fluid for which $\tau^{-}<\tau^{+}$ (subplot (a)) and $\tau^{-}>\tau^{+}$ (subplot (b)).}
%     \label{fig1}
% \end{figure}

A theoretical treatment of Quincke rotation of drops is more challenging than that of solid spheres because of the deforming interface and the associated straining flow. Some theoretical models retain charge convection but are restricted to axisymmetric drops, and so are unable to capture Quincke rotation~\citep{shutov2002shape,shkadov2002drop,das2017nonlinear}. The first effort to capture Quincke rotation in a theoretical model was made by~\citet{feng20022d}, who developed a 2D model valid for small values of the electric Reynolds number, which is the ratio of the conductive and convective time scales for charge redistribution. \citet{he2013electrorotation} developed a small-deformation theory capable of capturing the transition to Quincke rotation,  only including the rotational component of the flow in the charge convection. \citet{das2021three} extended that theory to include the straining flow in the charge convection. Experiments by~\citet{salipante2010electrohydrodynamics,salipante2013electrohydrodynamic} have also revealed the existence of hysteresis in the onset of Quincke rotation in drops, which has been reproduced in recent mathematical models~\citep{peng2025bistability,mcdougall2025nonlinear}, and chaotic dynamics at higher field strengths.

The combined effect of surfactant and an electric field on drop dynamics has only been examined to a limited extent and is the focus of this paper. The first such study was \citet{ha1995effects}, who developed a small-deformation theory accurate to second order in $\mathrm{Ca_E}$ with accompanying experiments. They found that surfactant enhances the deformation of both prolate and oblate drops, with a stronger effect for oblate drops owing to weaker surface convection in the prolate case.
\citet{teigen2010influence} later used a level-set method to study surfactant effects across the three classes identified by \citet{lac2007axisymmetric}. Their results clarified that surfactant increases the deformation of prolate A and oblate drops but decreases that of prolate B drops, because surfactant is convected away from the poles and towards the equator.
\citet{nganguia2013equilibrium} investigated nondiffusing surfactant using both a second-order small-deformation theory and a spheroidal model. As the surfactant concentration is varied, competing tip-stretching and surface-dilution effects produces a non-monotonic dependence of deformation on concentration.
In subsequent work, \citet{nganguia2019effects} incorporated surface diffusion and showed that increasing surfactant concentration strengthens (weakens) the deformation of prolate A drops and weakens (strengthens) that of prolate B drops when surface convection dominates (when diffusion dominates).
Recent related studies include the boundary-integral simulations of \citet{sorgentone20193d} on electrohydrodynamic interactions between surfactant-laden drops, those of \citet{han2022surfactant} incorporating dilational viscosity, and the hybrid lattice-Boltzmann/finite-difference simulations of \citet{zhang2021modeling} on drops subjected to the combined effects of a shear flow and an electric field. To the best of our knowledge, the only studies to have considered the effects of surfactant while retaining charge convection are those of~\citet{poddar2018sedimentation,poddar2019electrical,poddar2019electrorheology}, who have considered sedimenting drops and drops confined in Poiseuille flows. However, the effects of surfactant on Quincke rotation of drops have not yet been explored. Hence, in this work we develop a three-dimensional semi-analytical small-deformation theory that retains charge convection and is therefore able to capture the transition to Quincke rotation.

The article is organised as follows. In~\S\ref{formulation}, we present the relevant governing equations and boundary conditions of the problem to be solved and summarise the nondimensionalisation and scaling assumptions. In~\S\ref{solution} we present our solution to the problem. In~\S\ref{results}, we discuss results of our theory, first for a drop in the Taylor regime and then for a drop in the Quincke regime. Finally, in~\S\ref{conclusions}, we outline the main results of the theory and indicate some avenues for future research.

\section{Problem formulation}\label{formulation}

\begin{figure}
    \centering
    \includegraphics[width=0.9\linewidth]{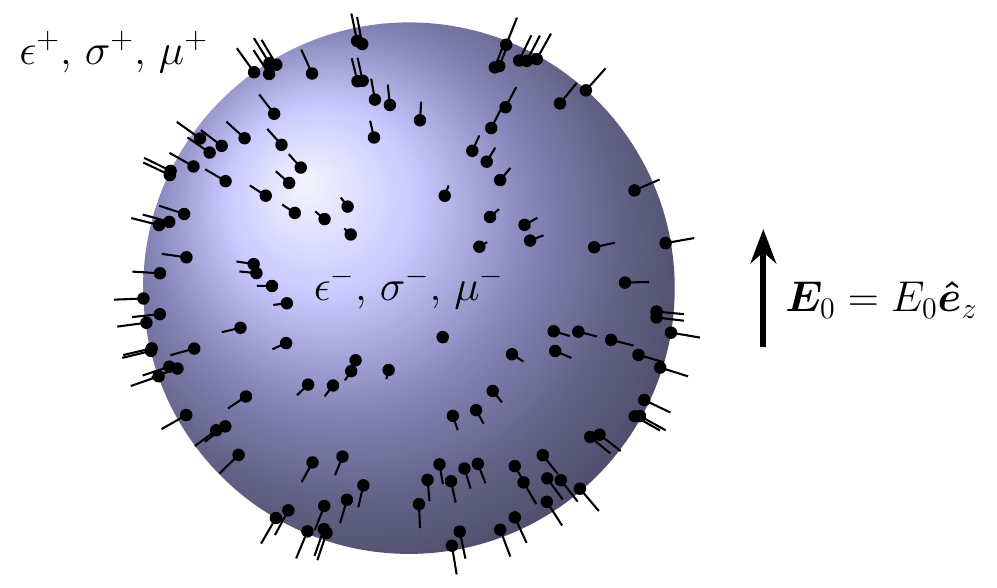}
    \caption{An uncharged, neutrally buoyant, leaky dielectric drop coated with a dilute monolayer of insoluble, apolar surfactant immersed in an immiscible leaky dielectric fluid. A uniform DC electric field is applied in the $z$ direction, $\bm{E}_{0}=E_{0}\bm{\hat{e}}_{z}.$}
    \label{problemdiagram}
\end{figure}

We consider an uncharged, neutrally buoyant leaky dielectric drop, immersed in an immiscible leaky dielectric fluid (Figure~\ref{problemdiagram}). The drop is coated with a dilute monolayer of insoluble, apolar surfactant. The electric permittivities of the fluids are denoted by $\varepsilon^{\pm}$, the conductivities by $\sigma^{\pm},$ and the viscosities by $\mu^{\pm},$ where the plus subscript refers to the properties of the surrounding fluid, and the minus subscript refers to the properties of the drop. The drop domain is denoted by $V^{-},$ the surrounding fluid domain is denoted by $V^{+},$ and the interface between them is denoted by $\partial V$. The drop is initially spherical with radius $a,$ but deforms upon the application of a uniform DC electric field $\bm{E}_{0}$ pointing in the $z$ direction. Upon the application of the field, free charge from both fluids accumulates at the drop interface, developing an interfacial charge distribution. 
Assuming that both of the fluids are electrically neutral, the electric field in each fluid domain, $\bm{E}^{\pm}(\bm{r},t),$ is divergence free. Further, in the electrostatic limit, the electric field is irrotational, allowing the introduction of an electric potential $\varphi^{\pm}(\bm{r},t),$ such that $\bm{E}^{\pm}=-\bnabla\varphi^{\pm}.$ The electric potential is governed by Laplace's equation in both the drop and the surrounding fluid,
\begin{equation}\label{laplaceeqn}
    \nabla^{2}\varphi^{\pm}(\bm{r},t)=0 \qquad \mbox{for } \bm{r}\in V^{\pm}.
\end{equation}
The electric potential is continuous over the interface, and consequently so is the tangential component of the electric field,
\begin{equation}
    \llbracket\varphi\rrbracket=0, \quad \llbracket\bm{E}\cdot(\tensor{I}-\bm{n}\bm{n})\rrbracket=\bm{0}  \quad \mbox{for }\bm{r}\in \partial V,
\end{equation}
where $\tensor{I}$ is the identity tensor, $\bm{n}$ is a unit vector normal to the drop surface, pointing into the surrounding fluid, and the double bracket notation denotes the jump across the interface of any field variable defined in both the drop and surrounding fluid, i.e. $\llbracket f\rrbracket=f^{+}(\bm{r})-f^{-}(\bm{r})$ for $\bm{r}\in \partial V.$
Conversely, the normal component of the electric field is discontinuous, and the interfacial charge density $q$ is given by Gauss' law,
\begin{equation}\label{gauss}
    q=\llbracket \varepsilon\bm{E}\rrbracket\cdot\bm{n}\qquad\mbox{for }\bm{r}\in \partial V.
\end{equation}
Two mechanisms contribute to the evolution of $q.$ The first is the Ohmic current jump across the interface, and the second is surface charge convection by the fluid flow. These are represented by the second and third terms, respectively, in the charge transport equation
\begin{equation}\label{chargecons}
    \frac{\partial q}{\partial t}+\llbracket\sigma\bm{E}\cdot\bm{n}\rrbracket+\bnabla_{s}\cdot(q\bm{v})=0 \qquad\mbox{for }\bm{r}\in \partial V.
\end{equation}
Here, $(\tensor{I}-\bm{n}\bm{n})\cdot\bnabla_{s}$ is the surface gradient operator. The presence of the fluid velocity $\bm{v}(\bm{r},t)$ renders~\eqref{chargecons} nonlinear, which makes analytical progress difficult. However, it is still amenable to numerical solution. In most of the relevant experiments and industrial applications, the drops are typically small (on the order of millimetres) and the fluids are rather viscous, so we assume the Reynolds number of the flow to be small and hence inertia is neglected. The fluid velocity $\mathbf{v}^{\pm}$ and dynamic pressure $p^{\pm}$ then satisfy the Stokes equations in both domains,
\begin{equation}\label{stokes}
    \mu^{\pm}\nabla^{2}\bm{v}^{\pm}=\bnabla p^{\pm}, \qquad \bnabla\cdot\bm{v}^{\pm}=0 \qquad \mbox{for }\bm{r}\in V^{\pm},
\end{equation}
and the velocity is continuous across the interface,
\begin{equation}\label{vtangcontinuous}
    \llbracket\bm{v}\rrbracket=\bm{0} \qquad\mbox{for }\bm{r}\in \partial V.
\end{equation}
 The interface is advected with the flow, so the kinematic boundary condition, evaluated on the perturbed drop interface, can be expressed as
\begin{equation}\label{kinematiccondition0}
    \frac{\partial f}{\partial t}=\bm{v}\bcdot\bm{n}\qquad\mbox{for }\bm{r}\in \partial V,
\end{equation}
where $f(\bm{r},t)$ is a shape function to be defined in~\S\ref{dropshapeandsurfactant}. 
The discontinuity of the material parameters between the drop and surrounding fluid gives rise to electric stresses on the interface, which are balanced by hydrodynamic, capillary and Marangoni stresses. This is expressed through a dynamic boundary condition given by
\begin{equation}\label{stressbalance}  \llbracket\bm{f}^{E}\rrbracket+\llbracket\bm{f}^{H}\rrbracket=\gamma(\bnabla_{s}\cdot\bm{n})\bm{n}-\bnabla_{s}\gamma\qquad\mbox{for }\bm{r}\in \partial V,
\end{equation}
where $\bnabla_{s}\cdot\bm{n}=2\kappa_{m}$ is twice the mean surface curvature, $\gamma=\gamma(\bm{r},t)$ is the surface tension and $\llbracket\bm{f}^{E}\rrbracket=\llbracket\tensor{\tau}^{E}\rrbracket\cdot\bm{n}$ and $\llbracket\bm{f}^{H}\rrbracket=\llbracket\tensor{\tau}^{H}\rrbracket\cdot\bm{n}$ are the electric and hydrodynamic traction jumps across the interface, where
\begin{equation}\label{elecstresstensor}
\tensor{\tau}^{E}=\varepsilon\left(\bm{E}\bm{E}-\frac{1}{2}E^{2}\tensor{I}\right)
\end{equation}
and
\begin{equation}\label{hdstresstensor}
    \tensor{\tau}^{H}=-p\tensor{I}+\mu\left(\bnabla\bm{v}+\bnabla\bm{v}^{T}\right)
\end{equation}
are the electric and hydrodynamic stress tensors, respectively.

We assume the surface tension varies around the drop depending on the local surfactant distribution $\Gamma,$ giving rise to tangential Marangoni stresses in the dynamic boundary condition, represented by the second term on the right hand side of~\eqref{stressbalance}. Gradients in $\Gamma$ arise from convection by the fluid flow and are counteracted by surface diffusion. The evolution of $\Gamma$ is therefore governed by the surface convection–diffusion equation,
\begin{equation}\label{Gammatransdimensional}
    \frac{\partial \Gamma}{\partial t}+\bnabla_{s}\cdot(\bm{v}_{s}\Gamma)+(\bnabla_{s}\bcdot\bm{n})(\bm{v\cdot n})\Gamma-D_{s}\nabla_{s}^{2}\Gamma=0 \qquad\mbox{for }\bm{r}\in \partial V,
\end{equation}
where $D_{s}$ is the surface diffusivity of the surfactant~\citep{stone1990simple}. The second and third terms correspond to surfactant convection. The precise relationship between the surface tension $\gamma$ and the surfactant distribution $\Gamma$ is not straightforward, and several equations of state have been proposed to capture various physical effects relevant in different regimes \citep{rosen2012surfactants,manikantan2020surfactant}. Most commonly used is the Langmuir equation,
\begin{equation}
\label{gamma(Gamma)}
\gamma=\gamma_{0}+RT\Gamma_{\infty}\ln\left(1-\frac{\Gamma}{\Gamma_{\infty}}\right),
\end{equation}
where $R$ is the gas constant, $T$ is the absolute temperature, $\Gamma_{\infty}$ is the maximum (i.e. the fully saturated) surfactant concentration and $\gamma_{0}$ is the surface tension of the drop in the absence of surfactant. The Langmuir equation is generally valid for a wide range of surfactant concentrations, including those approaching $\Gamma_{\infty}$. However, for dilute concentrations of surfactant like those considered in this work, it is appropriate to linearise it about $\Gamma=0$, in which case the surface tension is given by
\begin{equation}
\label{gamma(Gamma)lin}
\gamma=\gamma_{0}-RT\Gamma.
\end{equation}
In the absence of external forcing, or in the limit of rapid diffusion, the surfactant distribution is uniform around the drop and is given by the equilibrium concentration $\Gamma_{\rm{eq}}.$ The corresponding surface tension at $\Gamma=\Gamma_{\rm{eq}}$ is $\gamma=\gamma_{\rm{eq}}(<\gamma_{0}).$ 

\subsection{Nondimensionalisation and scaling assumptions}

Permittivities, conductivities, and viscosities are scaled with $\sigma^-$, $\varepsilon^+$, and $\mu^+$, which leads to three dimensionless numbers,
\begin{equation}
S =\frac{\sigma^{+}}{\sigma^{-}}, \quad Q=\frac{\varepsilon^{-}}{\varepsilon^{+}}, \quad \lambda=\frac{\mu^{-}}{\mu^{+}}.
\end{equation}
We take the initial drop radius $a$ as the characteristic length, the electrohydrodynamic flow time \begin{equation}
    \tau_{{\rm EHD}}=\frac{\mu^{+}}{\epsilon^{+}E_{0}^{2}}
\end{equation} as the characteristic time, and $\Gamma_{\rm{eq}}$ and $\gamma_{\rm{eq}}$ as the characteristic surfactant concentration and surface tension, respectively. Other choices for the time scale which are sometimes used in studies of drop electrohydrodynamics include the Maxwell–Wagner relaxation time $\tau_{\mathrm MW}=(\varepsilon^{-}+2\varepsilon^{+})/(\sigma^{-}+2\sigma^{+})$, which characterizes the timescale of dipole moment polarization and the angular velocity associated with Quincke rotation~\citep{das2017electrohydrodynamics, das2017nonlinear, das2021three}, the charge relaxation time of the outer fluid $\tau_{\rm{c}}^{+}$~\citep{lanauze2015nonlinear,peng2025bistability}, and the capillary relaxation time $\tau_{\gamma}=\mu^{+}a/\gamma_{0},$ which is the timescale for the drop to return to a spherical shape~\citep{esmaeeli2011transient,esmaeeli2020transient}.
The stress balance~\eqref{stressbalance} can be written in nondimensional form as
\begin{equation}\label{stressbalancenondim}
    \llbracket\bm{f}^{E}\rrbracket+\llbracket\bm{f}^{H}\rrbracket=\frac{1}{\mathrm{Ca_{E}}}\left[\gamma(\bnabla_{s}\cdot\bm{n})\bm{n}-\bnabla_{s}\gamma\right]\qquad\mbox{for }\bm{r}\in \partial V,
\end{equation}
which includes the fourth dimensionless number, the electric capillary number $\mathrm{Ca_{E}}$, defined as
\begin{equation}
 \mathrm{Ca_{E}}=\frac{\varepsilon^{+}E_{0}^{2}a}{\gamma_{\rm{eq}}}.
\end{equation}
It represents the ratio of electric to capillary stresses.  We assume that the deviation of the drop shape from sphericity is small, in which case it is appropriate to follow the classical domain perturbation approach in which the flow fields are calculated assuming that the drop remains perfectly spherical~\citep{taylor1934formation,cox1969deformation,barthes1973deformation}. These flow fields are then used in the normal stress balance at the drop interface to determine the drop shape to first order in $\delta,$ a small parameter quantifying the deviation of the drop shape from sphericity. It will be shown in~\S\ref{stressbalancesection} that $\delta\propto\mathrm{Ca_{E}},$ hence we require $\mathrm{Ca_{E}}\ll 1$ in order for the drop to remain nearly spherical. If desired, the $\mathcal{O}(\delta)$ drop shape can be used to obtain the corrections to the flow fields at the next order in $\delta,$ but in the present work we describe only the leading order solution. Note that the uniform reduction of surface tension from that of the clean drop $\gamma_{0}$ to that of the the uniformly coated drop $\gamma_{\rm{eq}}$ is implicitly accounted for by using $\gamma_{\rm{eq}}$ to define $\mathrm{Ca_{E}}$. Defining the capillary number this way allows us to focus on the effects of Marangoni stresses and spatial variations in the surface tension. 
The charge conservation equation~\eqref{chargecons} in nondimensional form is written
\begin{equation}\label{chargeconsnondim}
    \frac{\partial q}{\partial t}+\frac{1}{{\rm Re_{E}}}\llbracket\bm{j}\bcdot\bm{n}\rrbracket+\bnabla_{s}\bcdot(q\bm{v})=0\qquad\mbox{for }\bm{r}\in \partial V,
\end{equation}
where
\begin{equation}
    {\rm Re_{E}}=\frac{\epsilon^{+2}E_{0}^{2}}{\sigma^{+}\mu^{+}}
\end{equation}
is the electric Reynolds number, which quantifies the rates of charge convection and relaxation relative to the rate of charging of the interface by Ohmic currents from the bulk fluids. In addition, the discontinuity of the normal current across the interface is given in nondimensional form by
\begin{equation}\label{currentjump}\llbracket\bm{j}\bcdot\bm{n}\rrbracket=\bm{E}^{+}\bcdot\bm{n}-\frac{1}{S}\bm{E}^{-}\bcdot\bm{n},\end{equation}
while the charge distribution $q$ introduced in~\eqref{gauss} is
\begin{equation}\label{chargedensity}
    q=(1-Q)\bm{\hat{E}}_{0}\bcdot\bm{\hat{r}}+(2+Q)\bm{P}\bcdot\bm{\hat{r}}.
\end{equation}
The surfactant transport equation~\eqref{Gammatransdimensional} in nondimensional form is written
\begin{equation}\label{Gammatransnondimensional}
    \frac{\partial \Gamma}{\partial t}+\bnabla_{s}\cdot(\bm{v}_{s}\Gamma)+(\bnabla_{s}\bcdot\bm{n})(\bm{v\cdot n})\Gamma-\frac{1}{\rm{Pe_{s}}}\nabla_{s}^{2}\Gamma=0\qquad\mbox{for }\bm{r}\in \partial V,
\end{equation}
where
\begin{equation}
    \mathrm{Pe_{s}}=\frac{\epsilon^{+}E_{0}^{2}a^{2}}{D_{{\rm s}}\mu^{+}}
\end{equation}
is the surface P\'eclet number, which is the ratio of the advective and diffusive surfactant transport rates. However, instead of treating $\rm{Pe_{s}}$ as an independent parameter, we instead define
\begin{equation}
    \zeta=\frac{\mathrm{Pe_{s}}}{\mathrm{Re_{E}}},
\end{equation}
which is analogous to the parameter $\gamma$ used in studies of drops in extensional flows~\citep{stone1990effects,milliken1993effect}, and is convenient because it characterises the strength of the diffusion of the surfactant while remaining independent of the electric field strength. 
Finally, the equation of state~\eqref{gamma(Gamma)lin} in nondimensional form is written as
\begin{equation}
    \gamma=1+\mathrm{El}\left(1-\Gamma\right).
\end{equation}
Here, $\rm{El}$ denotes the elasticity number, defined as
\begin{equation}
    \mathrm{El}=\frac{RT\Gamma_{\rm{eq}}}{\gamma_{\rm{eq}}}=\frac{\gamma_{0}-\gamma_{{\rm eq}}}{\gamma_{\rm{eq}}},
\end{equation}
which quantifies the variation of the surface tension as a function of the surfactant concentration. Changing $\mathrm{El}$ alters the sensitivity of the surface tension to variations in surfactant concentration, thereby modifying the strength of Marangoni stresses.

In general, the convective terms in both the charge transport and surfactant transport equations introduce nonlinearities that prevent closed-form analytical solutions. One possible approach is to linearize these equations and perform asymptotic expansions in $\mathrm{Re_E}$ and $\mathrm{Ca_E}$ simultaneously, but this can be cumbersome. Here we instead retain the nonlinear terms and solve the resulting equations numerically. The numerical solution of the charge transport equation is accurate in parameter regimes where charge convection by the induced straining flow remains weak, which occurs when $\mathrm{Re_{E}}=\mathcal{O}(1)$ and $\lambda\gg1$. The solution of the surfactant transport remains accurate when the surfactant distribution remains almost uniform, which occurs either when $\zeta\ll1$ or when $\mathrm{Ca_{E}}/\rm{El}\ll 1$. In the present work, we adopt the latter assumption, and make no assumption about the magnitude of $\zeta.$ Thus, given that $\mathrm{Ca_{E}}\ll 1,$ we therefore require $\mathrm{El}=\mathcal{O}(1)$, consistent with~\citet{vlahovska2005deformation}.

\section{Problem solution}\label{solution}

We now describe the solution to the problem formulated in the previous section. In \S\ref{dropshapeandsurfactant}, we describe the definitions of the drop shape and surfactant distributions. In \S\ref{electrostatics}, we present the solution to Laplace's equation~\eqref{laplaceeqn} based on a multipole expansion. In \S\ref{hydrodynamics}, we present the solution to the Stokes equations for the fluid velocity and pressure. In \S\ref{kinematics}, we apply the kinematic boundary condition to relate the flow fields inside and outside the drop to one another and to the motion of the interface. In \S\ref{stressbalancesection}, we apply the stress boundary condition to relate the electric, hydrodynamic, capillary and Marangoni stresses at the interface. Finally, in \S\ref{chargetransport} and \S\ref{surfactanttransport}, we use the charge and surfactant transport equations~\eqref{chargeconsnondim} and~\eqref{surfactanttransport} to derive ordinary differential equations (ODEs) governing the evolutions of the dipole moment and surfactant distribution.

\subsection{Definitions of drop shape and surfactant distribution}\label{dropshapeandsurfactant}
In general, both the drop shape $f(\theta,\phi,t)$ and the surfactant distribution $\Gamma(\theta,\phi,t)$ can be expanded in a series of surface spherical harmonics. As mentioned previously, we assume the deviations of $f$ and $\Gamma$ from the spherical drop and the uniform surfactant distribution, respectively, are small. Then, in anticipation of the form of the electric stress and the resulting flow fields, we retain only the second-order harmonics, and so write
\begin{subequations}
\begin{align}
    f(\theta,\phi,t)&=1+\delta\tensor{Q}^{f}:\bm{\hat{r}\hat{r}},
    \label{fdefinition}\\\Gamma(\theta,\phi,t)&=1+\epsilon\tensor{Q}^{\Gamma}:\bm{\hat{r}\hat{r}}.\label{Gammadefinition}
    \end{align}
\end{subequations}The double dot notation represents the scalar contraction of two tensors, and $\bm{\hat{r}}=\bm{r}/r$ denotes the unit vector in the radial direction. Thus, for example, $\tensor{Q}\bcdoubledot \bm{\hat{r}\hat{r}}=Q_{ij}\hat{r}_{j}\hat{r}_{i}$. $\tensor{Q}^{f}$ and $\tensor{Q}^{\Gamma}$ are symmetric irreducible second-order tensors each containing five independent coefficients. These coefficients are directly related to the five coefficients of a sum of $\ell=2$ associated Legendre polynomials and are referred to as the shape coefficients and the surfactant coefficients, respectively. In addition, $\delta\ll 1$ and $\epsilon\ll 1$ are small parameters which quantify the deviations of the drop shape from sphericity and the deviations of the surfactant distribution from the uniform coverage $\Gamma_{\rm{eq}},$ respectively, both of which will be defined in \S\ref{stressbalancesection}. 

Using the definition in~\eqref{fdefinition}, the surface normal $\bm{n}$ and the mean surface curvature $\kappa_{m}$ of the interface are given by
\begin{equation}\label{normal and curvature}
    \bm{n}=\frac{\bm{\hat{r}}-\bnabla f}{|\bm{\hat{r}}-\bnabla f|}=\bm{\hat{r}}-2\delta(\tensor{Q}^{f}\bcdot\bm{\hat{r}})\bcdot(\tensor{I}-\bm{\hat{r}\hat{r}}),\quad
    \kappa_{m}=\frac{1}{2}\bnabla_{s}\bcdot\bm{n}=1+2\delta \tensor{Q}^{f}\bcdoubledot\bm{\hat{r}\hat{r}}.
\end{equation}

\subsection{Electrostatics}\label{electrostatics}
To satisfy Laplace's equation~\eqref{laplaceeqn}, the electric potentials inside and outside the drop are expressed as multipole expansions. In the present work, only the dipole moment is retained. However, higher-order multipoles can be induced by charge convection by the induced straining flow. It will be shown in \S\ref{stressbalancesection} that the leading-order straining flow, and hence the induced higher-order multipoles, appear at $\mathcal{O}(\lambda^{-1})$. However, the effect of these higher-order multipoles on the drop shape or flow fields appears only at $\mathcal{O}(\lambda^{-2}).$ Hence, by retaining terms up to and including $\mathcal{O}(\lambda^{-1}),$ the higher-order multipoles can be neglected and the potential inside and outside the drop can be written as
\begin{align}\label{potentialinsideoutside}
    \varphi^{-} = \varphi_{0} + \bm{P}\bcdot\bm{r}, \qquad  \varphi^{+}= \varphi_{0} + {r^{-3}}\bm{P}\bcdot\bm{r},
\end{align}
respectively, where $\varphi_{0}$ is the potential associated with the applied electric field and $\bm{P}(t)$ denotes the induced dipole moment of the drop. The potentials inside and outside the drop are given by growing and decaying harmonics, respectively. The electric fields inside and outside the drop are given by
\begin{equation}
    \bm{E}^{-}=\bm{\hat{E}}_{0}-\bm{P}, \qquad  \bm{E}_{1}=\bm{\hat{E}}_{0}-r^{-3}\bm{P}\bcdot(\tensor{I}-3\bm{\hat{r}\hat{r}}),
\end{equation}
respectively.
After calculating the electric stress tensor according to~\eqref{elecstresstensor}, the jump in electric traction across the drop interface can be written
\begin{equation}\label{elecstress}
\llbracket\bm{f}^{E}\rrbracket=\llbracket\tensor{\tau}^{E}\rrbracket\bcdot\bm{\hat{r}}=p^{E}\bm{\hat{r}}+\tensor{Q}_{\bm{\hat{r}}}^{E}\bcdoubledot\bm{\hat{r}\hat{r}\hat{r}}+\left(\tensor{Q}_{\bm{\hat{t}}}^{E}\bcdot\bm{\hat{r}}\right)\bcdot(\tensor{I}-\bm{\hat{r}\hat{r}})+\bm{T}^{E}\times\bm{\hat{r}},
\end{equation}
where
\begin{subequations}\label{elecstresscoefficients}
\begin{align}
        p^{E} =&\; -\frac{1-Q}{6}+\frac{4-Q}{3}\left(\bm{\hat{E}}_{0}\bcdot\bm{P}\right)+\frac{Q+2}{6}P^{2}, \\ 
        \begin{split}
            \tensor{Q}^{E}_{\bm{\hat{r}}} =&\; (1-Q)\bm{\hat{E}}_{0}\bm{\hat{E}}_{0}+\frac{1+2Q}{2}\left(\bm{\hat{E}}_{0}\bm{P}+\bm{P}\bm{\hat{E}}_{0}\right)+\frac{5-2Q}{2}\bm{P}\bm{P} \\
         & \;+\left[ -\frac{1-Q}{3}-\frac{1+2Q}{3}\left(\bm{\hat{E}}_{0}\bcdot\bm{P}\right)-\frac{5-2Q}{6}P^{2}\right]\tensor{I},
        \end{split}\label{elecstresscoefficients_b} \\
        \begin{split}
        \tensor{Q}^{E}_{\bm{\hat{t}}} =& \;(1-Q)\bm{\hat{E}}_{0}\bm{\hat{E}}_{0}+\frac{1+2Q}{2}\left(\bm{\hat{E}}_{0}\bm{P}+\bm{P}\bm{\hat{E}}_{0}\right)-(2+Q)\bm{P}\bm{P} \\
        &\;+\left[-\frac{1-Q}{3}-\frac{1+2Q}{3}\left(\bm{\hat{E}}_{0}\bcdot\bm{P}\right)+\frac{2+Q}{3}P^{2}\right]\tensor{I},
        \end{split}\label{elecstresscoefficients_c} \\
        \bm{T}^{E}=&\;\frac{3}{2}\bm{P}\times\bm{\hat{E}}_{0}, \label{elecstresscoefficients_d}
    \end{align}
    \end{subequations}
where $P=|\bm{P}|.$
\subsection{Hydrodynamics}\label{hydrodynamics}

We express the fluid velocity and pressure using the general solution of the Stokes equations~\eqref{stokes} derived by~\citet{lamb1924hydrodynamics}, given by
\begin{subequations}
\begin{align}
     p &= \sum_{n=-\infty}^{\infty}p_{n}, \\
    \bm{v} &= \sum_{\substack{n=-\infty, \\ n \neq -1}}^{\infty}\left[\bnabla\times(\bm{r}\chi_{n})+\bnabla\phi_{n}+\frac{n+3}{2\mu(n+1)(2n+3)}r^{2}\bnabla p_{n}-\frac{n}{\mu(n+1)(2n+3)}\bm{r}p_{n}\right].\label{lambv}
\end{align}
\end{subequations}
Here, $p_{n},$ $\chi_{n}$ and $\phi_{n}$ each denote a solid spherical harmonic of order $n$~\citep{kim2013microhydrodynamics,happel2012low}. Constant or growing harmonics (non-negative $n$) can be expressed as the contraction of an irreducible symmetric tensor of order $n$ and the tensor product of $n$ unit position vectors,
 \begin{equation}\label{harmonicsin}
     p_{n}=r^{n}T^{(n)}_{i_{1}i_{2}i_{3}...i_{n}}\hat{r}_{i_{1}}\hat{r}_{i_{2}}\hat{r}_{i_{3}}...\hat{r}_{i_{n}},
 \end{equation}
 with summation implied over $i_{1},i_{2},i_{3},...,i_{n},$ and where $T^{(n)}$ is an irreducible symmetric tensor. The decaying harmonics (negative $n$) are similarly given by
  \begin{equation}\label{harmonicsout}
     p_{-n-1}=r^{-n-1}T^{(n)}_{i_{1}i_{2}i_{3}...i_{n}}\hat{r}_{i_{1}}\hat{r}_{i_{2}}\hat{r}_{i_{3}}...\hat{r}_{i_{n}},
 \end{equation}
with the $\phi_{n}$ and $\chi_{n}$ harmonics defined the same way. The general hydrodynamic stress on the drop surface associated with the velocity~\eqref{lambv} is
\begin{equation}\label{lambstress}
    \bm{f}^{H}=\mu\sum_{n=-\infty}^{\infty}\left[(n-1)\bnabla\times(\bm{r}\chi_{n})+2(n-1)\bnabla\phi_{n}+\frac{n(n+2)\bnabla p_{n}-(2n^{2}+4n+3)\bm{r}p_{n}}{(n+1)(2n+3)}\right].
\end{equation}
We retain only growing harmonics in the solution inside the drop and decaying harmonics in the solution outside the drop, and then consider the resulting stresses on each side of the interface as given by~\eqref{lambstress}. Then, by comparing the various stress modes with those in~\eqref{elecstress}, we can deduce which harmonics need to be retained. After doing so, the velocities inside and outside the drop are
\begin{equation}
\begin{aligned}\label{interiorflow0}
        \bm{v}^{-}=\bm{d}^{\chi}\times\bm{r}+\bm{d}^{\phi}+2\tensor{q}^{\phi}\bcdot\bm{r}+\frac{r^{2}\tensor{q}^{p}\bcdot\bm{r}}{21\lambda}\bcdot(5\tensor{I}-2\bm{\hat{r}\hat{r}}),
    \end{aligned}
    \end{equation}
\begin{equation}\begin{aligned}\label{exteriorflow0}
\bm{v}^{+}=
r^{-3}\bm{D}^{\chi}\times\bm{r}+r^{-5}\frac{\tensor{Q}^{p}\bcdoubledot\bm{rrr}}{2}+r^{-5}\left(\tensor{Q}^{\phi}\bcdot\bm{r}\right)\bcdot(2\tensor{I}-5\bm{\hat{r}\hat{r}}),
\end{aligned}\end{equation}
respectively, where $\bm{d}^{\rm{\chi}}(t)$ and $\bm{D}^{\chi}(t)$ are vectors corresponding to the solid body rotational flow induced by Quincke rotation, $\bm{d}^{\phi}(t)$ is a uniform flow field inside the drop, and $\tensor{q}^{p}(t),$ $\tensor{Q}^{p}(t),$ $\tensor{q}^{\phi}(t),$ and $\tensor{Q}^{\phi}(t)$ are second-order irreducible symmetric tensors corresponding to the straining flows induced by the quadrupolar electric stresses $\tensor{Q}^{E}_{\bm{\hat{t}}}.$ These vectors and tensors will henceforth be referred to as the flow coefficients. The hydrodynamic traction jump associated with~\eqref{interiorflow0} and~\eqref{exteriorflow0} is
\begin{equation}\begin{split}\label{hdtractioninitial}
\llbracket\bm{f}^{H}\rrbracket =&\;\left(\tilde{p}^{-}-\tilde{p}^{+}\right)\bm{\hat{r}}+\left(24\tensor{Q}^{\phi}-3\tensor{Q}^{p}-4\lambda\tensor{q}^{\phi}+\frac{\tensor{q}^{p}}{7}\right)\bcdoubledot\bm{\hat{r}\hat{r}\hat{r}}
\\
&\;+\left[\left(-16\tensor{Q}^{\phi}+\tensor{Q}^{p}-\frac{16\tensor{q}^{p}}{21}-4\lambda\tensor{q}^{\phi}\right)\bcdot\bm{\hat{r}}\right]\bcdot(\tensor{I}-\bm{\hat{r}\hat{r}})-3\bm{D}^{\chi}\times\bm{\hat{r}},
\end{split}\end{equation}
where $\tilde{p}^{-}$ and $\tilde{p}^{+}$ are the isotropic pressures inside and outside the drop, given by setting $n=0$ in~\eqref{harmonicsin} and~\eqref{harmonicsout}, respectively.

\subsection{Kinematic boundary condition}\label{kinematics}

Next, we relate the flow coefficients in~\eqref{interiorflow0} and~\eqref{exteriorflow0} to one another and to the motion of the interface using the tangential no-slip condition~\eqref{vtangcontinuous} along with the kinematic boundary condition~\eqref{kinematiccondition0}. First, applying continuity of velocity tangential to the interface yields the relations
\begin{equation}\begin{aligned}\label{tangentialkcrelations}
\bm{D}^{\chi}=\bm{d}^{\chi},\qquad \bm{0}=\bm{d}^{\phi},\qquad 2\tensor{Q}^{\phi}=2\tensor{q}^{\phi}+\frac{5\tensor{q}^{p}}{21\lambda}.
\end{aligned}\end{equation}
Second, expanding~\eqref{kinematiccondition0} yields
\begin{equation}\begin{aligned}\label{radialkc}
\bm{v}^{-}\bcdot\bm{\hat{r}}-\delta\bm{v}^{-}\bcdot\bnabla \left(\tensor{Q}^{f}:\bm{\hat{r}\hat{r}}\right)=\bm{v}^{+}\bcdot\bm{\hat{r}}-\delta\bm{v}^{+}\bcdot\bnabla \left(\tensor{Q}^{f}:\bm{\hat{r}\hat{r}}\right)
=\delta\frac{\partial }{\partial t}\left(\tensor{Q}^{f}:\bm{\hat{r}\hat{r}}\right).
\end{aligned}\end{equation}
Since the solid body rotational flows $\bm{D}^{\chi}$ and $\bm{d}^{\chi}$ are independent of $\lambda$ while the straining flow is assumed to be comparatively weak (scaling as $\lambda^{-1}$), the velocities $\bm{v}^{\pm}$ appearing in the nonlinear convective products $\bm{v}^{\pm}\bcdot\bnabla f$ include only $\bm{D}^{\chi}$ and $\bm{d}^{\chi}$ at leading order. Given that $\bm{d}^{\chi}=\bm{D}^{\chi},$ then 
\begin{equation}\begin{aligned}\label{vnablafsymmetric}
\bm{v}^{+}\bcdot\bnabla f=\bm{v}^{-}\bcdot\bnabla f=\left(\varepsilon_{lkj}Q^{f}_{li}D^{\chi}_{k}+\varepsilon_{lki}Q^{f}_{lj}D^{\chi}_{k}\right)\hat{r}_{i}\hat{r}_{j}=\tensor{Q}^{\chi f}\bcdoubledot\bm{\hat{r}\hat{r}},
\end{aligned}
\end{equation} 
where $\varepsilon_{lkj}$ and $\varepsilon_{lki}$ are the Levi--Civita tensor and \begin{equation}\label{qchif}Q_{ij}^{\chi f} = \varepsilon_{lkj}Q^{f}_{li}D^{\chi}_{k}+\varepsilon_{lki}Q^{f}_{lj}D^{\chi}_{k}.\end{equation} In the Taylor regime, this term has no effect since there is no rotational flow. In the Quincke regime, it acts to reduce the drop deformation by convecting away the shape distortions caused by the electric stresses and the straining flow. The kinematic condition can thus be written
\begin{equation}\begin{aligned}
\bm{v}^{-}\bcdot\bm{\hat{r}}=\bm{v}^{+}\bcdot\bm{\hat{r}}
=\delta\left(\frac{\partial \tensor{Q}^{f}}{\partial t}+\tensor{Q}^{\chi f}\right),
\end{aligned}\end{equation}
 and substituting in the fluid velocities given in~\eqref{interiorflow0} and~\eqref{exteriorflow0} yields
 \begin{equation}\begin{aligned}\label{radialkcfin}
2\tensor{q}^{\phi}+\frac{1}{7\lambda}\tensor{q}^{p}=\frac{1}{2}\tensor{Q}^{p}-3\tensor{Q}^{\phi}
=\delta\left(\frac{\partial \tensor{Q}^{f}}{\partial t}+\tensor{Q}^{\chi f}\right).
\end{aligned}\end{equation}
Combining this with the relations given in~\eqref{tangentialkcrelations} allows us to express the flow coefficients $\tensor{Q}^{\phi},$ $\tensor{q}^{\phi}$ and $\tensor{q}^{p}$ in terms of $\tensor{Q}^{p}$ and $\delta(\partial \tensor{Q}^{f}/\partial t+\tensor{Q}^{\chi f})$:
\begin{subequations}
\begin{align}
\tensor{Q}^{\phi} &= \frac{1}{6}\tensor{Q}^{p}-\frac{\delta}{3}\left(\frac{\partial \tensor{Q}^{f}}{\partial t}+\tensor{Q}^{\chi f}\right), \label{eq:Qphi1} \\
\tensor{q}^{p} &= \frac{7\lambda}{2}\tensor{Q}^{p}-\frac{35\lambda\delta}{2}\left(\frac{\partial \tensor{Q}^{f}}{\partial t}+\tensor{Q}^{\chi f}\right), \label{eq:qp1} \\
\tensor{q}^{\phi} &= -\frac{1}{4}\tensor{Q}^{p}+\frac{7\delta}{4}\left(\frac{\partial \tensor{Q}^{f}}{\partial t}+\tensor{Q}^{\chi f}\right). \label{eq:qphi1}
\end{align}
\end{subequations}
The hydrodynamic traction~\eqref{hdtractioninitial} can be rewritten as\begin{equation}\label{finalhdstress}
    \llbracket\bm{f}^{H}\rrbracket=\llbracket\tensor{\tau}^{H}\rrbracket\bcdot\bm{\hat{r}}=p^{H}\bm{\hat{r}}+\tensor{Q}_{\bm{\hat{r}}}^{H}\bcdoubledot\bm{\hat{r}\hat{r}\hat{r}}+\left(\tensor{Q}_{\bm{\hat{t}}}^{H}\bcdot\bm{\hat{r}}\right)\bcdot(\tensor{I}-\bm{\hat{r}\hat{r}})+\bm{T}^{H}\times\bm{\hat{r}},
\end{equation}
where
\begin{subequations}
\begin{align}
p^{H} &= \tilde{p}^{-}-\tilde{p}^{+}, \label{hdpressure} \\
\tensor{Q}^{H}_{\bm{\hat{r}}} &= \frac{2+3\lambda}{2}\tensor{Q}^{p}-\frac{16+19\lambda}{2}\delta\left(\frac{\partial \tensor{Q}^{f}}{\partial t}+\tensor{Q}^{\chi f}\right), \label{hdstresscoeffQr} \\
\tensor{Q}^{H}_{\bm{\hat{t}}} &= -\frac{5(1+\lambda)}{3}\tensor{Q}^{p}+\frac{16+19\lambda}{3}\delta\left(\frac{\partial \tensor{Q}^{f}}{\partial t}+\tensor{Q}^{\chi f}\right), \label{hdstresscoeffQt} \\
\bm{T}^{H} &= -3\bm{D}^{\chi}. \label{hdtorque}
\end{align}
\end{subequations}

\subsection{Stress balance}\label{stressbalancesection}
We now apply the dynamic boundary condition~\eqref{stressbalancenondim}, enforcing a balance between the electric and hydrodynamic stresses~\eqref{elecstress} and~\eqref{finalhdstress}, the Marangoni stresses arising from nonuniform surface tension, and the usual capillary stresses. Expanding the right-hand side of~\eqref{stressbalancenondim} yields
\begin{equation}\begin{aligned}\label{stressbalance_final}
\llbracket\bm{f}^{E}\rrbracket+\llbracket\bm{f}^{H}\rrbracket=\frac{2}{\mathrm{Ca_{E}}}\bm{\hat{r}}-\frac{4}{\mathrm{Ca_{E}}}\delta(\tensor{Q}^{f}\cdot\bm{r})\cdot(\tensor{I}-2\bm{\hat{r}\hat{r}})+\frac{2\rm{El}}{\mathrm{Ca_{E}}}\epsilon(\tensor{Q}^{\Gamma}\cdot\bm{\hat{r}})\cdot(\tensor{I}-2\bm{\hat{r}\hat{r}}).
\end{aligned}\end{equation}
Considering the radial component of~\eqref{stressbalance_final}, we obtain a balance of isotropic pressures and quadrupolar stresses, which yields two equations; one for the pressures,
\begin{equation}
    \begin{aligned}
        p^{H}=-p^{E}+\frac{2}{\mathrm{Ca_{E}}},
    \end{aligned}
\end{equation}
and one for the quadrupolar stresses,
\begin{equation}
    \begin{aligned}
    \label{radquadrupolarstressbalance}\tensor{Q}_{\bm{\hat{r}}}^{H}=\frac{2}{\mathrm{Ca_{E}}}\left(-\rm{El}\epsilon\tensor{Q}^{\Gamma}+2\delta\tensor{Q}^{f}\right)-\tensor{Q}^{E}_{\bm{\hat{r}}}.
    \end{aligned}
\end{equation}

In order for the capillary stresses to balance the $\mathcal{O}(1)$ electric and hydrodynamic stresses, we require $\delta\propto\rm{Ca_{E}},$ and following previous work~\citep{taylor1966studies} we choose
\begin{equation}
    \delta=\frac{3\rm{Ca_{E}}}{4(1+2S)^{2}}.
\end{equation}
Analogously, for the Marangoni stresses to balance the electrostatic and hydrodynamic stresses, we require $\epsilon\propto\rm{Ca_{E}}/\rm{El},$ and so define
\begin{equation}
    \epsilon=\frac{\rm{Ca_{E}}}{\rm{El}},
\end{equation}
which is the reciprocal of the usual Marangoni number. 

Next, examining the tangential components of~\eqref{stressbalance_final} we have a balance of quadrupolar stresses and torques, which again yields two equations; one for the quadrupolar stresses,
\begin{equation}
    \begin{aligned}\label{tangquadstressbal}
\tensor{Q}_{\bm{\hat{t}}}^{H}=\frac{2\rm{El}}{\mathrm{Ca_{E}}}\epsilon\tensor{Q}^{\Gamma}-\tensor{Q}^{E}_{\bm{\hat{t}}},
    \end{aligned}
\end{equation}
and one for the torques,
\begin{equation}\label{torquebal}
    \bm{T}^{H}=-\bm{T}^{E}.
\end{equation}
Substituting the expressions for the quadrupolar hydrodynamic stresses~\eqref{hdstresscoeffQr} and~\eqref{hdstresscoeffQt} into~\eqref{radquadrupolarstressbalance} and~\eqref{tangquadstressbal} gives two equations for the flow coefficients $\tensor{Q}^{p}$ and the shape evolution $\partial\tensor{Q}^{f}/\partial t,$ which are then given by 
\begin{equation}\label{straining}
    \tensor{Q}^{p}=\frac{2}{3+2\lambda}\left[-\frac{3}{(1+2S)^{2}}\tensor{Q}^{f}+\tensor{Q}^{E}_{\bm{\hat{r}}}+\frac{3}{2}\tensor{Q}^{E}_{\bm{\hat{t}}}-\frac{\rm{El}}{\mathrm{Ca_{E}}}\epsilon\tensor{Q}^{\Gamma}\right]
\end{equation}
and
\begin{align}\label{shapetrans}\begin{split}
   \frac{\partial \tensor{Q}^{f}}{\partial t} = &\;\frac{10(1+\lambda)}{\delta(16+19\lambda)(3+2\lambda)}\left[-\frac{3}{(1+2S)^{2}}\tensor{Q}^{f}+\tensor{Q}^{E}_{\bm{\hat{r}}}\right.\\&\left. +\;\frac{6+9\lambda}{10(1+\lambda)}\tensor{Q}^{E}_{\bm{\hat{t}}}+\frac{(4+\lambda)\rm{El}}{5(1+\lambda)\mathrm{Ca_{E}}}\epsilon\tensor{Q}^{\Gamma}\right]-\tensor{Q}^{\chi f}.\end{split}
\end{align}
Substituting the electric and hydrodynamic torques given in~\eqref{elecstresscoefficients_d} and~\eqref{hdtorque} into~\eqref{torquebal}, we find
\begin{equation}\label{torquebalance}
    \bm{D}^{\chi}=\frac{1}{2}\bm{P}\times\bm{\hat{E}}_{0}.
\end{equation}
At this point, the flow fields and drop shape are known in terms of the dipole moment $\bm{P}$ and the surfactant coefficients in $\tensor{Q}^{\Gamma},$ which will be determined in the following sections via application of their respective transport equations.

\subsection{Charge transport}\label{chargetransport}

The dipole moment evolves according to the dimensionless charge transport equation~\eqref{chargecons} at the interface of the drop. Substituting the fluid velocity from either~\eqref{interiorflow0} or~\eqref{exteriorflow0}, the charge density~\eqref{chargedensity} and the normal current jump~\eqref{currentjump} into~\eqref{chargeconsnondim} yields an evolution equation for the dipole moment $\bm{P}$:
\begin{equation}\begin{split}\label{dipoleevolution}
    \frac{\mbox{d}\bm{P}}{\mbox{d} t}=&\;-\frac{1}{\mathrm{Re_{E}}}\frac{1+2S}{S(2+Q)}\bm{P}-\frac{1}{\mathrm{Re_{E}}}\frac {S-1}{S(2+Q)}\bm{\hat{E}}_{0}+\bm{D}^{\chi}\times\left(\frac{1-Q}{2+Q}\bm{\hat{E}}_{0}+\bm{P}\right)
        \\&\;+\left(\frac{1-Q}{2+Q}\bm{\hat{E}}_{0}+\bm{P}\right)\bcdot\left(\frac{1}{5}\tensor{Q}^{p}-\frac{6}{5}\delta\left(\frac{\partial \tensor{Q}^{f}}{\partial t}+\tensor{Q}^{\chi f}\right)\right).\end{split}
\end{equation}
A solution for which $\bm{P}$ is parallel or antiparallel to $\bm{\hat{E}}_{0}$ always exists, in which case Quincke rotation is impossible, is evident from~\eqref{torquebalance}. However, above a critical field strength, a solution in which the dipole moment has a transverse component also exists, which leads to Quincke rotation.
For a solid sphere, the critical field strength is given by
\begin{equation}\label{critfield}
 E_{\rm{c,s}}=\sqrt{\frac{2\mu^{+}(2+Q)(1+2S)}{3\varepsilon^{+}\tau_{\rm{MW}}(SQ-1)}},
\end{equation}
as first derived by~\citet{jones1984quincke}.

\subsection{Surfactant transport}\label{surfactanttransport}
The surfactant distribution evolves according to the dimensionless transport equation~\eqref{Gammatransnondimensional}. Substituting in $\Gamma$ from~\eqref{Gammadefinition}, $\bm{v}$ from~\eqref{interiorflow0} or~\eqref{exteriorflow0}, $\bm{n}$ and $\kappa_{m}$ from~\eqref{normal and curvature}, and retaining leading-order terms, we obtain a system of ODEs for the quadrupole $\tensor{Q}^{\Gamma},$
\begin{equation}\label{qgammaODE}
    \frac{\partial \tensor{Q}^{\Gamma}}{\partial t}=\frac{1}{\epsilon}\tensor{Q}^{p}-4\frac{\delta}{\epsilon}\left(\frac{\partial\tensor{Q}^{f}}{\partial t}+\tensor{Q}^{\chi f}\right)-\tensor{Q}^{\chi\Gamma}-\frac{6}{\zeta\rm{Re_{E}}}\tensor{Q}^{\Gamma},
\end{equation}
where 
\begin{equation}
\label{qchigamma} Q^{\chi\Gamma}_{ij}=Q^{\Gamma}_{il}\epsilon_{lkj}D_{k}^{\chi}+Q^{\Gamma}_{jl}\epsilon_{lki}D_{k}^{\chi}.
\end{equation}
Analogously to $\tensor{Q}^{\chi f}$ as defined in~\eqref{qchif}, $\tensor{Q}^{\chi \Gamma}$ acts to restore a uniform surfactant distribution after Quincke rotation begins. 
\subsection{Summary of differential equations}
The full system of nonlinear ODEs governing the system is written compactly as
\begin{equation}\label{eqnssummary}
    \frac{\mbox{d}}{\mbox{d}t}\begin{bmatrix}
        \bm{P}
        \\\tensor{Q}^{f}
        \\\tensor{Q}^{\Gamma}
\end{bmatrix}=\mathcal{F}\left(\bm{P},\tensor{Q}^{f},\tensor{Q}^{\Gamma};\mathrm{Ca_{E}},\mathrm{Re_{E}},\zeta,\rm{El},\epsilon,Q,S,\lambda\right).
\end{equation}
We use an explicit fourth-order Runge-Kutta method (RK4) to integrate the system of equations~\eqref{eqnssummary}. However, this procedure only allows us to obtain stable steady state solutions, and so in order to obtain unstable steady state solutions, we use a pseudo-arclength continuation algorithm. For the RK4 method, the initial conditions are those of an initially spherical drop with uniform surfactant distribution, $\tensor{Q}^{f}(0)=\tensor{Q}^{\Gamma}(0)=\bm{0}$,
and zero surface charge density $q(\bm{r},0)=0$. The charge density condition when substituted into~\eqref{chargedensity} gives the initial condition for the dipole moment,
\begin{equation}\label{dipoleICs}
\bm{P}(0)=\frac{Q-1}{2+Q}\bm{\hat{E}}_{0}. 
\end{equation}
For the pseudo-arclength continuation algorithm, we use the MATLAB nonlinear solver {\tt fsolve} as a corrector at each iteration and initial solutions are calculated using equations for the base state variables that will be given in \S\ref{lsa}.

\section{Results}\label{results}
In this section, we present results obtained from the model described in the previous sections.  In \S\ref{lsa}, we describe the linearisation of the differential equations derived in the previous section, which is used to examine the stability of the system. In \S\ref{validation}, we demonstrate that the predictions of the present model are in agreement with those of earlier models. In \S\ref{taylorregime}, we investigate the combined effects of surfactant and charge convection on the drop dynamics in the Taylor regime. In \S\ref{quinckeregime}, we explore how surfactant effects impact the stability and angular velocity of a drop in the Quincke regime.

\subsection{Linear stability analysis}\label{lsa}
To investigate the onset of Quincke rotation, we linearly perturb the system around a steady, axisymmetric base state corresponding to a drop in the Taylor regime which allows us to perform the analysis in the $xz$ plane without loss of generality. Denoting base state quantities by a zero subscript, axisymmetry implies
\begin{align}
    P_{0,x}=Q^{\Gamma}_{0,xz}=Q^{f}_{0,xz}=0, \quad Q^{\Gamma}_{0,xx}=Q^{\Gamma}_{0,yy},\quad Q^{f}_{0,xx}=Q^{f}_{0,yy}. 
\end{align}
Therefore, $\bm{D}_{0}^{\chi}=\bm{0}$ and so $\tensor{Q}_{0}^{\chi f}=\tensor{Q}_{0}^{\chi\Gamma}=\tensor{0}.$
Explicitly, we perturb the variables as
  \begin{equation}
      \begin{pmatrix}
          P_{x}\\P_{z}\\Q_{xx}^{f}\\Q_{zz}^{f}\\Q_{xz}^{f}\\Q_{xx}^{\Gamma}\\Q_{zz}^{\Gamma}\\Q_{xz}^{\Gamma}
      \end{pmatrix}=\begin{pmatrix}
          0\\P_{0,z}\\Q_{0,xx}^{f}\\Q_{0,zz}^{f}\\0\\Q_{0,xx}^{\Gamma}\\Q_{0,zz}^{\Gamma}\\0
      \end{pmatrix}+\Delta\begin{pmatrix}
              P_{\Delta,x}\\P_{\Delta,z}\\Q_{\Delta,xx}^{f}\\Q_{\Delta,zz}^{f}\\Q_{\Delta,xz}^{f}\\Q_{\Delta,xx}^{\Gamma}\\Q_{\Delta,zz}^{\Gamma}\\Q_{\Delta,xz}^{\Gamma}
      \end{pmatrix},
  \end{equation}
where the small parameter $\Delta\ll 1$ is a measure of the size of the perturbations to the variables, and the $\Delta$ subscript denotes the perturbed variables. 
%In order to analyse linear perturbations, we consider first order terms in $\Delta.$ 
The $z$ component of the base state dipole moment, $P_{0,z}$, is computed from a cubic equation derived from the charge transport equation~\eqref{chargeconsnondim}, namely,
  \begin{equation}
    \mathcal{A}P_{0,z}^{3}+\mathcal{B}P_{0,z}^{2}+\mathcal{C}P_{0,z}+\mathcal{D}=0,
  \end{equation}
where
\begin{subequations}
  \begin{align}
&\mathcal{A}=-\frac{2(2+Q)}{25(1+\lambda)+5\frac{\mathrm{El}}{\rm{Ca_{E}}}\epsilon},
  \\
  & \mathcal{B}=\frac{6Q}{25(1+\lambda)+5\frac{\mathrm{El}}{\rm{Ca_{E}}}\epsilon},
  \\
  &\mathcal{C}=-1+\frac{6(1-Q^2)}{25(2+Q)(1+\lambda)+5(2+Q)\frac{\mathrm{El}}{\rm{Ca_{E}}}\epsilon},
  \\
      &\mathcal{D}=\frac{1-S}{1+2S}+\frac{2(1-Q)^{2}}{25(2+Q)(1+\lambda)+5(2+Q)\frac{\mathrm{El}}{\rm{Ca_{E}}}\epsilon},
      \end{align}
  \end{subequations}
which must be solved numerically. The base state surfactant coefficients are given by
\begin{subequations}
\begin{align}
Q^{\Gamma}_{0,xx}=&\;-\frac{1-Q+(1+2Q)P_{0,z}-(2+Q)P_{0,z}^{2}}{\epsilon\left(30(1+\lambda)\frac{1}{\zeta\rm{Re_{E}}}+6\frac{\rm{El}}{\rm{Ca_{E}}}\right)},
\\
Q^{\Gamma}_{0,zz}=&\;-2Q^{\Gamma}_{0,xx},\label{QGamma0zz}
\end{align}    
\end{subequations}
and the base state shape coefficients are
\begin{subequations}
    \begin{align}\begin{split}
        Q^{f}_{0,xx}=&\;\frac{(1+2S)^2}{3}\left[-\frac{3}{2}P_{0,z}^{2}-\frac{(16+19\lambda)\frac{1}{\zeta\rm{Re_{E}}}+4\frac{\rm{El}}{\rm{Ca_{E}}}}{30(1+\lambda)\frac{1}{\zeta\rm{Re_{E}}}+6\frac{\rm{El}}{\rm{Ca_{E}}}}\right.\\ &\;\times\left(1-Q+(1+2Q)P_{0,z}-(2+Q)P_{0,z}^2\right)\Biggr],\end{split}
        \\Q^{f}_{0,zz}=&\;-2Q^{f}_{0,xx}.
    \end{align}
\end{subequations}
Ultimately, the resulting linear system can be written as

    \begin{equation}\label{linearsystem}
\frac{\mbox{d}}{\mbox{d}t}\begin{pmatrix}
              P_{\Delta,x}\\P_{\Delta,z}\\Q_{\Delta,xx}^{f}\\Q_{\Delta,zz}^{f}\\Q_{\Delta,xz}^{f}\\Q_{\Delta,xx}^{\Gamma}\\Q_{\Delta,zz}^{\Gamma}\\Q_{\Delta,xz}^{\Gamma}
      \end{pmatrix}=\begin{bmatrix}J_{11}&0&0&0&J_{15}&0&0&J_{18}
      \\0&J_{22}&0&J_{24}&0&0&J_{27}&0
      \\0&J_{32}&J_{33}&0&0&J_{36}&0&0
      \\0&J_{42}&0&J_{44}&0&0&J_{47}&0
      \\J_{51}&0&0&0&J_{55}&0&0&J_{58}
      \\0&J_{62}&J_{63}&0&0&J_{66}&0&0
      \\0&J_{72}&0&J_{74}&0&0&J_{77}&0
      \\J_{81}&0&0&0&J_{85}&0&0&J_{88}
      
      \end{bmatrix}\begin{pmatrix}
              P_{\Delta,x}\\P_{\Delta,z}\\Q_{\Delta,xx}^{f}\\Q_{\Delta,zz}^{f}\\Q_{\Delta,xz}^{f}\\Q_{\Delta,xx}^{\Gamma}\\Q_{\Delta,zz}^{\Gamma}\\Q_{\Delta,xz}^{\Gamma}
      \end{pmatrix},
  \end{equation}
where, for brevity, the elements of the Jacobian matrix are given in appendix A. The electric field strength is increased until one of the eigenvalues of the Jacobian has a positive real part, indicating instability.

\subsection{Validation}\label{validation}
In this section, we demonstrate that the present model recovers the linear term of the second-order small-deformation theory of~\citet{nganguia2013equilibrium} in the limits of zero electric Reynolds number, $\mathrm{Re_E} \rightarrow 0$, and zero surfactant diffusion, $\mathrm{Pe_s} \rightarrow \infty$ or equivalently $\zeta \rightarrow \infty$. Our model also reduces to that of~\citet{stone1990effects} if the electric field is switched off and the drop is instead exposed to an extensional flow.

First, we consider the behaviour of a drop in the Taylor regime in the limit of zero surfactant diffusion, as studied by~\citet{nganguia2013equilibrium}. In this limit, the flow is completely arrested once the surfactant has been redistributed such that the Marangoni stresses balance the electric stresses. Therefore, charge convection can have no impact on the steady drop shape or surfactant distribution, although it can have a transient impact while the surfactant is being redistributed. At steady state, in the absence of charge convection, the charge transport equation simplifies to
\begin{equation}\label{currentcontinuity}
    \llbracket\bm{j}\bcdot\bm{\hat{r}}\rrbracket=0,
\end{equation}
which yields a simple expression for the dipole moment,
\begin{equation}\label{simpledipole}
    \bm{P}=\frac{1-S}{1+2S}\bm{\hat{E}}_{0}.
\end{equation}
Hence, if charge convection is absent, then the dipole moment is either parallel or anti-parallel to the applied field and Quincke rotation is impossible, as is clear from~\eqref{torquebalance}. Then, the terms associated with rotation ($\tensor{Q}^{\chi f}$ and $\tensor{Q}^{\chi\Gamma}$) may be removed from the ODE for the surfactant quadrupole~\eqref{qgammaODE}, which in the limit $\zeta\rightarrow\infty$ becomes
\begin{equation}
    \frac{\partial \tensor{Q}^{\Gamma}}{\partial t}=\frac{1}{\epsilon}\tensor{Q}^{p}+4\frac{\delta}{\epsilon}\frac{\partial\tensor{Q}^{f}}{\partial t}.
\end{equation}
Evidently, at steady state we have $\tensor{Q}^{p}=\tensor{0}.$ Using this result together with~\eqref{simpledipole} allows us to obtain algebraic solutions of Eqs.~\eqref{straining} and~\eqref{shapetrans}, yielding
\begin{subequations}
\begin{align}\label{Qfnondiffusing}
    \tensor{Q}^{f}=\frac{3}{2}\left(4S(1-SQ)+(1-S)^{2}\right)\left(\bm{\hat{E}}_{0}\bm{\hat{E}}_{0}-\frac{1}{3}\tensor{I}\right)
    \end{align}
    and
\begin{align}
     \tensor{Q}^{\Gamma}=\frac{9S(1-SQ)}{2(1+2S)^2}\left(\bm{\hat{E}}_{0}\bm{\hat{E}}_{0}-\frac{1}{3}\tensor{I}\right).
     \end{align}
\end{subequations}
Thus, the sign of the discriminating function $4S(1-SQ)+(1-S)^{2}$ determines the prolate (positive) or oblate (negative) nature of the deformation, as found by~\citet{nganguia2013equilibrium}. To quantify the deformation, we use the parameter $D_{Q},$ defined as
\begin{equation}
    D_{Q}=\frac{L-B}{L+B},
\end{equation}
where $L$ and $B$ are the long and short axes of the drop in the $xz$ plane. In the Taylor regime, the drop adopts a steady spheroidal shape and $L$ and $B$ are parallel and perpendicular to the applied field, or vice versa. In the Quincke regime, the drop adopts a steady tilted shape with its longest axis misaligned with the $z$ axis (in the case of a prolate B drop) or the $x$ axis (in the case of an oblate drop) by an angle which we denote $\theta_{T}$. The deformation parameter $D_{Q}$ can be written as
\begin{equation}\label{dqdef}
    D_{Q} = \frac{1}{2} \delta |Q^{f}_{xx} - Q_{zz}^{f}| \sqrt{1 + \left(\frac{2Q^{f}_{xz}}{Q_{xx}^{f} - Q_{zz}^{f}}\right)^{2}} + \mathcal{O}(\delta^{2}).
\end{equation}
Note that $D_{Q}$ is always positive, unlike the more commonly used deformation parameter $D=(l_{\parallel}-l_{\perp})/(l_{\parallel}+l_{\perp}),$ where $l_{\parallel}$ and $l_{\perp}$ are the lengths of the axes of the drop parallel and perpendicular to the applied field, respectively. In the Taylor regime, $Q^{f}_{xz}=0$ so $D_{Q} = \frac{1}{2} \delta |Q^{f}_{xx} - Q_{zz}^{f}|$, in which case $D_{Q}=|D|.$ Using this expression, the steady drop shape~\eqref{Qfnondiffusing} corresponds to a deformation of
\begin{equation}\label{eq:deformation_nganguia}
    D_{Q}=\frac{9\rm{Ca_{E}}}{16(1+2S)^2}|4S(1-SQ)+(1-S)^{2}|,
\end{equation}
which is equivalent to the linear term of Eq.~(18) of~\citet{nganguia2013equilibrium}, up to the sign convention for oblate deformations.

Next, to demonstrate consistency with the results of the model of~\citet{stone1990effects}, we consider the case in which diffusion is strong, and replace the electric field with an imposed straining flow of the form
\begin{equation}\label{extensionalflow}
    \bm{v}^{\infty}=\tensor{S}_{e}\cdot\bm{r}=\pm k\begin{bmatrix}
         -\frac{1}{2}  & 0 & 0\\0 & -\frac{1}{2} & 0\\0&0&1
    \end{bmatrix}\cdot\bm{r},
\end{equation}
where $k$ is the shear rate. Equation~\eqref{extensionalflow} corresponds to a uniaxial extensional flow for positive $k,$ and a biaxial extensional flow for negative $k.$ These flow fields have been studied by~\citet{stone1990effects} for low $\rm{Pe}_{\rm{s}}$ and involve qualitatively similar convective fluxes at the drop surface to those induced by a prolate A drop (uniaxial extension) or an oblate or prolate B drop (biaxial extension) exposed an electric field. In either case, the surfactant transport equation~\eqref{qgammaODE} becomes

\begin{equation}\label{qgammatransstone}
    \frac{\partial \tensor{Q}^{\Gamma}}{\partial t}=\frac{1}{\epsilon}\tensor{Q}^{p}-4\frac{\rm{Ca}}{\epsilon}\frac{\partial\tensor{Q}^{f}}{\partial t}+\frac{5}{\epsilon}\tensor{S}_{e}-\frac{6}{\rm{Pe}_{\rm{s}}}\tensor{Q}^{\Gamma},
\end{equation}
where \begin{equation}{\rm Ca}=\frac{\mu^{+}ka}{\gamma_{\rm{eq}}}\ll 1\end{equation} is the capillary number, which represents the ratio of viscous to capillary stresses. Also, the P\'eclet number for the extensional flow is \begin{equation}{\rm Pe}_{\rm{s}}=\frac{ka^{2}}{D_{s}},\end{equation} and the flow coefficients in $\tensor{Q}^{p}$ are given by
\begin{equation}\tensor{Q}^{p}=\frac{2}{3+2\lambda}\left(-4\tensor{Q}^{f}
    -\frac{\rm{El}}{\mathrm{Ca}}\epsilon\tensor{Q}^{\Gamma}+5(1-\lambda)\tensor{S}_{e}\right).\end{equation}
In addition, the drop shape evolves according to 

\begin{equation}\label{shapetransstone}
   \frac{\partial \tensor{Q}^{f}}{\partial t}=\frac{1}{\rm{Ca}(3+2\lambda)}\left(-\frac{40(1+\lambda)}{(16+19\lambda)}\tensor{Q}^{f}+\frac{2(4+\lambda)}{(16+19\lambda)}\frac{\rm{El} }{\rm{Ca}}\epsilon\tensor{Q}^{\Gamma}+5\tensor{S}_{e}\right).
\end{equation}
Then, $\tensor{Q}^{\Gamma}$ can be eliminated between~\eqref{qgammatransstone} and~\eqref{shapetransstone} to form an equation for $\partial \tensor{Q}^{\Gamma}/\partial t$ in terms of $\partial \tensor{Q}^{f}/\partial t,$ $\tensor{Q}^{f}$ and $\tensor{S}_{e}.$ Differentiating~\eqref{shapetransstone} and substituting in this equation gives a second order ODE for $\tensor{Q}^{f}$ which is decoupled from $\tensor{Q}^{\Gamma},$
\begin{align}
\begin{split}
    \frac{(3+2\lambda)(16+19\lambda)}{2}\frac{\rm{Ca}^{2}}{\rm{El}}\frac{\partial^{2} \tensor{Q}^{f}}{\partial t^{2}}+\left[(32+23\lambda)\rm{Ca}+20(1+\lambda)\frac{\rm{Ca}}{\rm{El}}\right.\\\left.+3(3+2\lambda)(16+19\lambda)\frac{\rm{Ca}^{2}}{\rm{El}\rm{Pe}_{\rm{s}}}\right]\frac{\partial \tensor{Q}^{f}}{\partial t}+24\left[1+5(1+\lambda)\frac{\rm{Ca}}{\rm{Pe}_{\rm{s}}}\right]\tensor{Q}^{f} 
    \\=15\left[4+(16+19\lambda)\frac{\rm{Ca}}{\rm{El}\rm{Pe}_{\rm{s}}}\right]\tensor{S}_{e},
\end{split}
\end{align}
which is equivalent to Eq.~(22) of~\citet{stone1990effects}. A similar decoupled ODE can be written for $\tensor{Q}^{\Gamma}.$

\subsection{Taylor regime}\label{taylorregime}

\begin{figure}
    \centering
    \includegraphics[width=\linewidth]{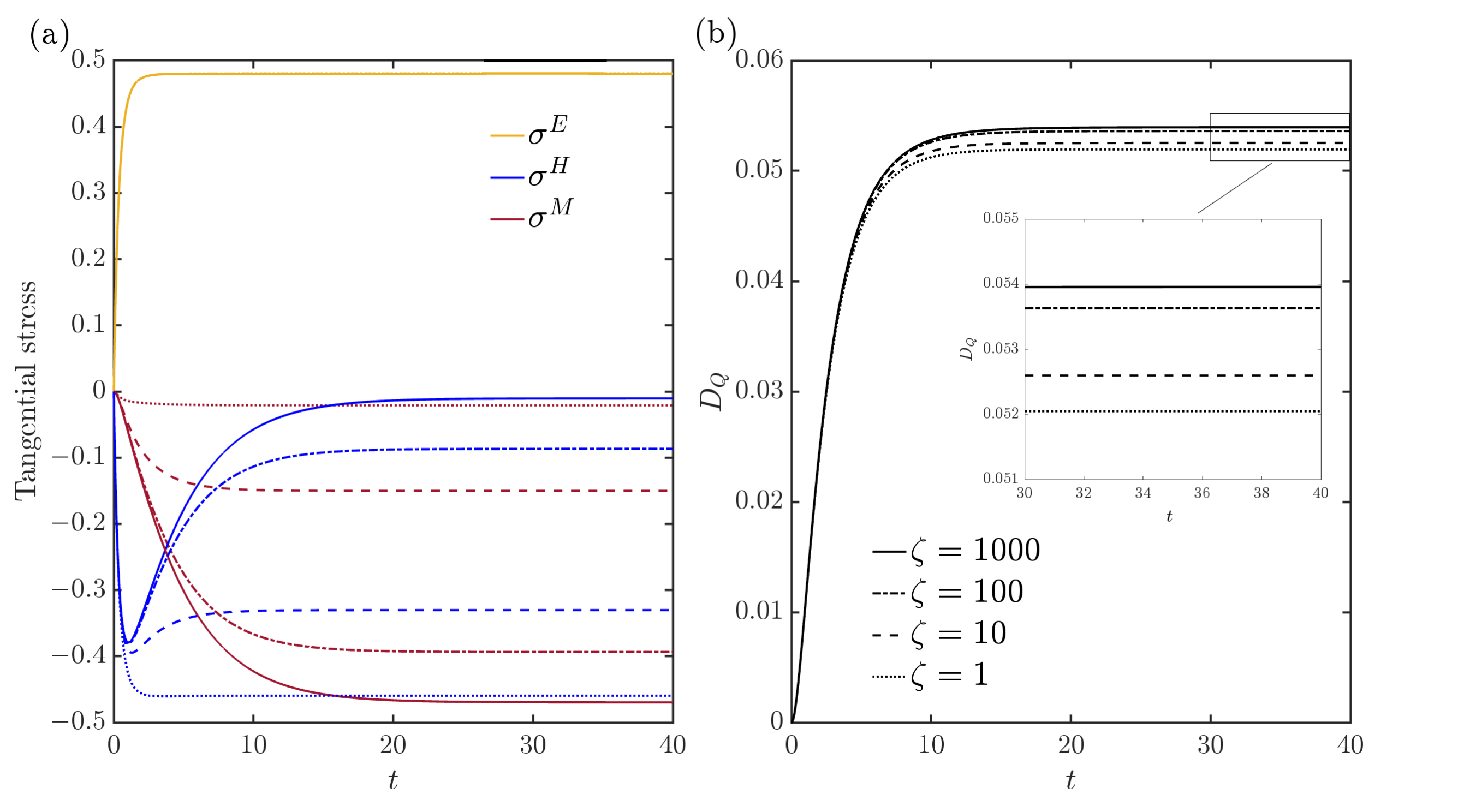}
    \caption{Evolution of (a) the tangential electric (yellow lines), hydrodynamic (blue lines) and Marangoni (dark red lines) stresses and (b) the deformation of a prolate A drop, for $\zeta=1$ (dotted lines), $\zeta=10$ (dashed lines), $\zeta=100$ (dash-dotted lines) and $\zeta=1000$ (solid lines). In these figures, $S=1/3$, $Q=1,$ $\lambda=10,$ ${\rm El}=0.5$, ${\rm Re_{E}=1},$ $\rm{Ca_{E}}=0.2.$}
    \label{transientfigure}
\end{figure}
In this section, we study the combined effects of surfactant and charge convection on the drop dynamics in the Taylor regime.
\subsubsection{Transient behaviour}First, Figure~\ref{transientfigure}(a) shows the evolution of the tangential electric (yellow lines), hydrodynamic (blue lines) and Marangoni (dark red lines) stresses on the surface of a prolate A drop for which $S=1/3$, $Q=1,$ $\lambda=10.$ Specifically, we plot $\sigma^{E},$ $\sigma^{H}$ and $\sigma^{M},$ defined as 
\begin{subequations}
\begin{align}
    \sigma^{E}=Q^{E}_{\bm{\hat{t}},zz},
    \quad\sigma^{H}=Q^{H}_{\bm{\hat{t}},zz},
    \quad\sigma^{M}=-2\frac{{\rm El}\epsilon}{{\rm Ca_{E}}}Q^{\Gamma}_{zz}.
    \end{align}
\end{subequations}

In Figure~\ref{transientfigure}, ${\rm El}=0.5$, ${\rm Re_{E}=1},$ $\rm{Ca_{E}}=0.2,$ and results are shown for $\zeta=1$ (dotted lines), $\zeta=10$ (dashed lines), $\zeta=100$ (dash-dotted lines), and $\zeta=1000$ (solid lines). The electric stress does not change much between these values of $\zeta,$ suggesting that charge convection does not impact the steady state dipole moment values from equation \eqref{dipoleevolution} significantly and so the four yellow lines are indistinguishable. When $\zeta=1,$ the electric stress shears the fluid into motion and is largely balanced by the resulting hydrodynamic stress, while the Marangoni stress remains relatively weak due to relatively strong surfactant diffusion. Each of $\sigma^{E},$ $\sigma^{H},$ and $\sigma^{M}$ monotonically approaches its steady value with no overshoot as time increases. When $\zeta=10,$ the flow is more suppressed than when $\zeta=1,$ and the Marangoni stress plays a stronger role in balancing the electric stress than it does when $\zeta=1,$ although the hydrodynamic stress is still the stronger of the two. In this case, the hydrodynamic stress slightly overshoots its steady value as the steady surfactant distribution is established. Consistent with previous studies~\citep{nganguia2013equilibrium,mandal2016dielectrophoresis,nganguia2019effects}, the flow is suppressed substantially in the cases $\zeta=100$ and $\zeta=1000$, and the Marangoni stress plays a much stronger role in balancing the electric stress than does the hydrodynamic stress. The hydrodynamic stress initially increases quickly in response to the electric stresses, convecting surfactant around the drop. This creates surface tension gradients that lead to a relatively strong Marangoni stress that suppresses the flow and decreases the hydrodynamic stress. Figure~\ref{transientfigure}(b) shows the corresponding evolution of the deformation parameter $D_{Q}$, and the inset displays an enlarged view of the steady values. As expected, the deformation increases with $\zeta,$ and approaches its steady value monotonically as time increases for all of the values of $\zeta$ shown.
\subsubsection{Steady behaviour}

\begin{figure}
    \centering
    \includegraphics[width=\linewidth]{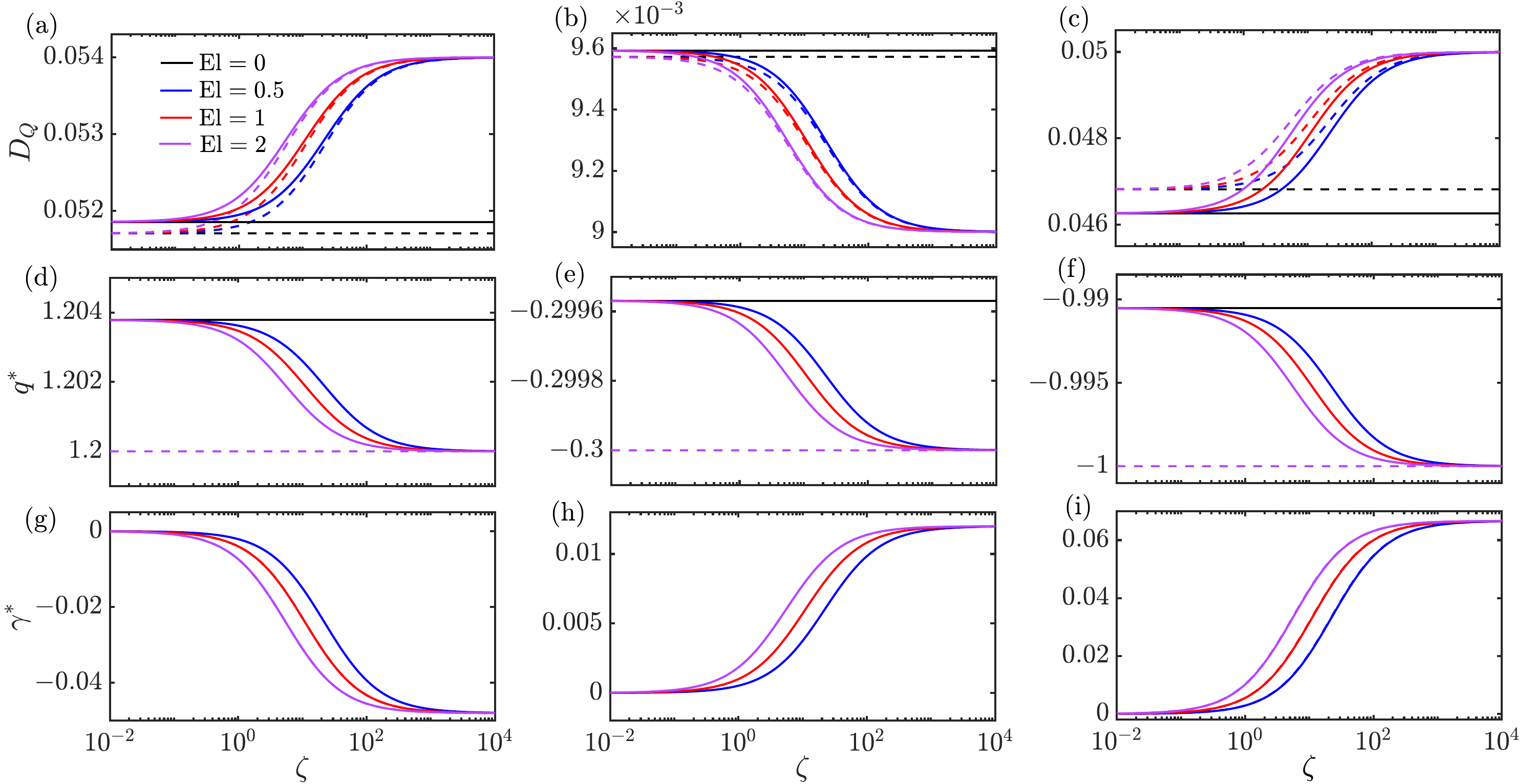}
    \caption{Dependence of the steady deformation parameter $D_{Q}$ of (a) a prolate A drop with $S=1/3$, $Q=1$, (b) a prolate B drop with $S=1/3$, $Q=7/2,$ and (c) an oblate drop with $S=1$, $Q=2$ on $\zeta$ for $\rm{El}=$ 0 (black lines), 0.5 (blue lines), 1 (red lines) and 2 (purple lines). The solid lines are the results obtained when retaining charge convection, while the dashed lines are the results obtained when neglecting charge convection. (d--f) show the strength of the surface charge distribution, and (g--i) show the strength of the surface tension variations. In all cases, $\rm{Ca_{E}}=0.2,$ $\rm{Re_{E}}=1,$ $\lambda=10.$}
    \label{DvsPs}
\end{figure}
Figures~\ref{DvsPs}(a--c) show the dependence of the steady deformation parameter $D_{Q}$ of the three drop classes (prolate A, prolate B, and oblate) on $\zeta$ for four values of $\rm{El}.$ Figure~\ref{DvsPs}(a) shows results for a prolate A drop with $S=1/3,$ $Q=1$, Figure~\ref{DvsPs}(b) shows results for a prolate B drop with $S=1/3$, $Q=7/2$, and Figure~\ref{DvsPs}(c) shows results for an oblate drop with $S=1,$ $Q=2$. In the Taylor regime, the charge distribution $q$ can be written as
\begin{equation}
    q=q^{*}\cos\theta,
\end{equation}
where $q^{*}=1-Q+(2+Q)P_{z},$ while the surface tension can be written as
\begin{equation}
    \gamma=1+\gamma^{*}P_{2}(\cos\theta),
\end{equation}
where $\gamma^{*}=-{\rm Ca_{E}}Q^{\Gamma}_{zz}$ and $P_{2}(\cos\theta)=(3\cos^{2}\theta-1)/2$ is the second Legendre polynomial. Figures~\ref{DvsPs}(d--f) show $q^{*}$, and Figures~\ref{DvsPs}(g--i) show $\gamma^{*}$, again as functions of $\zeta$. In all cases in Figure~\ref{DvsPs}, $\rm{Ca_{E}}=0.2,$ $\rm{Re_{E}}=1,$ and $\lambda=10.$ The solid lines are the results obtained when retaining charge convection, while the dashed lines are the results obtained when neglecting charge convection. Black lines correspond to a uniform surface tension drop, i.e., $\rm{El}=0,$ while the blue, red and purple lines correspond to $\rm{El}=0.5, 1, 2,$ respectively. Note that in Figures~\ref{DvsPs}(g--i), each dashed line lies almost on top of the solid line of the same colour indicating that charge convection has a minimal effect. It is helpful to first examine the dashed lines in Figure~\ref{DvsPs}, which reveal how the drop dynamics depend on $\zeta$ in the absence of charge convection as studied previously~\citep{nganguia2019effects}. For all three classes of drops, in the limit of strong surfactant diffusion ($\zeta\rightarrow 0$), Figures~\ref{DvsPs}(g--i) show that the surfactant distribution remains uniform regardless of the value of $\mathrm{El},$ and Figures~\ref{DvsPs}(a--c) show that all drops of each class thus attain the same deformation as a uniform surface tension drop of the relevant class. As the value of $\zeta$ increases, the effect of varying the elasticity number $\mathrm{El}$ becomes stronger. In the cases of prolate A and oblate drops, the deformation increases with increasing $\zeta$ or $\mathrm{El}$, while both trends are reversed for prolate B drops. However, in the case of prolate B and oblate drops, the surface tension at the poles increases with increasing $\zeta$ or $\mathrm{El},$ while the surface tension at the poles of a prolate A drop decreases with increasing $\zeta$ or $\mathrm{El}.$ As we further increase $\zeta$ towards the limit of zero surfactant diffusion ($\zeta\rightarrow\infty$), the flow is completely arrested at steady state. Within each class, all drops converge to the same surface tension profile and deformation. Under these circumstances, the surfactant is redistributed by the initial fluid flow until the Marangoni stresses alone balance the tangential electric stresses, and the degree of deformation depends solely on $\rm{Ca_{E}},$ $S,$ and $Q,$ as can be seen from~\eqref{eq:deformation_nganguia}. Note that in Figures~\ref{DvsPs}(d--f), the dashed lines of all four colours lie exactly on top of each other. This is because in the absence of charge convection, the charge distribution is decoupled from the flow and is found by taking the simplified expression for the dipole moment~\eqref{simpledipole}, yielding
\begin{equation}\label{chargedistribution_noconvection}
    q=\left(1-Q+\frac{(2+Q)(1-S)}{1+2S}\right)\cos\theta
\end{equation} 
for the charge. Thus, the surfactant can affect the charge distribution only through modifying the fluid flow appearing in the convective term in the full dipole evolution equation~\eqref{dipoleevolution}. Next, we focus on the solid lines in Figure~\ref{DvsPs}, which correspond to results obtained when charge convection is retained in the model.  Previous studies have found that in the absence of surfactants charge convection increases deformation for prolate drops and decreases deformation for oblate drops~\citep{feng1999electrohydrodynamic,lanauze2015nonlinear,das2017nonlinear}. We observe the same effects for surfactant covered drops as well for all values of $\mathrm{El}$ and $\zeta$. This is clearly shown in Figures~\ref{DvsPs}(a,b), in which the solid lines lie above the dashed lines for the prolate drops, and Figure~\ref{DvsPs}(c), in which the solid lines lie below the dashed lines for the oblate drops. As noted in previous studies, charge convection has a greater impact on the deformation of oblate drops than prolate ones. The impact on prolate B drops is particularly small; the hydrodynamic stress, which promotes oblate deformation, is slightly weakened compared to the dominant electric stress, which promotes prolate deformation and is strengthened by charge convection.  In the limit of strong surfactant diffusion ($\zeta\rightarrow 0$), the surfactant distribution remains uniform regardless of the value of $\mathrm{El}.$ The effect of charge convection is strongest in this limit due to the absence of Marangoni stresses. In the limit of zero surfactant diffusion ($\zeta\rightarrow\infty$), the opposite happens and the flow is completely arrested at steady state, hence charge convection has no effect on drop deformation. All surfactant-laden drops of each class attain the same degree of deformation, regardless of the value of $\rm{El}.$ 

Figures~\ref{DvsPs}(d--f) show that for prolate A drops, low values of $\zeta$ lead to a stronger positive charge distribution at the north pole (relative to the direction of the applied field) of the drop, while for oblate and prolate B drops, low values of $\zeta$ lead to a weaker negative charge distribution at the north pole of the drop. In the prolate A class (Figure~\ref{DvsPs}(d)), the charge distribution is parallel to the field, and since the interfacial velocity carries charges from the equator to the poles, the charge distribution is strengthened by convection. In the prolate B (Figure~\ref{DvsPs}(e)) and oblate classes (Figure~\ref{DvsPs}(f)), the charge distribution is antiparallel to the field, and since the interfacial velocity carries charges from the poles to the equator, the charge distribution is weakened by convection. For all three classes, as $\zeta$ is increased, the effect of charge convection is diminished because the flow is suppressed, and the value of $q^{*}$ converges to that in the absence of convection given in~\eqref{chargedistribution_noconvection}. 

Figures~\ref{DvsPs}(g--i) show that charge convection has very little impact on the surface tension profile for drops in all three classes for the parameter values used. This is explained by the fact that the flow is suppressed when surfactant diffusion is weak, while the surface tension profile is almost uniform regardless of charge convection when surfactant diffusion is strong. 

\subsection{Quincke regime}\label{quinckeregime}

In the Quincke regime, the applied electric field exceeds a critical threshold, and the axisymmetric steady state of the drop becomes unstable. When the axisymmetry is broken by a perturbation to the drop shape or dipole moment, the drop tilts relative to the applied field and undergoes steady rotation with a constant angular velocity. In this section, we discuss the effects of surfactant on the drop dynamics in the Quincke regime.

\subsubsection{Hysteresis}
\begin{figure}
    \centering
    \includegraphics[width=\linewidth]{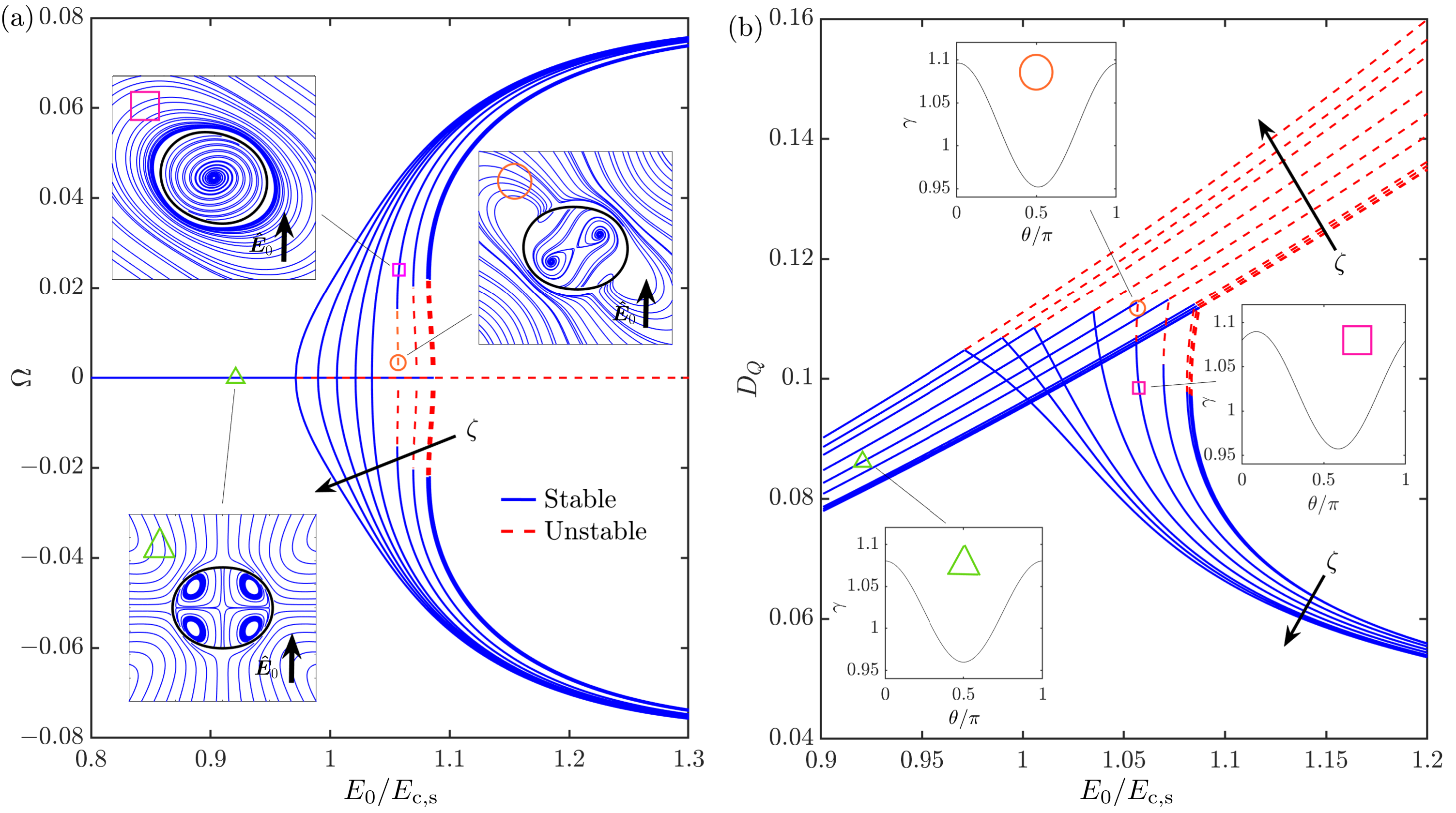}
    \caption{Bifurcation diagrams showing (a) the angular velocity $\Omega$ and (b) the deformation $D_{Q}$ of the drop as functions of the scaled electric field strength $E_{0}/E_{\mathrm{c,s}}$ for eleven values of $\zeta=0.01,$ $0.1,$ $0.5,$ $1,$ $5,$ $10,$ $20,$ $30,$ $50,$ $100,$ $1000,$ where solid blue lines denote stable solutions and dashed red lines denote unstable solutions. The arrows indicate the directions of increasing $\zeta.$ The insets in (a) show streamlines around the drop, and the insets in (b) show the corresponding surface tension as a function of $\theta.$ In all cases, $\rm{El}=0.5$ and the other parameter values are those used by~\citet{salipante2010electrohydrodynamics}, given in Table~\ref{parametervalues}. }
    \label{bifurcations}
\end{figure}

The experiments of~\citet{salipante2010electrohydrodynamics} demonstrated hysteresis in the angular velocity of drops undergoing Quincke rotation; this behaviour has also been captured theoretically~\citep{peng2025bistability,mcdougall2025nonlinear}. We first show that the present model reproduces this hysteresis and then examine how surfactant alters it. The bifurcation diagrams in Figure~\ref{bifurcations} show the angular velocity $\Omega$ (Figure~\ref{bifurcations}(a)) and the deformation $D_{Q}$ (Figure~\ref{bifurcations}(b)) of the drop as functions of the scaled electric field strength $E_{0}/E_{\mathrm{c,s}}.$ In Figure~\ref{bifurcations}, solid blue lines denote stable solutions and dashed red lines denote unstable solutions. We show solutions for eleven values of $\zeta=0.01,$ $0.1,$ $0.5,$ $1,$ $5,$ $10,$ $20,$ $30,$ $50,$ $100,$ $1000.$ Note that in both Figure~\ref{bifurcations}(a) and~\ref{bifurcations}(b), the curves for $\zeta=0.01,$ $0.1,$ $0.5,$ $1$ lie almost on top of one another and so appear as a single thick line. 
The insets in Figure~\ref{bifurcations}(a) show streamlines around the drop for three different solutions for the case $\zeta=10,$ and the insets in Figure~\ref{bifurcations}(b) show the corresponding surface tension as a function of $\theta.$ In all cases, $\rm{El}=0.5$ and the other parameter values are those used by~\citet{salipante2010electrohydrodynamics}, given in Table~\ref{parametervalues}. 

Figure~\ref{bifurcations}(a) shows that a stable solution with a straining flow but zero rotational flow (streamlines shown in the inset with the green triangle, with maximum velocity $0.0058a/\tau_{{\rm EHD}}$) exists up to a critical electric field strength, at which it becomes unstable and a perturbation to the dipole moment or shape of the drop will lead to steady rotation (streamlines shown in the inset with the pink square, with maximum velocity $0.045a/\tau_{{\rm EHD}}$). The present model predicts that hysteresis occurs for the cases in which $\zeta=10$ or less, and that the critical field strength for rotation decreases substantially as $\zeta$ is increased. The suppression of the straining flow at high values of $\zeta$ may contribute to this effect; previous work on drops without surfactant have shown that the critical field strength decreases as the strength of the straining flow decreases~\citep{das2021three}. Remarkably, for $\zeta=100$ and $\zeta=1000$, the model predicts the onset of rotation at a lower critical field strength than that for a solid sphere. In the cases exhibiting hysteresis, the onset of rotation corresponds to a subcritical pitchfork bifurcation, and there is an unstable solution branch on which the drop is tilted and rotates (streamlines shown in the inset with the orange circle, with maximum velocity $0.011a/\tau_{{\rm EHD}}$). At higher $\zeta,$ the pitchfork bifurcation becomes supercritical, and no unstable rotating solution exists. 
\begin{table}
    \centering
    \begin{tabular}{c c c c c c c c}

         $\epsilon^{+}/\epsilon_{0}$ & $\epsilon^{-}/\epsilon_{0}$ & $\sigma^{+}$ & $\sigma^{-}$ & $\mu^{+}$& $\mu^{-}$ & $\gamma$ &$a$   \\ \hline
         $5.3$  & $3$ & \SI{4.5e-11}{\siemens/\meter} & \SI{1.23e-12}{\siemens/\meter} & \SI{0.69}{\Pa/\second}& \SI{9.73}{\Pa/\second} & \SI{4.5e-3}{\N/\meter} &\SI{0.9e-3}{\meter}\\ 
    \end{tabular}
    \caption{Material properties used by~\citet{salipante2010electrohydrodynamics}. $\epsilon_{0}=$\SI{8.8542e-12}{\farad/\meter} is the permittivity of free space.}
    \label{parametervalues}
\end{table}

\begin{figure}
    \centering
    \includegraphics[width=0.7\linewidth]{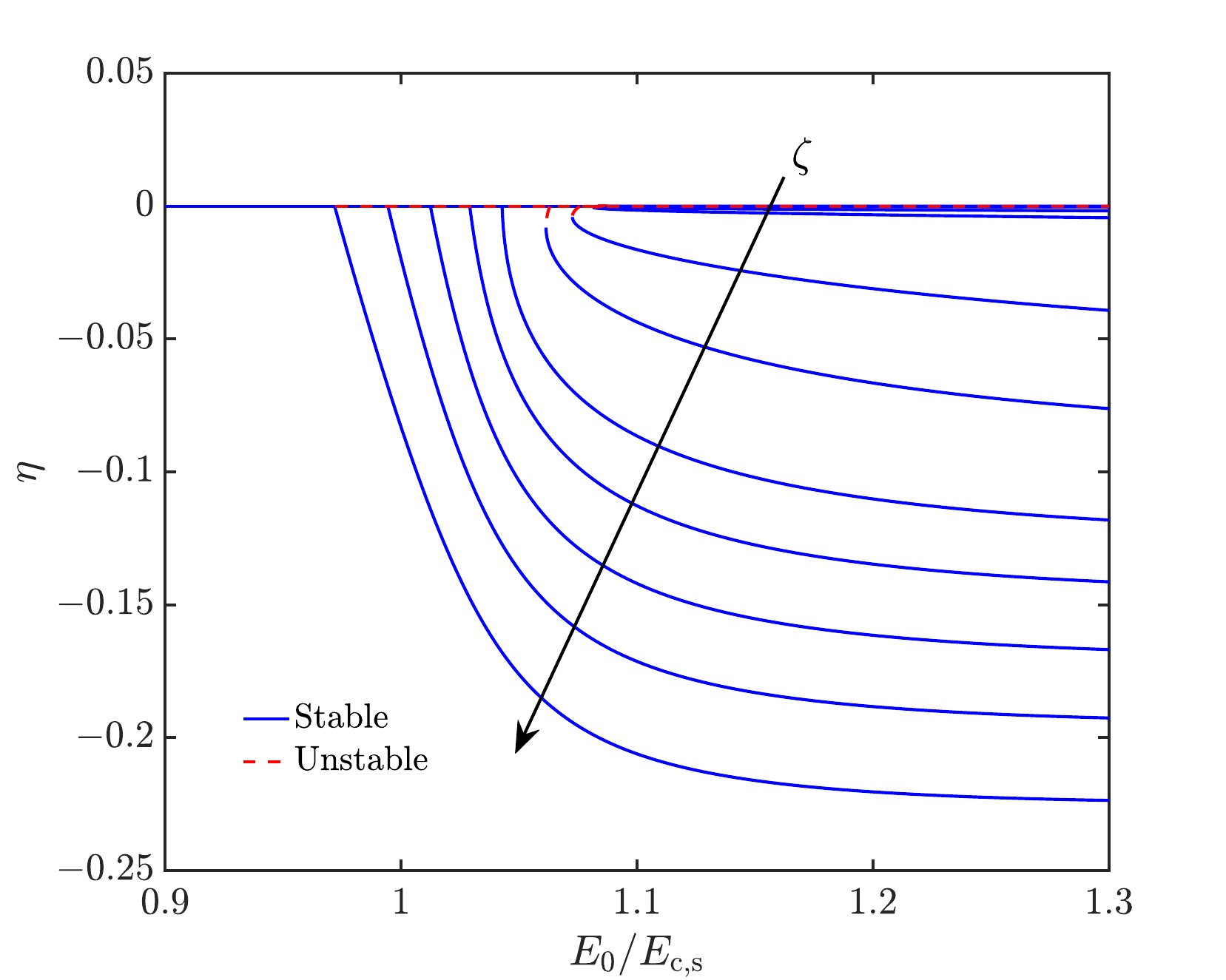}
    \caption{Parameter $\eta$ given by~\eqref{etadef} as a function of the scaled electric field strength $E_{0}/E_{\mathrm{c,s}}$ for the same parameter values as those used in Figure~\ref{bifurcations}. The arrow indicates the direction of increasing $\zeta.$}
    \label{etafig}
\end{figure}

Figure~\ref{bifurcations}(b) shows the drop deformation corresponding to the angular velocities in Figure~\ref{bifurcations}(a). For the given set of parameters, the drop attains an oblate shape in the Taylor regime. Before Quincke rotation begins, the deformation is greater for higher values of $\zeta$ than for lower ones, consistent with Figure~\ref{DvsPs}(c), and surfactant accumulates at the drop equator and decreases the surface tension there (see the inset marked with the green triangle), as discussed previously. After the onset of rotation, the deformation decreases due to the rotational flow through the term $\tensor{Q}^{\chi f},$ as discussed in~\S\ref{kinematics}. The surface tension gradients are largest for solutions on the unstable branch; while increasing the electric field strength increases the surface tension gradients, the rotational component of the flow tends to restore a uniform surfactant distribution through the term $\tensor{Q}^{\chi \Gamma}$ in~\eqref{qgammaODE}. This is analogous to the effect of the term $\tensor{Q}^{\chi f},$ which acts to restore a spherical drop shape once rotation begins.
In the Quincke regime, the inset in Figure~\ref{bifurcations}(b) marked with the pink square indicates that there is more surfactant at the equator of the drop than at the poles, just as in the Taylor regime. However, counterintuitively, we find that higher values of $\zeta$ lead to less deformation than do lower values of $\zeta$. This is a result of only considering the deformation of the drop in the $xz$ plane; at high $\zeta,$ surfactant is transported out of the $xz$ plane and towards the tips of the drops along the axis of rotation. The convection of surfactant out of the $xz$ plane is quantified by defining the parameter
\begin{equation}
\begin{split}
    \eta=&\;\frac{1}{2\pi}\int_{0}^{\pi}\Gamma(\theta,\phi=0)+\Gamma(\theta,\phi=\pi)d\theta\\&\;-\frac{1}{2\pi}\int_{0}^{\pi}\Gamma(\theta,\phi=\pi/2)+\Gamma(\theta,\phi=3\pi/2)d\theta,
\end{split}\end{equation}
which measures the difference between the average amount of surfactant in the $xz$ plane and that in the $yz$ plane, and takes the simple form
\begin{equation}\label{etadef}
    \eta=\frac{1}{2}\epsilon\left(Q^{\Gamma}_{xx}-Q^{\Gamma}_{yy}\right).
\end{equation}Figure~\ref{etafig} shows $\eta$ as a function of $E_{0}/E_{\mathrm{c,s}}$ for the same parameter values as those used in Figure~\ref{bifurcations}. In the Taylor regime, $\eta$ is zero because the surfactant distribution is axisymmetric. However, once rotation begins, $\eta$ becomes negative, indicating that surfactant is moving out of the $xz$ plane and accumulating in the $yz$ plane. This effect, which is stronger for higher values of $\zeta,$ leads to an increase of the surface tension in the $xz$ plane, which in turn leads to decreased deformation. 
 
\begin{figure}
    \centering
    \includegraphics[width=\linewidth]{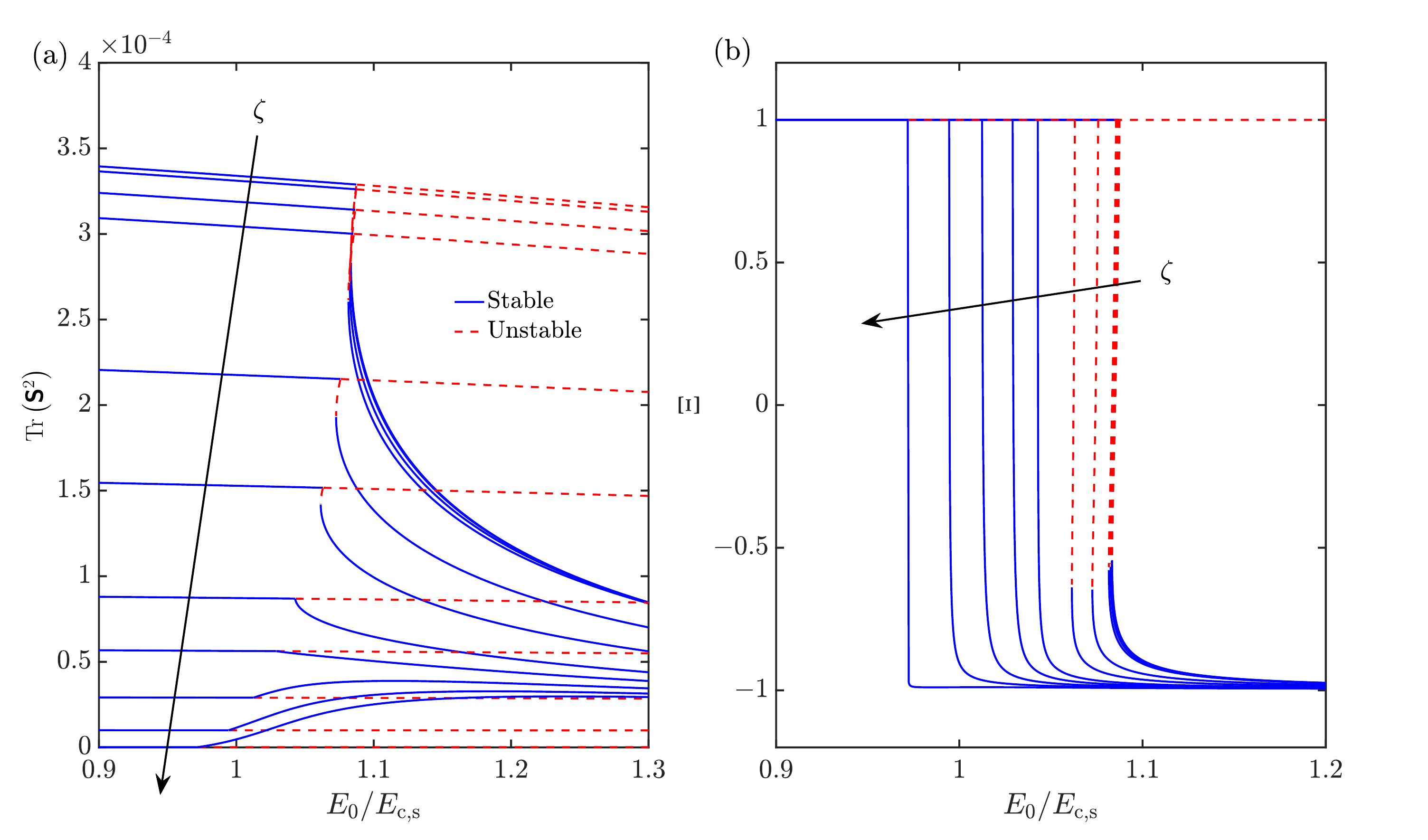}
    \caption{Flow field strength $\mbox{Tr}(\tensor{S}^2)$ given by~\eqref{trS2def} and flow assessment parameter $\Xi$ given by~\eqref{Xidef} as functions of the scaled electric field strength $E_{0}/E_{\mathrm{c,s}}$ for the same parameter values as those used in Figures~\ref{bifurcations} and~\ref{etafig}. The arrows indicate the directions of increasing $\zeta.$}
    \label{strainingandXi}
\end{figure}

As previously discussed, the strength of the straining flow in the Taylor regime depends on the value of $\zeta.$ For high values of $\zeta,$ the flow is suppressed by the Marangoni stresses. However, in the Quincke regime, the surfactant distribution becomes more uniform, so the Marangoni stresses are diminished and the straining flow may be relatively unsuppressed. The strength of the straining flow is quantified by the trace of $\tensor{S}^{2},$ where $\tensor{S}=(\bnabla\bm{v}+\bnabla\bm{v}^{T})/2$ is the rate-of-strain tensor~\citep{das2017electrohydrodynamics}. We evaluate $\mbox{Tr}(\tensor{S}^2)$ at the center of the drop, $\bm{r}=\bm{0}$, where it is given by
\begin{equation}\label{trS2def}
    \mbox{Tr}(\tensor{S}^2)=8\left(q^{\phi 2}_{xx}+q^{\phi 2}_{zz}+q^{\phi 2}_{xz}+q^{\phi}_{xx}q^{\phi}_{zz}\right).
\end{equation}
Figure~\ref{strainingandXi}(a) shows $\mbox{Tr}(\tensor{S}^2)$ as a function of $E_{0}/E_{\mathrm{c,s}}$ for the same parameter values as those used in Figures~\ref{bifurcations} and~\ref{etafig}. As expected, in the Taylor regime, the straining flow becomes weaker for larger values of $\zeta.$ In the cases corresponding to $\zeta\leq30,$ the straining flow is weaker in the Quincke regime than it is in the Taylor regime. This is also the case for a clean drop. However, for the cases $\zeta=50,$ $\zeta=100,$ and $\zeta=1000,$  the straining flow is stronger in the Quincke regime than it is in the Taylor regime. This is a direct result of the flow suppression effect in the Taylor regime which becomes less prominent in the Quincke regime.  As the electric field strength is increased further, the dependence of strength of the straining flow on $\zeta$ decreases because the surfactant distribution becomes more uniform as a result of the strong rotational flow through the term $\tensor{Q}^{\chi\Gamma}$ in~\eqref{qgammaODE}.

The relative strengths of the straining and rotational components of the flow field can be compared by defining the flow assessment parameter~\citep{astarita1979objective,oliveira2009purely,poole2023inelastic}
\begin{equation}\label{Xidef}
    \Xi=\frac{\mbox{Tr}(\tensor{S}^{2})-\mbox{Tr}(\tensor{W}^{2})}{\mbox{Tr}(\tensor{S}^{2})+\mbox{Tr}(\tensor{W}^{2})},
\end{equation}
where $\tensor{W}=(\bnabla\bm{v}-\bnabla\bm{v}^{T})/2$ is the rate-of-rotation tensor~\citep{das2017electrohydrodynamics}. Thus, $\Xi=1$ corresponds to purely straining flow and $\Xi=-1$ corresponds to purely rotational flow. Figure~\ref{strainingandXi}(b) shows $\Xi$ as a function of $E_{0}/E_{\mathrm{c,s}}$ for the same parameter values as in Figures~\ref{bifurcations} and~\ref{etafig}. In the Taylor regime, $\Xi=1$ for all values of $\zeta.$ Even though the straining flow is strongly suppressed for large values of $\zeta,$ the rotational flow is zero below the critical electric field strength for the bifurcation. Beyond the critical field strength, the value of $\Xi$ drops towards $\Xi=-1,$ indicating that the rotational component of the flow is dominant. For large values of $\zeta,$ this change is particularly sharp because the straining flow is relatively weak in the Quincke regime compared to that for small values of $\zeta$, consistent with Figure~\ref{strainingandXi}(a). Conversely, the transition from $\Xi=1$ towards $\Xi=-1$ is more gradual for smaller values of $\zeta,$ for which the straining flow is relatively strong in the Quincke regime. However, as the field strength is increased further, the value of $\Xi$ converges towards $-1$ for all values of $\zeta,$ reflecting the increasing dominance of the rotational flow. 

The destabilising effect of a weakly-diffusing surfactant on the Quincke rotation of drops may be useful in practical settings, in which rotation could be induced with a smaller electric field. It is important to bear in mind that this destabilising effect is relative to a uniformly coated drop with constant surface tension $\gamma_{\rm{eq}},$ rather than to a clean drop with the higher surface tension $\gamma_{0}.$ In other words, it is the spatial variations in the surfactant distribution, rather than simply its presence, that provide the effect. However, the critical field for a drop with lower constant surface tension is higher than that of one with higher surface tension, so one might expect this effect to be offset by the global reduction of surface tension arising from the addition of surfactant to a clean drop with surface tension $\gamma_{0}$. However, for the values $\gamma_{\rm{eq}}=4.5\times 10^{-3}$ and $\rm{El}=0.5,$ as used here, the corresponding constant surface tension of the same drop without surfactant is $\gamma_{0}=6.75\times 10^{-3}$. Rescaling the capillary number with this surface tension, the critical field for rotation is around $1.065E_{\rm{c,s}},$ while the critical field at $\zeta=0$ and a capillary number based on $\gamma_{\rm{eq}}$ is around $1.088E_{\rm{c,s}}$. The critical field at $\zeta=1000$ and a capillary number based on $\gamma_{\rm{eq}}$ is around $0.97E_{\rm{c,s}}$. Hence, the increase in the critical field owing to the global reduction in surface tension is much smaller than the decrease in the critical field due to the spatial variations in surface tension at high $\zeta.$ Moreover, the present model predicts that this destabilising effect becomes even more significant if ${\rm El}$ is decreased further, although this prediction should be interpreted with caution since the present asymptotic analysis formally assumes that ${\rm El}=\mathcal{O}(1).$ 

\subsubsection{Effects of $\rm{El}$ and $\zeta$ on Quincke rotation}
\begin{figure}
    \centering
    \includegraphics[width=\linewidth]{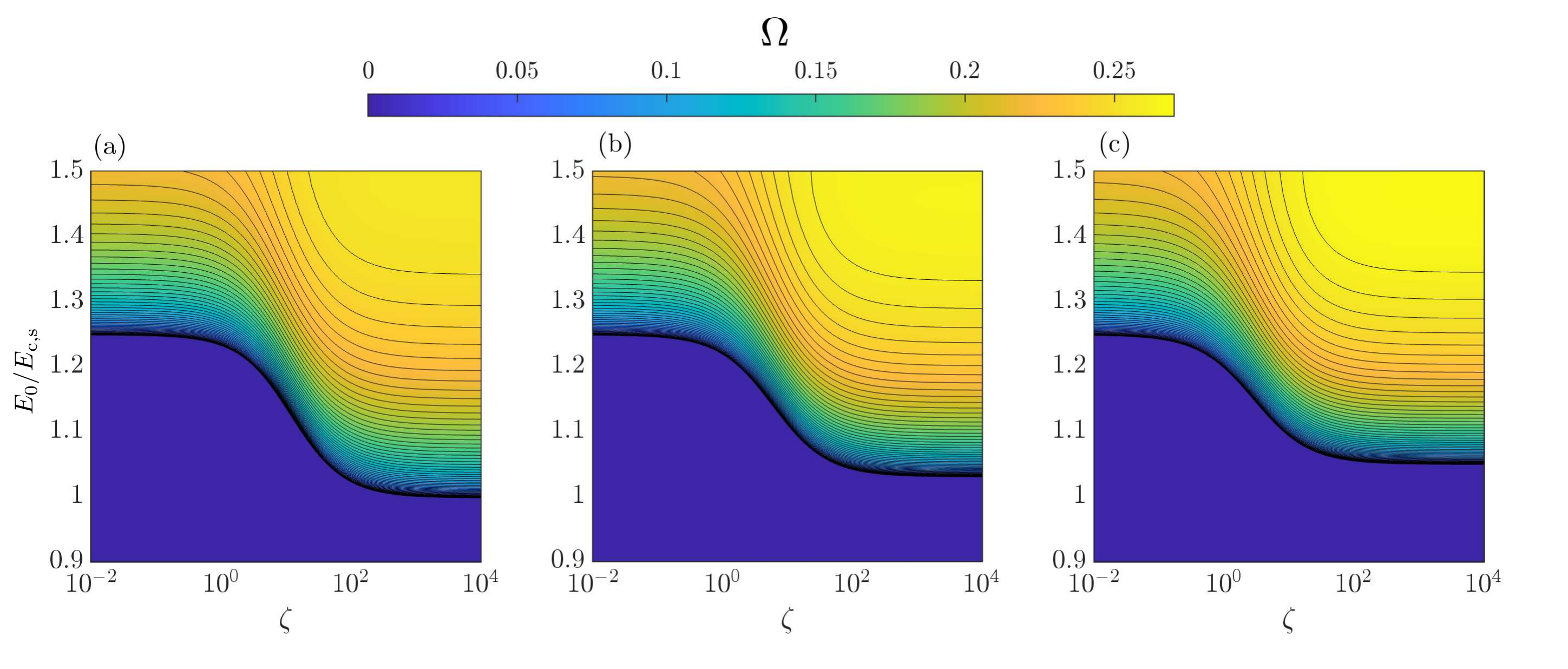}
    \caption{Contour plots showing the angular velocity $\Omega$ of a drop as a function of the scaled electric field strength $E_{0}/E_{\rm{c,s}}$ and $\zeta$ for (a) $\rm{El}=0.5$, (b) $\rm{El}=1$, and (c) $\rm{El}=2.$ In all cases, $Q=5,$ $S=30,$ $\lambda=10,$ and $0.04\leq\mathrm{Ca_{E}}\leq 0.1$ and $0.45\leq\mathrm{Re_{E}}\leq1.25$ in the range of field strengths shown.}
    \label{fixedElcontours}
\end{figure}
Next, we systematically explore the effects of varying $\rm{El}$ and $\zeta$ on the Quincke rotation of a drop, considering a drop-fluid system for which $Q=5,$ $S=30,$ and $\lambda=10$. The model does not predict hysteresis for these parameter values. Figure~\ref{fixedElcontours} shows contour plots of the angular velocity $\Omega$ of the drop as a function of the scaled electric field strength $E_{{\rm 0}}/E_{\rm{c,s}}$ and $\zeta$ at (a) $\rm{El}=0.5$, (b) ${\rm El}=1$ and (c) $\rm{El}=2.$ Over the range of field strengths shown, namely $0.9\leq E_{0}/E_{{\rm c,s}}\leq 1.5$, the other dimensionless parameters vary in the ranges $0.04\leq\mathrm{Ca_{E}}\leq 0.12$ and $0.45\leq\mathrm{Re_{E}}\leq1.25.$ In the limit of strong diffusion (i.e., $\zeta\rightarrow 0$), the critical field is the same for all three values of $\rm{El}$ because the surfactant distribution remains uniform. As $\zeta$ increases, the critical field decreases and approaches a limiting value as $\zeta$ becomes large. This behaviour is qualitatively similar at all three values of ${\rm El}.$  It is noted that the critical field in the limit of weak diffusion (i.e., $\zeta\rightarrow \infty$) is highest for $\rm{El}=2$ and lowest for $\rm{El}=0.5.$ 

\begin{figure}
    \centering
    \includegraphics[width=\linewidth]{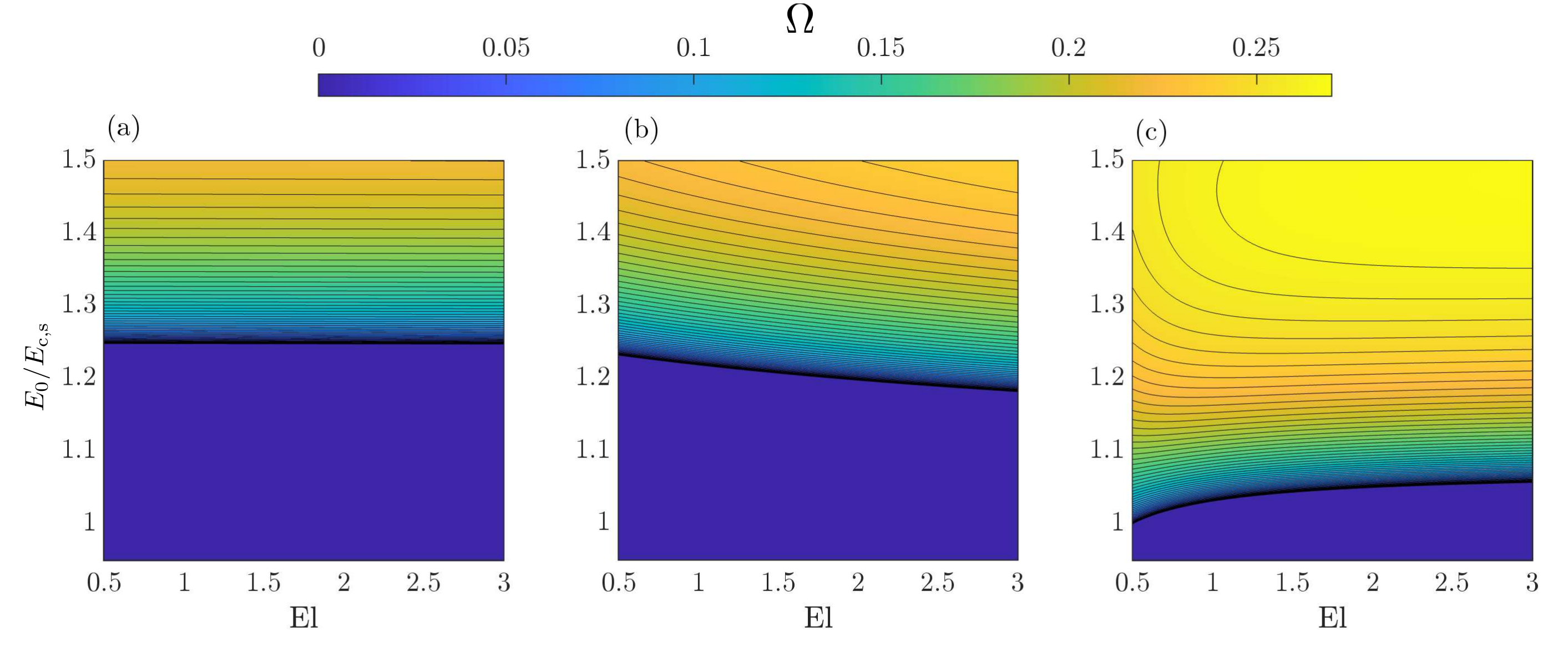}
    \caption{Contour plots showing the angular velocity $\Omega$ of a drop as a function of the scaled electric field strength $E_{0}/E_{\rm{c,s}}$ and the elasticity number $\rm{El}$ for (a) $\zeta=0.01$, (b) $\zeta=1$, and (c) $\zeta=1000.$ All the other dimensionless parameters are the same as those used in Figure~\ref{fixedElcontours}.}
    \label{fixedPscontours}
\end{figure}
Next, we fix $\zeta$ and study the effect of changing the scaled electric field strength $E_{0}/E_{{\rm c,s}}$ and the elasticity number ${\rm El}$ on the Quincke rotation of the drop. To this end, Figure~\ref{fixedPscontours} shows contour plots of the angular velocity $\Omega$ of the same drop as in Figure~\ref{fixedElcontours}, but now plotted as a function of $E_{0}/E_{\rm{c,s}}$ and $\rm{El}$ at (a) $\zeta=0.01,$ (b) $\zeta=1,$ and (c) $\zeta=1000.$ We denote $E_{\mathrm{c,d}}$ as the value of electric field strength, $E_\mathrm{0}$, at which the drop starts to rotate. In the case of strong diffusion (Figure~\ref{fixedPscontours}(a)), the drop behaves essentially like a uniform surface tension drop, and varying $\mathrm{El}$ has little effect on $E_{\mathrm{c,d}}$ or on $\Omega$. When diffusion is moderate (Figure~\ref{fixedPscontours}(b)), increasing $\mathrm{El}$ leads to a decrease in $E_{\mathrm{c,d}}$ and an increase in $\Omega$. However, when diffusion is weak (Figure~\ref{fixedPscontours}(c)), $E_{\mathrm{c,d}}$ instead increases as $\mathrm{El}$ increases. In this case, the dependence of $\Omega$ on ${\rm El}$ depends on the field strength. At field strengths only just strong enough to induce rotation (i.e., $E_{0}/E_{{\rm c,s}}$ less than approximately $1.1$ in Figure~\ref{fixedPscontours}(c)), an increase in ${\rm El}$ leads to a decrease in $\Omega.$ At slightly greater field strengths (i.e., $E_{0}/E_{{\rm c,s}}$ around $1.2$ in Figure~\ref{fixedPscontours}(c)), an increase in ${\rm El}$ initially leads to an increase in $\Omega,$ before the trend reverses and further increases in ${\rm El}$ lead to a decrease in $\Omega.$ At relatively large field strengths (i.e., $E_{0}/E_{{\rm c,s}}$ greater than approximately $1.3$ in Figure~\ref{fixedPscontours}(c)), an increase in ${\rm El}$ initially leads to an increase in $\Omega,$ before the effect vanishes and $\Omega$ becomes insensitive to further increases in ${\rm El}.$

\begin{figure}
    \centering
    \includegraphics[width=\linewidth]{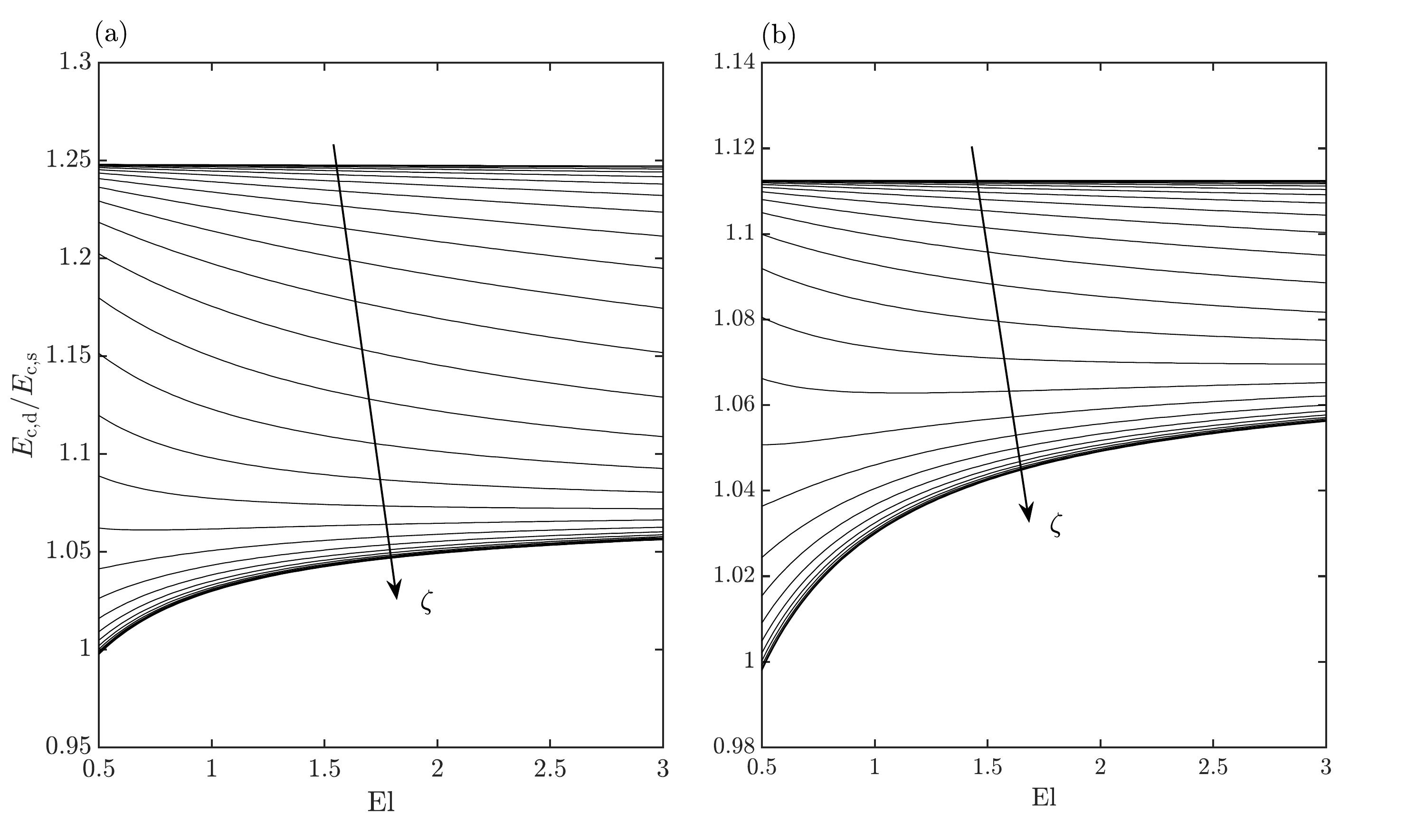}
    \caption{Scaled critical electric field strengths for the onset of Quincke rotation of a drop $E_{\mathrm{c,d}}/E_{\mathrm{c,s}}$ as functions of $\rm{El}$ for thirty logarithmically spaced values of $\zeta$ between $10^{-2}$ and $10^{4},$ specifically for $\zeta=10^{(-2+6n/29)}$ for $n=0,1,2,..,28,29.$ (a) shows results for $\lambda=10,$ and (b) shows results for $\lambda=50.$ The dimensionless parameters are the same as those used in Figures~\ref{fixedElcontours} and~\ref{fixedPscontours}.}
    \label{EcsvsE}
\end{figure}

The relationship between the elasticity number $\mathrm{El}$ and the critical electric field strength $E_{\mathrm{c,d}}$ can be explored further using the linear stability analysis described in \S\ref{lsa}. Figure~\ref{EcsvsE} shows the critical field strengths for rotation of a drop scaled by that for a solid sphere $E_{\mathrm{c,d}}/E_{\mathrm{c,s}}$ as functions of $\rm{El}$ for thirty logarithmically spaced values of $\zeta$ between $10^{-2}$ and $10^{4},$ specifically for $\zeta=10^{(-2+6n/29)}$ for $n=0,1,2,..,28,29.$ The curves in Figure~\ref{EcsvsE} generated from the linear stability analysis are the same as the marginal curves in Figure~\ref{fixedPscontours} that separate the Taylor and Quincke regimes.
Figure~\ref{EcsvsE}(a) shows results for $\lambda=10,$ and Figure~\ref{EcsvsE}(b) shows results for $\lambda=50.$ The other parameters are the same as those used in Figures~\ref{fixedElcontours} and~\ref{fixedPscontours}. Figure~\ref{EcsvsE} indicates that the effect of the surfactant is destabilising over the full range shown of $0.5\leq \rm{El}\leq3$, and that the effect is more pronounced for $\lambda=10$ than for $\lambda=50.$ It also shows that increasing $\rm{El}$ leads to a decrease in $E_{\mathrm{c,d}}/E_{\mathrm{c,s}}$ at lower values of $\zeta,$ and that the trend reverses at higher values of $\zeta,$ consistent with the results shown in Figure~\ref{fixedPscontours}. 

\begin{figure}
    \centering
    \includegraphics[width=\linewidth]{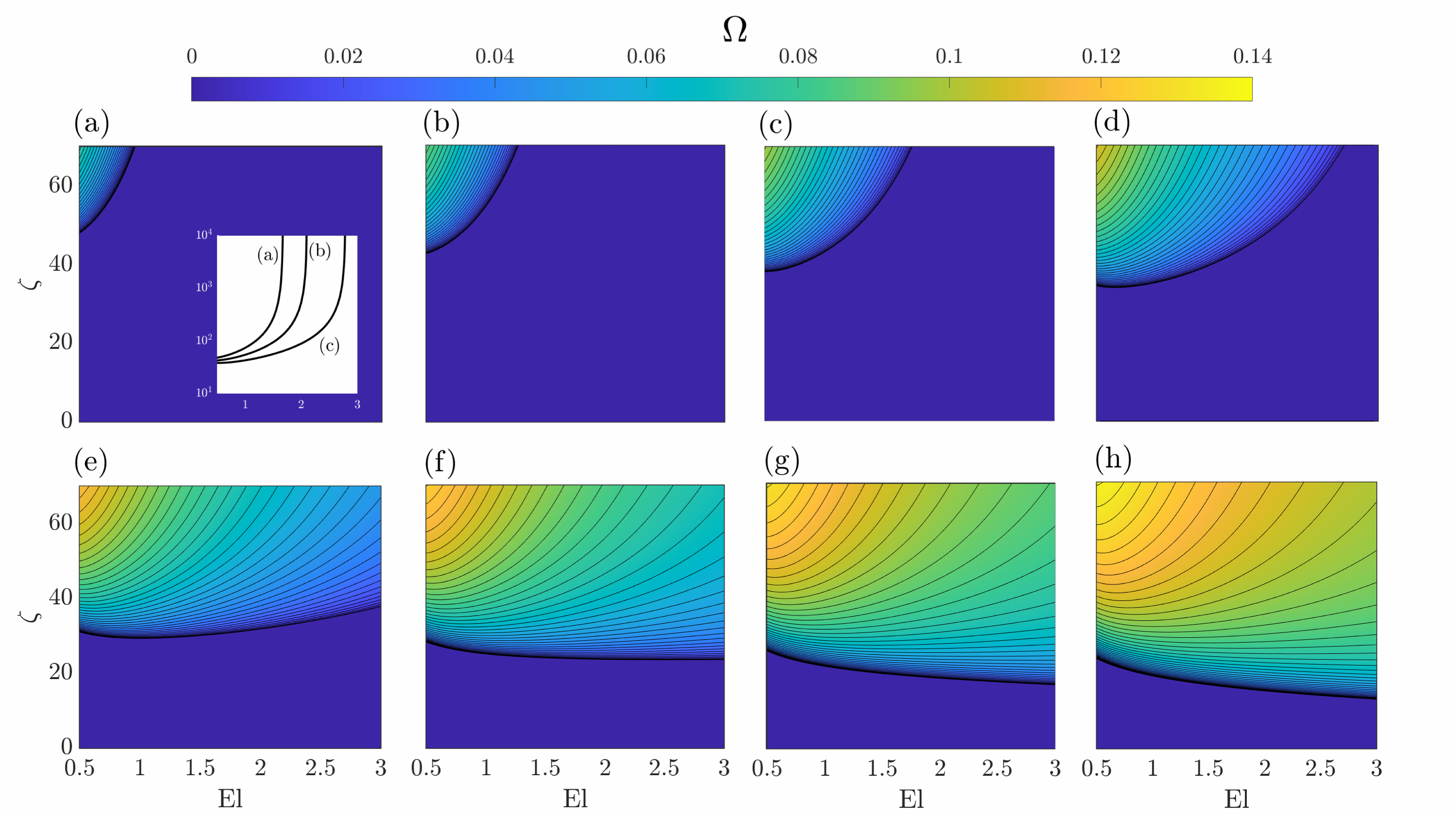}
    \caption{Contour plots showing the angular velocity $\Omega$ of a drop as a function of $\zeta$ and the elasticity number $\rm{El}$ for eight electric field strengths $E_{0}/E_{\rm{c,s}}=1.045,$ $1.05,$ $...,$ $1.075,$ $1.08.$ All the other dimensionless parameters are the same as those used in Figures~\ref{fixedElcontours},~\ref{fixedPscontours}, and~\ref{EcsvsE}. The inset in (a) displays the marginal curves between the axisymmetric and rotational solution regions as functions of $\rm{El}$ for (a) $E_{0}/E_{\rm{c,s}}=1.045,$ (b) $E_{0}/E_{\rm{c,s}}=1.05,$ and (c) $E_{0}/E_{\rm{c,s}}=1.055$ using a logarithmic scale.}
    \label{pvselcontours}
\end{figure}

Next, we focus our attention on the effects of varying $\zeta$ and the elasticity number $\rm{El}$ at fixed electric field strengths close to the critical field for Quincke rotation. Firstly, Figure~\ref{pvselcontours} shows contour plots of the angular velocity $\Omega$ of the drop as a function of $\zeta$ and $\rm{El}$ for eight field strengths increasing in increments of $0.005$ from $E_{0}/E_{\rm{c,s}}=1.045$ (Figure~\ref{pvselcontours}(a)) to $E_{0}/E_{\rm{c,s}}=1.080$ (Figure~\ref{pvselcontours}(h)). The inset in Figure~\ref{pvselcontours}(a) displays the marginal curves between the axisymmetric and rotational solution regions as functions of $\rm{El}$ for (a) $E_{0}/E_{\rm{c,s}}=1.045,$ (b) $E_{0}/E_{\rm{c,s}}=1.05,$ and (c) $E_{0}/E_{\rm{c,s}}=1.055$ using a logarithmic scale. At each field strength shown, the angular velocity is greatest for the highest values of $\zeta$ and the lowest values of $\rm{El}$ shown. As in Figure~\ref{fixedElcontours}, the critical field strength $E_{\mathrm{c,d}}/E_{\mathrm{c,s}}$ at which rotation occurs for a given value of ${\rm El}$ is decreased as $\zeta$ is increased. Hence, for a given field strength that is greater than the critical field strength for rotation in the limit $\zeta\rightarrow\infty$ (i.e., the limit of weak surfactant diffusion) but smaller than that in the limit $\zeta\rightarrow 0$ (i.e., the limit of strong surfactant diffusion), there exists a critical value of $\zeta$ at which the drop becomes unstable and Quincke rotation occurs. This critical value corresponds to the marginal curves between the axisymmetric and rotational solution regions. At the lowest field strength considered, $E_{0}/E_{\rm{c,s}}=1.045$ (Figure~\ref{pvselcontours}(a)), the critical value increases sharply as ${\rm El}$ is increased. As shown in the inset, the critical value diverges as ${\rm El}$ approaches $\rm{El}\approx 1.67.$ Above $\rm{El}\approx 1.67,$ the field is too weak to induce rotation at any value of $\zeta$. The critical values at the second and third field strengths shown, $E_{0}/E_{\rm{c,s}}=1.05$ and $E_{0}/E_{\rm{c,s}}=1.055$ (Figures~\ref{pvselcontours}(b) and (c)), diverge as the elasticity number approaches $\rm{El}\approx 2.1$ and $\rm{El}\approx 2.78,$ respectively. For field strengths at or above that shown in Figure~\ref{pvselcontours}(d), corresponding to $E_{0}/E_{\rm{c,s}}=1.06,$ the critical value of $\zeta$ is smaller than $\zeta=40$ for all values of ${\rm El}$ in the range shown. Furthermore, when the field strength is at or above $E_{0}/E_{\rm{c,s}}=1.07,$ the critical value of $\zeta$ decreases with increasing $\rm{El}$ in the range shown. Hence, the critical value of $\zeta$ and the angular velocity of the drop are significantly more sensitive to changes in the field strength for $\rm{El}=3$ than they are for $\rm{El}=0.5$. On the other hand, the contour lines indicate that for the field strengths shown, $\Omega$ is more sensitive to changes in $\zeta$ when $\rm{El}=0.5$ than when $\rm{El}=3.$ 
\begin{figure}
    \centering
    \includegraphics[width=\linewidth]{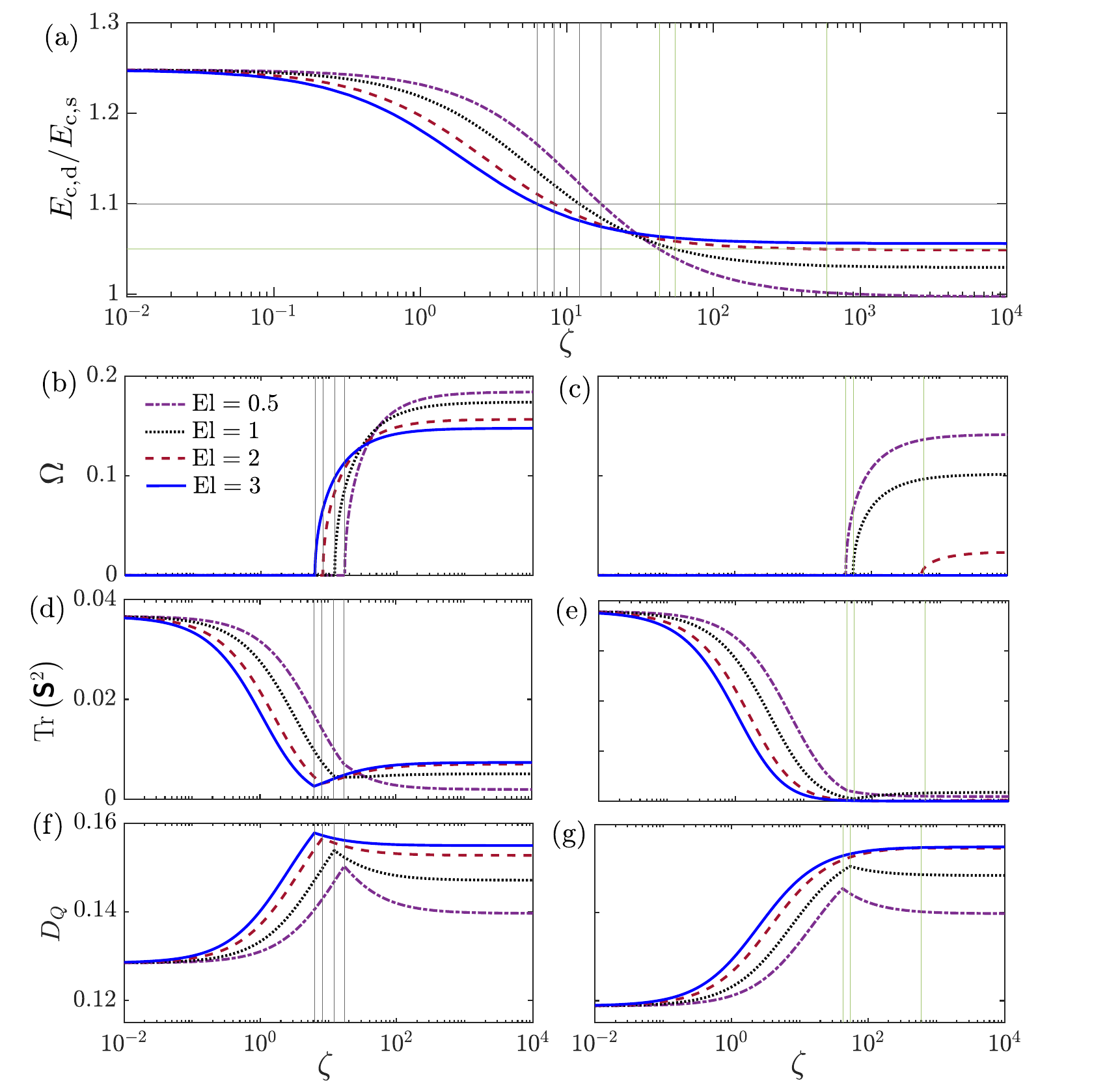}
    \caption{(a) Scaled critical electric field strength for Quincke rotation of a drop $E_{\mathrm{c,d}}/E_{\mathrm{c,s}}$, (b,c) angular velocity $\Omega$, (d,e) straining flow strength $\mbox{Tr}(\tensor{S}^2)$, and (f,g) deformation parameter $D_{Q}$ as functions of $\zeta$ for ${\rm El}=0.5$ (purple dash-dotted lines), $1$ (black dotted lines), $2$ (red dashed lines), and $3$ (blue solid lines). (b), (d) and (f) correspond to $E_{0}/E_{{\rm c,s}}=1.1,$ while (c), (e) and (g) correspond to $E_{0}/E_{{\rm c,s}}=1.05.$ These field strengths are indicated by the thin black and green horizontal lines in (a), respectively, and the predicted critical values of $\zeta$ are denoted in the relevant subplots by the thin vertical green and black lines. We again use $Q=5$ and $S=30,$ and the other relevant parameters are ${\rm Ca_{E}}=0.06$ and ${\rm Re_{E}}=0.6$ at $E_{0}/E_{\rm{c,s}}=1.05,$ and ${\rm Ca_{E}}=0.07$ and ${\rm Re_{E}}=0.67$ at $E_{0}/E_{\rm{c,s}}=1.1.$ }
    \label{DvsPsQuincke}
\end{figure}

Figure~\ref{DvsPsQuincke} shows the scaled critical electric field strength for Quincke rotation of a drop $E_{\mathrm{c,d}}/E_{\mathrm{c,s}}$ (Figure~\ref{DvsPsQuincke}(a)), angular velocity $\Omega$ (Figures~\ref{DvsPsQuincke}(b) and (c)), straining flow strength $\mbox{Tr}(\tensor{S}^2)$ ((d) and (e)), and deformation parameter $D_{Q}$ (Figures~\ref{DvsPsQuincke}(f) and (g)) as functions of $\zeta$ for ${\rm El}=0.5$ (purple dash-dotted lines), $1$ (black dotted lines), $2$ (red dashed lines), and $3$ (blue solid lines). Figures~\ref{DvsPsQuincke}(b), (d) and (f) correspond to $E_{0}/E_{{\rm c,s}}=1.1,$ while Figures~\ref{DvsPsQuincke}(c), (e) and (g) correspond to $E_{0}/E_{{\rm c,s}}=1.05.$
We again use $Q=5$ and $S=30,$ and the other relevant parameters are ${\rm Ca_{E}}=0.06$ and ${\rm Re_{E}}=0.6$ at $E_{0}/E_{\rm{c,s}}=1.05,$ and ${\rm Ca_{E}}=0.07$ and ${\rm Re_{E}}=0.67$ at $E_{0}/E_{\rm{c,s}}=1.1.$ These field strengths are marked by the thin horizontal lines in Figure~\ref{DvsPsQuincke}(a), in which the green line marks $E_{0}/E_{\rm{c,s}}=1.05$ and the black line marks $E_{0}/E_{\rm{c,s}}=1.1.$ The predicted critical values of $\zeta$ at the given field strengths are given by the $\zeta$ coordinates of the intersections of these horizontal lines and the solution curves in Figure~\ref{DvsPsQuincke}(a), and are denoted in the relevant subplots by the thin vertical green and black lines. The critical field strengths $E_{\mathrm{c,d}}/E_{\mathrm{c,s}}$ shown in Figure~\ref{DvsPsQuincke}(a) are obtained from the linear stability analysis described in \S\ref{lsa}, and are the same as the marginal curves separating the Taylor and Quincke regimes in Figure~\ref{fixedElcontours}. Similarly, Figures~\ref{DvsPsQuincke}(b) and (c) correspond to horizontal slices of Figure~\ref{fixedElcontours}. At $E_{0}/E_{{\rm c,s}}=1.05,$ the critical value of $\zeta$ is lower for lower values of ${\rm El}$ than for higher ones, and indeed no bifurcation occurs at all for the case ${\rm El}=3.$ However, the trend reverses at $E_{0}/E_{{\rm c,s}}=1.1,$ and the critical value of $\zeta$ decreases as $\rm{El}$ is increased. Figures~\ref{DvsPsQuincke}(b) and (c) show that at both of these field strengths, the angular velocity is larger for lower values of ${\rm El}$ than for higher ones as $\zeta$ grows large, even though drops with higher $\rm{El}$ destabilise at lower values of $\zeta$ at $E_{0}/E_{{\rm c,s}}=1.1.$ 
Figures~\ref{DvsPsQuincke}(d) and (e) show that as $
\zeta$ is increased, the straining flow is weakened due to the increasing Marangoni stresses. However, beyond the critical value of $\zeta,$ the straining flow is strengthened with further increases in $
\zeta$ for the cases ${\rm El}=1,$ $2,$ and $3$ at $E_{0}/E_{\mathrm{c,s}}=1.1$ (Figure~\ref{DvsPsQuincke}(d)) and for the cases ${\rm El}=1$ and $2$ at $E_{0}/E_{\mathrm{c,s}}=1.05$ (Figure~\ref{DvsPsQuincke}(e)). On the other hand, for ${\rm El}=0.5,$ the straining flow continues decreasing in strength as $\zeta$ is increased beyond its critical value at both field strengths considered. Figures~\ref{DvsPsQuincke}(f) and (g) show that in the limit of strong diffusion (i.e., in the limit $\zeta\rightarrow 0$), in which all of the drops are in the Taylor regime, all of the drops have the same deformation at a given field strength because the surfactant distribution remains uniform in all cases, as is the case for the results shown in Figure~\ref{DvsPs}. As $\zeta$ is increased, the deformation increases due to the accumulation of surfactant around the equator, with the deformation becoming greater for greater values of $\rm{El}$. However, above the critical value of $\zeta$, the deformation begins to decrease as $\zeta$ increases. The reduction of the deformation in the Quincke regime becomes more pronounced as ${\rm El}$ is decreased. Thus, in the limit of zero diffusion (i.e., in the limit $\zeta\rightarrow\infty$) the deformation of surfactant-laden oblate drops increases as $\rm{El}$ is increased in the Quincke regime, unlike in the Taylor regime, in which the deformation is the same for all values of $\rm{El}.$  

\section{Conclusions}\label{conclusions}

In the present work, we have described a small-deformation theory for a leaky dielectric drop coated with a dilute monolayer of insoluble apolar surfactant and subjected to a uniform DC electric field. The novelty of our work lies in the retention of charge convection in the transport equation for the surface charge, which allows the theory to capture Quincke rotation. We have examined the combined effects of surfactant and charge convection on drop deformation in the Taylor regime, and found that the two effects work cooperatively to promote increased deformation of prolate A drops, but that they work antagonistically in the cases of prolate B and oblate drops. Our theory predicts that the presence of a weakly-diffusing surfactant can eliminate the hysteresis that occurs in the angular velocity of drops undergoing Quincke rotation and can significantly lower the critical electric field strength for the onset of rotation from that for a drop of constant surface tension. We find that for an oblate drop in the Quincke regime, the surfactant is transported out of the rotational plane and accumulates at the tips of the drop along the axis of rotation.

A linear stability analysis of the system reveals that for strongly-diffusing surfactants (i.e., for low values of $\zeta$), an increase in $\rm{El}$ leads to an decrease in the critical electric field strength for the onset of rotation $E_{\mathrm{c,d}}$, while the opposite is true for weakly-diffusing surfactants (i.e., for high values of $\zeta$). We have explored the effects of varying $\zeta$ and the elasticity number $\rm{El}$ on the deformation and angular velocity of an oblate drop at fixed electric field strengths. We find that larger values of $\rm{El}$ lead to larger deformation than do smaller ones across the whole range of $\zeta$ studied, and smaller values of $\rm{El}$ lead to larger angular velocity than do larger ones in the limit of large $\zeta$. At intermediate values of $\zeta,$ the relationship between the angular velocity and ${\rm El}$ depends on the field strengths: at larger field strengths, drops with larger values of ${\rm El}$ begin rotating at a smaller critical value of $\zeta$ than do drops with smaller values of ${\rm El}.$ At smaller field strengths, the opposite is true, and drops with smaller values of ${\rm El}$ begin rotating at smaller critical values of $\zeta,$ while drops with larger values of ${\rm El}$ may not rotate at all, even in the limit of large $\zeta$.

While the linearisation of the Langmuir equation~\eqref{gamma(Gamma)} is convenient for the spherical harmonic representation used in the present theory, it restricts the present theory to dilute concentrations of surfactant, and the theory thus fails to capture some of the qualitative features of surfactant effects at higher concentrations as found by~\citet{nganguia2019effects}. For example, their spheroidal theory predicts that the deformation of prolate A drops decreases and that of prolate B drops increases with increasing surfactant concentration at low $\zeta,$ while the opposite is true at high $\zeta.$ However, since the present theory is valid for dilute concentrations, this complexity is lost. Full simulations using a finite or boundary element method may be used to explore the effects of nondilute concentrations. Additionally, the small-deformation theory of an isolated drop developed here is a first step towards understanding the electrorheology of emulsions of drops covered with surfactants, which could be achieved either using particle-based simulations~\citep{das2023absence} or by coarse-graining the governing equations~\citep{vlahovska2011rheology}. Finally, the framework developed here can be extended to incorporate interfacial rheological effects, such as surface viscosities~\citep{scriven1960dynamics,mandal2017influence,nganguia2023influence}, while retaining charge convection in three dimensions, to study their influence on Quincke rotation.

\backsection[Supplementary data]{\label{SupMat}MATLAB codes to reproduce the figures in this article are available in the Supplementary Material.}

%\backsection[Acknowledgements]{Acknowledgements may be included at the end of the paper, before the References section or any appendices. Several anonymous individuals are thanked for contributions to these instructions.}

\backsection[Funding]{The authors gratefully acknowledge the financial support of the University of Strathclyde in the form of a Research
Excellence Award.}

\backsection[Declaration of interests]{ The authors report no conflict of interest.}

% \backsection[Data availability statement]{The data that support the findings of this study are openly available in [repository name] at http://doi.org/[doi], reference number [reference number]. See JFM's \href{https://www.cambridge.org/core/journals/journal-of-fluid-mechanics/information/journal-policies/research-transparency}{research transparency policy} for more information}

\backsection[Author ORCIDs]{M. A. McDougall, https://orcid.org/0009-0005-2967-9437;\\ S. K. Wilson, https://orcid.org/0000-0001-7841-9643;\\ D. Das, https://orcid.org/0000-0003-2365-4720}

% \backsection[Author contributions]{Authors may include details of the contributions made by each author to the manuscript'}

\appendix

\section{}\label{appA}
In this appendix, for completeness, we give the elements of the Jacobian matrix in the linear system of equations~\eqref{linearsystem}.
\begin{align}\begin{split}
          J_{11}=&\;-\frac{1+2S}{S(2+Q){\rm Re_{E}}}-\frac{1-Q}{2(2+Q)}+\frac{(1-Q)(1-4Q)}{30(3+2\lambda)(2+Q)}\\&\;+P_{0,z}\left(-\frac{1}{2}+\frac{1}{5(3+2\lambda)}\left[\frac{5-8Q}{6}-\frac{9(14+11\lambda)(1-Q)}{(16+19\lambda)(2+Q)}\right]\right)
          \\&\;+P_{0,z}^{2}\frac{2}{15(3+2\lambda)}\left(2+Q-\frac{9(14+11\lambda)}{16+19\lambda}\right)\\&\;+Q^{f}_{0,xx}\frac{6(14+11\lambda)}{5(16+19\lambda)(3+2\lambda)(1+2S)^2}-Q^{\Gamma}_{0,xx}\frac{2(8+5\lambda)\epsilon{\rm El}}{\mathrm{Ca_{E}}(16+19\lambda)(3+2\lambda)},\raisetag{2ex}\end{split}
          \\J_{15}=&\;\left(P_{0,z}+\frac{1-Q}{2+Q}\right)\frac{6(14+11\lambda)}{5(16+19\lambda)(3+2\lambda)(1+2S)^2},
          \\J_{18}=&\;-\left(P_{0,z}+\frac{1-Q}{2+Q}\right)\frac{2(8+5\lambda)\epsilon{\rm El}}{\mathrm{Ca_{E}}(16+19\lambda)(3+2\lambda)},
          \\\begin{split}J_{22}=&\;-\frac{1+2S}{S(2+Q){\rm Re_{E}}}-\frac{6(1-Q^{2})}{15(3+2\lambda)(2+Q)}\\&\;+P_{0,z}\frac{4}{15(3+2\lambda)}\left(-3Q-\frac{9(14+11\lambda)(1-Q)}{(16+19\lambda)(2+Q)}\right)\\&\;+P_{0,z}^{2}\frac{2}{5(3+2\lambda)}\left[2+Q-\frac{9(14+11\lambda)}{16+19\lambda}\right]\\&\;+Q^{f}_{0,zz}\frac{6(14+11\lambda)}{5(16+19\lambda)(3+2\lambda)(1+2S)^2}-Q^{\Gamma}_{0,zz}\frac{2(8+5\lambda)\epsilon{\rm El}}{\mathrm{Ca_{E}}(16+19\lambda)(3+2\lambda)},\end{split}
          \\
          J_{24}=&\;J_{15},
          \quad \quad J_{27}=J_{18},
          \\J_{32}=&\;-\frac{1+2Q}{3\delta(3+2\lambda)}-\frac{2}{3\delta(3+2\lambda)}\left(\frac{45(1+\lambda)}{16+19\lambda}-(2+Q)\right)P_{0,z},
       \\J_{33}=&\;-\frac{30(1+\lambda)}{\delta(16+19\lambda)(3+2\lambda)(1+2S)^2},\\J_{36}=&\;\frac{2(4+\lambda)\epsilon{\rm El}}{\mathrm{Ca_{E}}\delta(16+19\lambda)(3+2\lambda)},\\
      J_{42}=&\;-2J_{32},
          \quad \quad J_{44}=J_{33},
          \quad \quad J_{47}=J_{36},      
          \\ J_{51}=&\;-\frac{3}{2}J_{32}+\frac{1}{2}\left(Q^{f}_{0,xx}-Q^{f}_{0,zz}\right),
          \\J_{55}=&\;J_{44}=J_{33},
          \quad \quad J_{58}=J_{47}=J_{36},\\
          J_{62}=&\;-\frac{(1+2Q)}{3\epsilon(3+2\lambda)}+\frac{2(68+47\lambda+Q(16+19\lambda))}{3\epsilon(16+19\lambda)(3+2\lambda)}P_{0,z},\\J_{63}=&\;\frac{6(4+\lambda)}{\epsilon(16+19\lambda)(3+2\lambda)(1+2S)^2},\\\displaybreak 
          J_{66}=&\;-\left(\frac{6}{\zeta\rm{Re_{E}}}+\frac{2(32+23\lambda){\rm El}}{\mathrm{Ca_{E}}(16+19\lambda)(3+2\lambda)}\right),
          \\J_{72}=&\;-2J_{62},
          \quad\quad J_{74}=J_{63},
          \quad\quad J_{77}=J_{66},
          \\J_{81}=&\;-\frac{3}{2}J_{62}+\frac{1}{2}\left(Q^{\Gamma}_{0,xx}-Q^{\Gamma}_{0,zz}\right),
          \\J_{85}=&\;J_{74}=J_{63},
          \quad\quad J_{88}=J_{77}=J_{66}.
      \end{align}

\bibliographystyle{jfm}
\bibliography{jfm}

@article{taylor1932viscosity,
  title={The viscosity of a fluid containing small drops of another fluid},
  author={Taylor, G. I.},
  journal={Proc. R. Soc. Lond. A},
  volume={138},
  number={834},
  pages={41--48},
  year={1932},
  publisher={The Royal Society London}
}

@article{esmaeeli2011transient,
  title={Transient electrohydrodynamics of a liquid drop},
  author={Esmaeeli, A. and Sharifi, P.},
  journal={Phys. Rev. E},
  volume={84},
  number={3},
  pages={036308},
  year={2011},
  publisher={APS}
}

@article{esmaeeli2020transient,
  title={Transient electrohydrodynamics of a liquid drop at finite {R}eynolds numbers},
  author={Esmaeeli, A. and Behjatian, A.},
  journal={J. Fluid Mech.},
  volume={893},
  pages={A26},
  year={2020},
  publisher={Cambridge University Press}
}

@article{kralova2009surfactants,
  title={Surfactants used in food industry: a review},
  author={Kralova, I. and Sj{\"o}blom, J.},
  journal={J. Disper. Sci. Technol.},
  volume={30},
  number={9},
  pages={1363--1383},
  year={2009},
  publisher={Taylor \& Francis}
}

@article{kovalchuk2021surfactant,
  title={Surfactant-mediated wetting and spreading: {R}ecent advances and applications},
  author={Kovalchuk, N. M. and Simmons, M. J. H.},
  journal={Curr. Opin. Colloid Interface Sci.},
  volume={51},
  pages={101375},
  year={2021},
  publisher={Elsevier}
}

@article{bureiko2015current,
  title={Current applications of foams formed from mixed surfactant--polymer solutions},
  author={Bureiko, A. and Trybala, A. and Kovalchuk, N. and Starov, V.},
  journal={Adv. Colloid Interface Sci.},
  volume={222},
  pages={670--677},
  year={2015},
  publisher={Elsevier}
}

@article{moody2000perfluorinated,
  title={Perfluorinated surfactants and the environmental implications of their use in fire-fighting foams},
  author={Moody, C. A. and Field, J. A.},
  journal={Environ. Sci. \& Technol.},
  volume={34},
  number={18},
  pages={3864--3870},
  year={2000},
  publisher={ACS Publications}
}

@article{vaisman2006role,
  title={The role of surfactants in dispersion of carbon nanotubes},
  author={Vaisman, L. and Wagner, H. D. and Marom, G.},
  journal={Adv. Colloid Interface Sci.},
  volume={128},
  pages={37--46},
  year={2006},
  publisher={Elsevier}
}

@article{lotya2010high,
  title={High-concentration, surfactant-stabilized graphene dispersions},
  author={Lotya, M. and King, P. J. and Khan, U. and De, S. and Coleman, J. N.},
  journal={ACS Nano},
  volume={4},
  number={6},
  pages={3155--3162},
  year={2010},
  publisher={ACS Publications}
}

@article{taylor1934formation,
  title={The formation of emulsions in definable fields of flow},
  author={Taylor, G. I.},
  journal={Proc. R. Soc. Lond. A},
  volume={146},
  number={858},
  pages={501--523},
  year={1934},
  publisher={The Royal Society London}
}

@article{rumscheidt1961particle,
  title={Particle motions in sheared suspensions {XII}. {D}eformation and burst of fluid drops in shear and hyperbolic flow},
  author={Rumscheidt, F. D. and Mason, S. G.},
  journal={J. Colloid Sci.},
  volume={16},
  number={3},
  pages={238--261},
  year={1961},
  publisher={Elsevier}
}

@article{poddar2018sedimentation,
  title={Sedimentation of a surfactant-laden drop under the influence of an electric field},
  author={Poddar, A. and Mandal, S. and Bandopadhyay, A. and Chakraborty, S.},
  journal={J. Fluid Mech.},
  volume={849},
  pages={277--311},
  year={2018},
  publisher={Cambridge University Press}
}

@article{poddar2019electrical,
  title={Electrical switching of a surfactant coated drop in Poiseuille flow},
  author={Poddar, A. and Mandal, S. and Bandopadhyay, A. and Chakraborty, S.},
  journal={J. Fluid Mech.},
  volume={870},
  pages={27--66},
  year={2019},
  publisher={Cambridge University Press}
}

@article{barthes1973deformation,
  title={Deformation and burst of a liquid droplet freely suspended in a linear shear field},
  author={Barth\`es-Biesel, D. and Acrivos, A.},
  journal={J. Fluid Mech.},
  volume={61},
  number={1},
  pages={1--22},
  year={1973},
  publisher={Cambridge University Press}
}

@article{zhang2021modeling,
  title={Modeling the deformation of a surfactant-covered droplet under the combined influence of electric field and shear flow},
  author={Zhang, J. and Liu, H. and Zhang, X.},
  journal={Phys. Fluids},
  volume={33},
  number={4},
  year={2021},
  pages={042109},
  publisher={AIP Publishing}
}

@article{han2022surfactant,
  title={Surfactant and dilatational viscosity effects on the deformation of liquid droplets in an electric field},
  author={Han, Y. and Koplik, J. and Maldarelli, C.},
  journal={J. Colloid Interface Sci.},
  volume={607},
  pages={900--911},
  number={1},
  year={2022},
  publisher={Elsevier}
}

@article{cox1969deformation,
  title={The deformation of a drop in a general time-dependent fluid flow},
  author={Cox, R. G.},
  journal={J. Fluid Mech.},
  volume={37},
  number={3},
  pages={601--623},
  year={1969},
  publisher={Cambridge University Press}
}

@article{nawab1958viscosity,
  title={The viscosity of dilute emulsions},
  author={Nawab, M. A. and Mason, S. G.},
  journal={Trans. Faraday Soc.},
  volume={54},
  pages={1712--1723},
  year={1958},
  publisher={Royal Society of Chemistry}
}

@article{bartok1959particle,
  title={Particle motions in sheared suspensions: {VIII}. {S}inglets and doublets of fluid spheres},
  author={Bartok, W. and Mason, S. G.},
  journal={J. Colloid Sci.},
  volume={14},
  number={1},
  pages={13--26},
  year={1959},
  publisher={Elsevier}
}

@article{chaffey1967second,
  title={A second-order theory for shear deformation of drops},
  author={Chaffey, C. E. and Brenner, H.},
  journal={J. Colloid Interface Sci.},
  volume={24},
  number={2},
  pages={258--269},
  year={1967},
  publisher={Elsevier}
}

@article{flumerfelt1980effects,
  title={Effects of dynamic interfacial properties on drop deformation and orientation in shear and extensional flow fields},
  author={Flumerfelt, R. W.},
  journal={J. Colloid Interface Sci.},
  volume={76},
  number={2},
  pages={330--349},
  year={1980},
  publisher={Elsevier}
}

@article{stone1990effects,
  title={The effects of surfactants on drop deformation and breakup},
  author={Stone, H. A. and Leal, L. G.},
  journal={J. Fluid Mech.},
  volume={220},
  pages={161--186},
  year={1990},
  publisher={Cambridge University Press}
}

@article{mandal2016dielectrophoresis,
  title={Dielectrophoresis of a surfactant-laden viscous drop},
  author={Mandal, S. and Bandopadhyay, A. and Chakraborty, S.},
  journal={Phys. Fluids},
  volume={28},
  number={6},
  year={2016},
pages={062006},
  publisher={AIP Publishing}
}

@article{mandal2016effect,
  title={Effect of surfactant on motion and deformation of compound droplets in arbitrary unbounded {S}tokes flows},
  author={Mandal, S. and Ghosh, U. and Chakraborty, S.},
  journal={J. Fluid Mech.},
  volume={803},
  pages={200--249},
  year={2016},
  publisher={Cambridge University Press}
}

@article{mandal2017effect,
  title={Effect of {M}arangoni stress on the bulk rheology of a dilute emulsion of surfactant-laden deformable droplets in linear flows},
  author={Mandal, S. and Das, S. and Chakraborty, S.},
  journal={Phys. Rev. Fluids},
  volume={2},
  number={11},
  pages={113604},
  year={2017},
  publisher={APS}
}

@article{mandal2017influence,
  title={Influence of interfacial viscosity on the dielectrophoresis of drops},
  author={Mandal, S. and Chakraborty, S.},
  journal={Phys. Fluids},
  volume={29},
  number={5},
  year={2017},
pages={052002},
  publisher={AIP Publishing}
}

@article{poddar2019electrorheology,
  title={Electrorheology of a dilute emulsion of surfactant-covered drops},
  author={Poddar, A. and Mandal, S. and Bandopadhyay, A. and Chakraborty, S.},
  journal={J. Fluid Mech.},
  volume={881},
  pages={524--550},
  year={2019},
  publisher={Cambridge University Press}
}

@article{ha1995effects,
  title={Effects of surfactant on the deformation and stability of a drop in a viscous fluid in an electric field},
  author={Ha, J.-W. and Yang, S.-M.},
  journal={J. Colloid Interface Sci.},
  volume={175},
  number={2},
  pages={369--385},
  year={1995},
  publisher={Elsevier}
}

@article{teigen2010influence,
  title={Influence of surfactant on drop deformation in an electric field},
  author={Teigen, K. E. and Munkejord, S. T.},
  journal={Phys. Fluids},
  volume={22},
  number={11},
  year={2010},
pages={112104},
  publisher={AIP Publishing}
}

@article{vlahovska2005deformation,
  title={Deformation of a surfactant-covered drop in a linear flow},
  author={Vlahovska, P. M. and Loewenberg, M. and Blawzdziewicz, J.},
  journal={Phys. Fluids},
  volume={17},
  number={10},
  year={2005},
  publisher={AIP Publishing},
  pages={103103}
}

@article{nganguia2013equilibrium,
  title={Equilibrium electro-deformation of a surfactant-laden viscous drop},
  author={Nganguia, H. and Young, Y.-N. and Vlahovska, P. M. and Blawzdziewicz, J. and Zhang, J. and Lin, H.},
  journal={Phys. Fluids},
  volume={25},
  number={9},
  year={2013},
  pages={092106},
  publisher={AIP Publishing}
}

@article{nganguia2019effects,
  title={Effects of surfactant transport on electrodeformation of a viscous drop},
  author={Nganguia, H. and Pak, O. S. and Young, Y.-N.},
  journal={Phys. Rev. E},
  volume={99},
  number={6},
  pages={063104},
  year={2019},
  publisher={APS}
}

@article{okonski_thacher,
  title={The Distortion of Aerosol Droplets by an Electric Field},
  author={O'Konski, C. T. and Thacher, H. C.},
  journal={J. Phys. Chem.},
  year={1953},
  volume={57},
  pages={955-958}
}

@article{allan1962particle,
  title={Particle behaviour in shear and electric fields {I}. {D}eformation and burst of fluid drops},
  author={Allan, R. S. and Mason, S. G.},
  journal={Proc. R. Soc. Lond. A},
  volume={267},
  number={1328},
  pages={45--61},
  year={1962},
  publisher={The Royal Society London}
}

@article{taylor1966studies,
  title={Studies in electrohydrodynamics. {I}. {T}he circulation produced in a drop by an electric field},
  author={Taylor, G. I.},
  journal={Proc. R. Soc. Lond. A},
  volume={291},
  number={1425},
  pages={159--166},
  year={1966},
  publisher={The Royal Society London}
}

@article{torza1971electrohydrodynamic,
  title={Electrohydrodynamic deformation and bursts of liquid drops},
  author={Torza, S. and Cox, R. G. and Mason, S. G. },
  journal={Philos. Trans. R. Soc. Lond. A},
  volume={269},
  number={1198},
  pages={295--319},
  year={1971},
  publisher={The Royal Society London}
}

@article{ajayi1978note,
  title={A note on {T}aylor’s electrohydrodynamic theory},
  author={Ajayi, O. O.},
  journal={Proc. R. Soc. Lond. A},
  volume={364},
  number={1719},
  pages={499--507},
  year={1978},
  publisher={The Royal Society London}
}

@article{shutov2002shape,
  title={The shape of a drop in a constant electric field},
  author={Shutov, A. A.},
  journal={Tech. Phys.},
  volume={47},
  pages={1501--1508},
  year={2002},
  publisher={Springer}
}

@article{shkadov2002drop,
  title={Drop and bubble deformation in an electric field},
  author={Shkadov, V. Y. and Shutov, A. A.},
  journal={Fluid Dyn.},
  volume={37},
  number={5},
  pages={713--724},
  year={2002},
  publisher={Springer}
}

@article{feng1999electrohydrodynamic,
  title={Electrohydrodynamic behaviour of a drop subjected to a steady uniform electric field at finite electric {R}eynolds number},
  author={Feng, J. Q.},
  journal={Proc. R. Soc. Lond. A},
  volume={455},
  number={1986},
  pages={2245--2269},
  year={1999},
  publisher={The Royal Society}
}

@article{vizika1992electrohydrodynamic,
  title={The electrohydrodynamic deformation of drops suspended in liquids in steady and oscillatory electric fields},
  author={Vizika, O. and Saville, D. A.},
  journal={J. Fluid Mech.},
  volume={239},
  pages={1--21},
  year={1992},
  publisher={Cambridge University Press}
}

@article{lanauze2015nonlinear,
  title={Nonlinear electrohydrodynamics of slightly deformed oblate drops},
  author={Lanauze, J. A. and Walker, L. M. and Khair, A. S.},
  journal={J. Fluid Mech.},
  volume={774},
  pages={245--266},
  year={2015},
  publisher={Cambridge University Press}
}

@article{feng20022d,
  title={A 2{D} electrohydrodynamic model for electrorotation of fluid drops},
  author={Feng, J. Q.},
  journal={J. Colloid Interface Sci.},
  volume={246},
  number={1},
  pages={112--121},
  year={2002},
  publisher={Elsevier}
}

@article{das2017electrohydrodynamics,
  title={Electrohydrodynamics of viscous drops in strong electric fields: numerical simulations},
author={Das, D. and Saintillan, D.},
  journal={J. Fluid Mech.},
  volume={829},
  pages={127--152},
  year={2017},
  publisher={Cambridge University Press}
}

@article{supeene2008deformation,
  title={Deformation of a droplet in an electric field: {N}onlinear transient response in perfect and leaky dielectric media},
  author={Supeene, G. and Koch, C. R. and Bhattacharjee, S.},
  journal={J. Colloid Interface Sci.},
  volume={318},
  number={2},
  pages={463--476},
  year={2008},
  publisher={Elsevier}
}

@article{lopez2011charge,
  title={A charge-conservative approach for simulating electrohydrodynamic two-phase flows using volume-of-fluid},
  author={L{\'o}pez-Herrera, J. M. and Popinet, S. and Herrada, M. A.},
  journal={J. Comput. Phys.},
  volume={230},
  number={5},
  pages={1939--1955},
  year={2011},
  publisher={Elsevier}
}

@article{quincke1896ueber,
  title={Ueber {R}otationen im constanten electrischen {F}elde},
  author={Quincke, G.},
  journal={Ann. Phys. Chem.},
  volume={295},
  number={11},
  pages={417--486},
  year={1896}
}

@article{weiler,
    author = {Weiler, W.},
    title ={Zur darstellung elektrischer kraftlinien} ,
    journal = {Z. Phys. Chem. Unterricht},
    volume={6},
    year = {1893},
    pages={194--195}

}

@article{lac2007axisymmetric,
  title={Axisymmetric deformation and stability of a viscous drop in a steady electric field},
  author={Lac, E. and Homsy, G. M.},
  journal={J. Fluid Mech.},
  volume={590},
  pages={239--264},
  year={2007},
  publisher={Cambridge University Press}
}

@article{jones1984quincke,
  title={Quincke rotation of spheres},
  author={Jones, T. B.},
  journal={IEEE Trans. Ind. Appl.},
  number={4}, 
  volume={IA-20},
  pages={845--849},
  year={1984},
  publisher={IEEE}
}

@article{lampa,
    author = {Lampa, A. S.},
    title = {Dielectric hysteresis},
    journal = {Sitzungberichte d. k. Akademie der Wissenshaften},
    year ={1906},
volume={115},
  pages={1657--1690},
}

@article{bricard2013,
  title={Emergence of macroscopic directed motion in populations of motile colloids},
  author={Bricard, A. and Caussin, J.-B. and Desreumaux, N. and Dauchot, O. and Bartolo, D.},
  journal={Nature},
  volume={503},
  number={7474},
  pages={95--98},
  year={2013},
  publisher={Nature Publishing Group UK London}
}

@article{bricard2015,
  title={Emergent vortices in populations of colloidal rollers},
  author={Bricard, A. and Caussin, J.-B. and Das, D. and Savoie, C. and Chikkadi, V. and Shitara, K. and Chepizhko, O. and Peruani, F. and Saintillan, D. and Bartolo, D.},
  journal={Nat. Commun.},
  volume={6},
  number={1},
  pages={7470},
  year={2015},
  publisher={Nature Publishing Group UK London}
}

@article{das2019active,
  title={Active particles powered by {Q}uincke rotation in a bulk fluid},
  author={Das, D. and Lauga, E.},
  journal={Phys. Rev. Lett.},
  volume={122},
  number={19},
  pages={194503},
  year={2019},
  publisher={APS}
}

@article{han2021low,
  title={Low-{R}eynolds-number, biflagellated {Q}uincke swimmers with multiple forms of motion},
  author={Han, E. and Zhu, L. and Shaevitz, J. W. and Stone, H. A.},
  journal={Proc. Natl. Acad. Sci.},
  volume={118},
  number={29},
  pages={e2022000118},
  year={2021},
  publisher={National Academy of Sciences}
}

@article{das2017nonlinear,
  title={A nonlinear small-deformation theory for transient droplet electrohydrodynamics},
  author={Das, D. and Saintillan, D.},
  journal={J. Fluid Mech.},
  volume={810},
  pages={225--253},
  year={2017},
  publisher={Cambridge University Press}
}

@article{turcu1987electric,
  title={Electric field induced rotation of spheres},
  author={Turcu, I},
  journal={J. Phys. A: Math. Gen.},
  volume={20},
  number={11},
  pages={3301--3307},
  year={1987},
  publisher={IOP Publishing}
}

@article{vlahovska2011rheology,
  title={On the rheology of a dilute emulsion in a uniform electric field},
  author={Vlahovska, P. M.},
  journal={J. Fluid Mech.},
  volume={670},
  pages={481--503},
  year={2011},
  publisher={Cambridge University Press}
}

@book{lamb1924hydrodynamics,
  title={Hydrodynamics},
  author={Lamb, H.},
  year={1932},
  edition={6},
  publisher={Cambridge University Press}
}

@book{kim2013microhydrodynamics,
  title={Microhydrodynamics: principles and selected applications},
  author={Kim, S. and Karrila, S. J.},
  year={1991},
  publisher={Butterworth-Heinemann}
}

@article{pan1997characteristics,
  title={Characteristics of electrorheological responses in an emulsion system},
  author={Pan, X.-D. and McKinley, G. H.},
  journal={J. Colloid Interface Sci.},
  volume={195},
  number={1},
  pages={101--113},
  year={1997},
  publisher={Elsevier}
}

@book{happel2012low,
  title={Low {R}eynolds number hydrodynamics: with special applications to particulate media},
  author={Happel, J. and Brenner, H.},
  year={1983},
  publisher={Martinus Nijhoff}
}

@article{bandopadhyay2016uniform,
  title={Uniform electric-field-induced lateral migration of a sedimenting drop},
  author={Bandopadhyay, A. and Mandal, S. and Kishore, N.K. and Chakraborty, S.},
  journal={J. Fluid Mech.},
  volume={792},
  pages={553--589},
  year={2016},
  publisher={Cambridge University Press}
}

@article{pawar1996marangoni,
  title={Marangoni effects on drop deformation in an extensional flow: {T}he role of surfactant physical chemistry. {I}. {I}nsoluble surfactants},
  author={Pawar, Y. and Stebe, K. J.},
  journal={Phys. Fluids},
  volume={8},
  number={7},
  pages={1738--1751},
  year={1996},
  publisher={American Institute of Physics}
}

@techreport{savic1953circulation,
    author = {Savic, P.},
    title ={Circulation and distortion of liquid drops falling through a viscous medium} ,
    institution ={Division of Mechanical Engineering, National Research Council Canada} ,
    year = {1953},
number={MT-22}
}

@article{davis1966influence,
  title={The influence of surfactants on the creeping motion of bubbles},
  author={Davis, R. E. and Acrivos, A.},
  journal={Chem. Eng. Sci.},
  volume={21},
  number={8},
  pages={681--685},
  year={1966},
  publisher={Elsevier}
}

@article{sadhal1983stokes,
  title={Stokes flow past bubbles and drops partially coated with thin films. {P}art 1. {S}tagnant cap of surfactant film--exact solution},
  author={Sadhal, S. S. and Johnson, R. E.},
  journal={J. Fluid Mech.},
  volume={126},
  pages={237--250},
  year={1983},
  publisher={Cambridge University Press}
}

@article{griffith1962effect,
  title={The effect of surfactants on the terminal velocity of drops and bubbles},
  author={Griffith, R. M.},
  journal={Chem. Eng. Sci.},
  volume={17},
  number={12},
  pages={1057--1070},
  year={1962},
  publisher={Elsevier}
}

@book{levich,
    author = {Levich, V. G.},
    title = {Physiochemical Hydrodynamics},
    publisher = {Prentice-Hall},
    year = {1962}
}

@article{haber1972hydrodynamics,
  title={Hydrodynamics of a drop submerged in an unbounded arbitrary velocity field in the presence of surfactants},
  author={Haber, S. and Hetsroni, G.},
  journal={Appl. Sci. Res.},
  volume={25},
  number={1},
  pages={215--233},
  year={1972},
  publisher={Springer}
}

@article{eow2003motion,
  title={Motion, deformation and break-up of aqueous drops in oils under high electric field strengths},
  author={Eow, J. S. and Ghadiri, M.},
  journal={Chem. Eng. Process.: Process Intensif.},
  volume={42},
  number={4},
  pages={259--272},
  year={2003},
  publisher={Elsevier}
}

@article{taylor1964disintegration,
  title={Disintegration of water drops in an electric field},
  author={Taylor, G. I.},
  journal={Proc. R. Soc. Lond. A},
  volume={280},
  number={1382},
  pages={383--397},
  year={1964},
  publisher={The Royal Society London}
}

@article{sadhal1986deformation,
  title={On the deformation of drops and bubbles with varying interfacial tension},
  author={Sadhal, S. S. and Johnson, R. E.},
  journal={Chem. Eng. Commun.},
  volume={46},
  number={1-3},
  pages={97--109},
  year={1986},
  publisher={Taylor \& Francis}
}

@article{milliken1994influence,
  title={The influence of surfactant on the deformation and breakup of a viscous drop: {T}he effect of surfactant solubility},
  author={Milliken, W. J. and Leal, L. G.},
  journal={J. Colloid Interface Sci.},
  volume={166},
  number={2},
  pages={275--285},
  year={1994},
  publisher={Elsevier}
}

@article{greenspan1977deformation,
  title={On the deformation of a viscous droplet caused by variable surface tension},
  author={Greenspan, H. P. },
  journal={Stud. Appl. Math.},
  volume={57},
  number={1},
  pages={45--58},
  year={1977},
  publisher={Wiley Online Library}
}

@article{milliken1993effect,
  title={The effect of surfactant on the transient motion of {N}ewtonian drops},
  author={Milliken, W. J. and Stone, H. A. and Leal, L. G.},
  journal={Phys. Fluids},
  volume={5},
  number={1},
  pages={69--79},
  year={1993},
  publisher={American Institute of Physics}
}

@article{eggleton1998adsorption,
  title={An adsorption--desorption-controlled surfactant on a deforming droplet},
  author={Eggleton, C. D. and Stebe, K. J.},
  journal={J. Colloid Interface Sci.},
  volume={208},
  number={1},
  pages={68--80},
  year={1998},
  publisher={Elsevier}
}

@article{eggleton1999insoluble,
  title={Insoluble surfactants on a drop in an extensional flow: a generalization of the stagnated surface limit to deforming interfaces},
  author={Eggleton, C. D. and Pawar, Y. P. and Stebe, K. J.},
  journal={J. Fluid Mech.},
  volume={385},
  pages={79--99},
  year={1999},
  publisher={Cambridge University Press}
}

@article{he2013electrorotation,
  title={Electrorotation of a viscous droplet in a uniform direct current electric field},
  author={He, H. and Salipante, P. F. and Vlahovska, P. M.},
  journal={Phys. Fluids},
  volume={25},
  number={3},
  year={2013},
pages={032106},
  publisher={AIP Publishing}
}

@article{das2021three,
  title={A three-dimensional small-deformation theory for electrohydrodynamics of dielectric drops},
  author={Das, D. and Saintillan, D.},
  journal={J. Fluid Mech.},
  volume={914},
  pages={A22},
  year={2021},
  publisher={Cambridge University Press}
}

@article{basu2024role,
  title={Role of surface charge convection on oblate droplets in different conductivity regimes},
  author={Basu, H. S. and Jena, S. K. and Kondaraju, S.},
  journal={Phys. Fluids},
  volume={36},
  number={10},
  year={2024},
pages={102125},
  publisher={AIP Publishing}
}

@article{dong2023unsteady,
  title={Unsteady electrorotation of a viscous drop in a uniform electric field},
  author={Dong, Q. and Sau, A.},
  journal={Phys. Fluids},
  volume={35},
  number={4},
  year={2023},
  publisher={AIP Publishing},
pages={047116}
}

@article{sherwood1988breakup,
  title={Breakup of fluid droplets in electric and magnetic fields},
  author={Sherwood, J. D.},
  journal={J. Fluid Mech.},
  volume={188},
  pages={133--146},
  year={1988},
  publisher={Cambridge University Press}
}

@article{allan1961effects,
  title={Effects of electric fields on coalescence in liquid + liquid systems},
  author={Allan, R. S. and Mason, S. G.},
  journal={Trans. Faraday Soc.},
  volume={57},
  pages={2027--2040},
  year={1961},
  publisher={Royal Society of Chemistry}
}

@article{eow2003drop,
  title={Drop--drop coalescence in an electric field: the effects of applied electric field and electrode geometry},
  author={Eow, J. S. and Ghadiri, M.},
  journal={Colloids Surf. A: Physicochem. Eng. Asp.},
  volume={219},
  number={1-3},
  pages={253--279},
  year={2003},
  publisher={Elsevier}
}

@article{ristenpart2009non,
  title={Non-coalescence of oppositely charged drops},
  author={Ristenpart, W. D. and Bird, J. C. and Belmonte, A. and Dollar, F. and Stone, H. A.},
  journal={Nature},
  volume={461},
  number={7262},
  pages={377--380},
  year={2009},
  publisher={Nature Publishing Group UK London}
}

@article{baygents1991electrophoresis,
  title={Electrophoresis of drops and bubbles},
  author={Baygents, J. C. and Saville, D. A.},
  journal={J. Chem. Soc. Faraday Trans.},
  volume={87},
  number={12},
  pages={1883--1898},
  year={1991},
  publisher={The Royal Society of Chemistry}
}

@article{firouznia2023spectral,
  title={A spectral boundary integral method for simulating electrohydrodynamic flows in viscous drops},
  author={Firouznia, M. and Bryngelson, S. H. and Saintillan, D.},
  journal={J. Comput. Phys.},
  volume={489},
  pages={112248},
  year={2023},
  publisher={Elsevier}
}

@article{peng2024equatorial,
  title={Equatorial blowup and polar caps in drop electrohydrodynamics},
  author={Peng, G. G. and Brand{\~a}o, R. and Yariv, E. and Schnitzer, O.},
  journal={Phys. Rev. Fluids},
  volume={9},
  number={8},
  pages={083701},
  year={2024},
  publisher={APS}
}

@article{salipante2010electrohydrodynamics,
  title={Electrohydrodynamics of drops in strong uniform dc electric fields},
  author={Salipante, P. F. and Vlahovska, P. M.},
  journal={Phys. Fluids},
  volume={22},
  number={11},
  year={2010},
pages={112110},
  publisher={AIP Publishing}
}

@article{salipante2013electrohydrodynamic,
  title={Electrohydrodynamic rotations of a viscous droplet},
  author={Salipante, P. F. and Vlahovska, P. M.},
  journal={Phys. Rev. E},
  volume={88},
  number={4},
  pages={043003},
  year={2013},
  publisher={APS}
}

@article{sorgentone20193d,
  title={A 3{D} boundary integral method for the electrohydrodynamics of surfactant-covered drops},
  author={Sorgentone, C. and Tornberg, A.-K. and Vlahovska, P. M.},
  journal={J. Comput. Phys.},
  volume={389},
  pages={111--127},
  year={2019},
  publisher={Elsevier}
}

@article{stone1990simple,
  title={A simple derivation of the time-dependent convective-diffusion equation for surfactant transport along a deforming interface},
  author={Stone, H. A.},
  journal={Phys. Fluids},
  volume={2},
  number={1},
  pages={111--112},
  year={1990}
}

@article{peng2025bistability,
  title={Bistability and charge-density blowup in the onset of drop {Q}uincke rotation},
  author={Peng, G. G. and Schnitzer, O.},
  journal={Phys. Rev. Fluids},
  volume={10},
  number={8},
  pages={081701},
  year={2025},
  publisher={APS}
}

@article{zhang2021quincke,
  title={Quincke oscillations of colloids at planar electrodes},
  author={Zhang, Z. and Yuan, H. and Dou, Y. and De La Cruz, M. O. and Bishop, K. J. M.},
  journal={Phys. Rev. Lett.},
  volume={126},
  number={25},
  pages={258001},
  year={2021},
  publisher={APS}
}

@article{mauleon2023dynamics,
  title={Dynamics and interactions of Quincke roller clusters: {F}rom orbits and flips to excited states},
  author={Mauleon-Amieva, A. and Allen, M. P. and Liverpool, T. B. and Royall, C. P.},
  journal={Sci. Adv.},
  volume={9},
  number={20},
  pages={eadf5144},
  year={2023},
  publisher={American Association for the Advancement of Science}
}

@article{reyes2023magnetic,
  title={Magnetic {Q}uincke rollers with tunable single-particle dynamics and collective states},
  author={Reyes Garza, R. and Kyriakopoulos, N. and Cenev, Z. M. and Rigoni, C. and Timonen, J. V. I.},
  journal={Sci. Adv.},
  volume={9},
  number={26},
  pages={eadh2522},
  year={2023},
  publisher={American Association for the Advancement of Science}
}

@article{fitzgerald2025rolling,
  title={Rolling at right angles: magnetic anisotropy enables dual-anisotropic active matter.},
  author={Fitzgerald, E. and Clavaud, C. and Das, D. and Lenton, I. C. D. and Waitukaitis, S. R.},
  note={{\it arXiv preprint arXiv:2508.05643}},
  year={2025}
}

@article{raju2021diversity,
  title={Diversity of non-equilibrium patterns and emergence of activity in confined electrohydrodynamically driven liquids},
  author={Raju, G. and Kyriakopoulos, N. and Timonen, J. V. I.},
  journal={Sci. Adv.},
  volume={7},
  number={38},
  pages={eabh1642},
  year={2021},
  publisher={American Association for the Advancement of Science}
}

@book{rosen2012surfactants,
  title={Surfactants and interfacial phenomena},
  author={Rosen, M. J. and Kunjappu, J. T.},
  year={2012},
  publisher={John Wiley \& Sons}
}

@article{manikantan2020surfactant,
  title={Surfactant dynamics: hidden variables controlling fluid flows},
  author={Manikantan, H. and Squires, T. M.},
  journal={J. Fluid Mech.},
  volume={892},
  pages={P1},
  year={2020},
  publisher={Cambridge University Press}
}

@article{das2023absence,
  title={On the absence of collective motion in a bulk suspension of spontaneously rotating dielectric particles},
  author={Das, D. and Saintillan, D.},
  journal={Soft Matter},
  volume={19},
  number={35},
  pages={6825--6837},
  year={2023},
  publisher={Royal Society of Chemistry}
}

@article{mcdougall2025nonlinear,
  title={Nonlinear Three-Dimensional Electrohydrodynamic Interactions of Viscous Dielectric Drops.},
  author={McDougall, M. A. and Wilson, S. K. and Das, D.},
  note={{\it arXiv preprint arXiv:2505.10986}},
  year={2025}
}

@article{nganguia2023influence,
  title={Influence of surface viscosities on the electrodeformation of a prolate viscous drop},
  author={Nganguia, H. and Das, D. and Pak, O. S. and Young, Y.-N.},
  journal={Soft Matter},
  volume={19},
  number={4},
  pages={776--789},
  year={2023},
  publisher={Royal Society of Chemistry}
}

@article{scriven1960dynamics,
  title={Dynamics of a fluid interface equation of motion for Newtonian surface fluids},
  author={Scriven, L. E.},
  journal={Chem. Eng. Sci.},
  volume={12},
  number={2},
  pages={98--108},
  year={1960},
  publisher={Elsevier}
}

@article{astarita1979objective,
  title={Objective and generally applicable criteria for flow classification},
  author={Astarita, G.},
  journal={J. Non-Newton. Fluid Mech.},
  volume={6},
  number={1},
  pages={69--76},
  year={1979},
  publisher={Elsevier}
}

@article{oliveira2009purely,
  title={Purely elastic flow asymmetries in flow-focusing devices},
  author={Oliveira, M. S. N. and Pinho, F. T. and Poole, R. J. and Oliveira, P. J. and Alves, M. A.},
  journal={J. Non-Newton. Fluid Mech.},
  volume={160},
  number={1},
  pages={31--39},
  year={2009},
  publisher={Elsevier}
}

@article{poole2023inelastic,
  title={Inelastic and flow-type parameter models for non-Newtonian fluids},
  author={Poole, R. J.},
  journal={J. Non-Newton. Fluid Mech.},
  volume={320},
  pages={105106},
  year={2023},
  publisher={Elsevier}
}

\end{document}